\documentclass[authoryear,3p]{elsarticle}
\usepackage{lipsum}
\def\ps@pprintTitle{%
	\let\@oddhead\@empty
	\let\@evenhead\@empty
	\def\@oddfoot{}%
	\let\@evenfoot\@oddfoot}
\makeatother
\usepackage{float}
\usepackage{graphicx}
\usepackage{epsfig}
\usepackage{epstopdf}
\usepackage{amssymb}
\usepackage{amsthm}
\usepackage{framed}
\usepackage{multirow}
\usepackage{amsmath}
\usepackage{setspace}
\usepackage[justification=centering]{caption}
\usepackage{subcaption}
\usepackage{xcolor}
\usepackage{soul}
\usepackage{lineno}
\usepackage{stmaryrd}
\usepackage[colorlinks=true,citecolor=black,linkcolor=black]{hyperref}

\date{\today}

\newcommand{\subsubsubsection}[1]{\paragraph{#1}\mbox{}\\} 

\begin{document}
	
	\title{Elastic-gap free strain gradient crystal plasticity model that effectively account for plastic slip gradient and grain boundary dissipation}
	\author{Anjan Mukherjee}
	\author[mymainaddress]{Biswanath Banerjee\corref{mycorrespondingauthor}}
	\cortext[mycorrespondingauthor]{Corresponding author}
	\ead{biswanath@civil.iitkgp.ac.in}
	\address[mymainaddress]{Department of Civil Engineering, Indian Institute of Technology, Kharagpur 721302, West Bengal, India}
	\begin{abstract}
		This paper proposes an elastic-gap free strain gradient crystal plasticity model that addresses dissipation caused by plastic slip gradient and grain boundary (GB) Burger tensor. The model involves splitting plastic slip gradient and GB Burger tensor into energetic dissipative quantities. Unlike conventional models, the bulk and GB defect energy are considered to be a quadratic functional of the energetic portion of slip gradient and GB Burgers tensor. The higher-order stresses for each individual slip systems and GB stresses are derived from the defect energy, following a similar evolution as the Armstrong-Frederick type backstress model in classical plasticity. The evolution equations consist of a hardening and a relaxation term. The relaxation term brings the nonlinearity in hardening and causes an additional dissipation. The applicability of the proposed model is numerically established with the help of two-dimensional finite element implementation. Specifically, the bulk and GB relaxation coefficients are critically evaluated based on various circumstances, considering single crystal infinite shear layer, periodic bicrystal shearing, and bicrystal tension problem. In contrast to the Gurtin-type model, the proposed model smoothly captures the apparent strengthening at saturation without causing any abrupt stress jump under non-proportional loading conditions. Moreover, when subjected to cyclic loading, the stress-strain curve maintains its curvature during reverse loading. The numerical simulation reveals that the movement of geometrically necessary dislocation (GND) towards the GB is influenced by the bulk recovery coefficient, while the dissipation and amount of accumulation of GND near the GB are controlled by the GB recovery coefficient. 
	\end{abstract}
	\maketitle
	\section{Introduction}
	The size effect of microstructured material has been investigated in the last few decades, as they are extensively used in micro-electronic or microstructure-oriented high-performance metallic materials. At the size range of a few micrometers to a few hundreds nanometers, these crystalline materials show the smaller being stronger type response. With the size reduction, nonuniform plastic strain develops inside the body, which leads to formation of geometrically necessary dislocation (GND) \citep{nye1953}. An additional strengthening is observed when these GND movement is hindered. Different small-scale experiments shows similar strengthening; for example, nanoindentation \citep{ABUALRUB2004,Pharr2010}, micro-bending \citep{KREINS2021142027}, micro-torsion \citep{fleck1994,LIU2012406}, micro-compression \citep{dimiduk2005size}, combined torsion and bending \citep{zhang_pnas}  etc. 
	
	In recent years, significant focus has been placed on strain gradient plasticity (SGP) and physics-driven strain gradient crystal plasticity (SGCP) in order to address size effects. These developments involve introducing a characteristic length scale into their constitutive equations to explain the size effect observed. \cite{aifantis1987physics} was the first to include length scale within the continuum plasticity framework. Inspired by this seminal work, many SGP theories are  proposed \citep{muhlhaus1991,fleck1994,nix1998indentation,gao1999mechanism,fleck2001,haque2003strain,gurtin2004gradient,gudmundson2004,gurtin2005,lele2008small,gurtin2009thermodynamics,kuroda2010alternative,voyiadjis2012thermo,fleck2014strain,bardella2015modelling,mukherjee2023strain}. Most of the SGP models can be showed to be related to the micromorphic approach \citep{forest2009micromorphic}. A comprehensive review of different SGP theories can be found in \cite{voyiadjis2019strain}.
	
	In the realm of SGCP theory, the pioneering work by \cite{shu1999strain}, a crystal plasticity extension of the phenomenological SGP theory proposed by \cite{hutchinson1997strain}, in elucidating size effects has diverted attention toward the advancement of SGCP theory. Several lower-order theories have emerged in response, eliminating the need for additional boundary conditions. These lower order theories also account for additional hardening attributed to GND formation alongside statistically stored dislocations (SSD) \citep{acharya2000lattice,bassani2001incompatibility,ohashi2005crystal,dunne2007lengthscale}. The mechanism based SGCP theory \cite{han2005mechanism,han2005mechanism_ii}, based on Taylor-hardening model, also belongs to this class. 
	
	Higher-order SGCP theories introduce additional boundary conditions concerning crystallographic plastic slip. These theories can generally be sub-categorized into two types: work-conjugate and non-work conjugate type theories \citep{kuroda2008formulations}. The work-conjugate theories considers higher-order stresses that are inherently related to the plastic slip gradient.
	\cite{gurtin2000plasticity,gurtin2002} introduced a work-conjugate type single crystal plasticity theory by considering the power expenditure and free energy augmentation caused by the plastic slip gradient. This addition necessitates additional microforce balance and boundary conditions for each slip system. Building on this framework, \cite{gurtin2007gradient} expanded the preceding single crystal plasticity theory by incorporating dissipation due to plastic slip gradient, alongside free energy changes due to plastic slip.
	Subsequently, various higher-order work-conjugate type theories have been proposed, sharing similar power expenditure characteristics with the Gurtin-type model but featuring different constitutive relations for evaluating higher-order stress quantities \citep{borg2007strain_a,yalcinkaya2011deformation,yalccinkaya2012non,klusemann2013plastic,anand2015stored,nellemann2017incremental,jebahi2020strain}. Additionally, \cite{wulfinghoff2012equivalent} proposed a scalar-based SGCP theory wherein the higher-order stress quantities are conjugate to the equivalent plastic slip gradient.  
	\cite{niordson2014computational} considered energetic version of the Gurtin's theory \citep{gurtin2007gradient} and dissipative theory due to \cite{borg2007strain_b} to suggest a finite element solution based on the minimization technique as derived in  \cite{fleck2009,fleck2009mathematical}. 
	
	The non-work-conjugate type theories do not consider additional higher-order stress and employ the conventional virtual work equation. However, the boundary condition on plastic slip can be imposed with the help of an additional partial differential equation relating GND density with plastic slip gradient. This class of theory considers a backstress that depends on the gradient of  GND density, which influences the evolution of plastic slip. Beginning with \cite{arsenlis2004evolution,evers2004non}, substantial investigation into this type of theory has been conducted, as evidenced by works such as \cite{yefimov2004comparison,bayley2006comparison,geers2006second,kuroda2008finite,geers2014coupled,wulfinghoff2015gradient,ELNAAMAN201976}. \cite{kuroda2006studies,kuroda2008formulations} established the equivalence between these two types of theories despite their distinct origins and mathematical formulations.
	
	The distribution of dislocations and their interactions with nearby crystals significantly impact the plastic behavior of polycrystalline materials. This interaction can happen in several ways at the microscopic level, such as by creating new dislocations, moving dislocations, or absorbing and releasing dislocations at the grain boundary \citep{shen1986dislocation}. In a classical experiment by  \citep{hall1951deformation,petch1953cleavage}, an additional material strengthening was observed due to grain boundary presence (known as the Hall-Petch effect). In later experiments, it is observed that during the low strain value, the dislocation makes a pile up close to the grain boundary \citep{shen1988dislocation,feaugas1999origin,feaugas2003grain}. At the higher strain values, the grain boundary behaves as a sink, which dissipates away the pile-ups and causes uniform grain level dislocation density \citep{sun1998mesoscale,sun2000observations}. In nano-indentation experiment by \citep{soer2005detection,wang2004indentation,britton2009nanoindentation} have observed a second strain burst (grain boundary burst) and an increase in hardness as the indenter approach close to the grain boundary. This grain boundary burst is considered due to grain boundary yielding  \citep{aifantis2007modeling,wang2004indentation}. When the indenter is near the grain boundary, the excess energy before the burst is considered free energy, and excess energy during the burst is considered grain boundary dissipation (for detail, refer to Fig. 2 of \cite{voyiadjis2014theory}). The study by  \citep{kalidindi2014mechanical} can be referred to for a comprehensive review of grain boundary characterization utilizing nano-indentation experiments. These experimental studies suggest an energetic and dissipative contribution of grain boundary, which need to be accounted for while modeling the grain boundary.
	
	The simplest approach to account for the grain boundary effect is to fully restrict it to dislocation motion \citep{gurtin2002,borg2007strain_b,wulfinghoff2012equivalent}. Such a boundary is termed a micro hard boundary \citep{gurtin2005boundary}. Conversely, the micro-free condition indicates that plastic slip occurs with complete unconstraint, resulting in zero micro traction. Over the past few decades, significant attention has been directed towards modeling the grain boundary under conditions ranging from micro-free to micro-hard. \cite{WULFINGHOFF201333} introduced a grain boundary model incorporating grain boundary strength, based on the equivalent plastic slip theory \citep{wulfinghoff2012equivalent}. This model enables the prediction of plastic slip initiation in the grain boundary, along with proposing a corresponding flow rule. Subsequently, \cite{bayerschen2015equivalent} further enhanced the model by considering grain boundary hardening. Additionally, \cite{erdle2017gradient} incorporated an equivalent plastic slip discontinuity across grain boundaries into the previous model. \cite{wulfinghoff2017generalized} proposed an interface model based on surface-related considerations. Their model can accommodate grain boundary isotropic and kinematic hardening. Furthermore, \cite{alipour2019grain} presented a similar grain boundary model tailored for the finite strain regime, while \cite{alipour2020grain} estimated the grain boundary strength of the previous model by incorporating grain misorientation. An excellent review for different model from slip transfer criteria can be found in \cite{bayerschen2016review}.
	
	Most of the previously mentioned grain boundary models, except \cite{erdle2017gradient}, have typically assumed a continuous equivalent slip profile, which contrasts with experimental observations of discontinuous plastic slip across grain boundaries \citep{lee1990tem}. The models presented by Gurtin \citep{gurtin2008theory} have the capability to include grain misorientation and grain boundary orientation. It can also address grain boundary slip discontinuity for individual slip systems using a grain boundary Burgers tensor. This theory allows for the modeling of grain boundary strengthening (due to dissipation) and hardening (due to free energy accumulation). The \cite{gurtin2008theory} model has been applied to various two-dimensional \citep{ozdemir2014modeling} and three-dimensional \citep{gottschalk2016computational,mcbride2016computational,yalccinkaya2021misorientation} problems, elucidating key aspects of the model. Subsequently, \cite{van2013grain} proposed a similar grain boundary model, capable of modeling both GB hardening and dissipation. Recent investigations by \cite{erdle2023analytical} indicate that models akin to \cite{gurtin2008finite} may struggle to maintain single crystal consistency when adjacent grains possess similar slip systems.
	
	The splitting of higher-order stress quantities into energetic and dissipative components is a key aspect of SCGP theories of the work conjugate type \citep{gurtin2007gradient}. The energetic part is associated to the free energy formation and the dissipative part participates in dissipation due to higher order kinematic variables (slip gradient or plastic strain gradient). However, the dissipative higher-order stress brings a constitutive singularity during certain non-proportional loading conditions 
	\citep{fleck2014strain}, also known as \emph{elastic-gap} in literature. Recent discrete dislocation studies suggest that no such elastic-gap exists during non-proportional loading history \citep{amouzou2023elastic}. One way to circumvent this singularity is to disregard the dissipative higher-order stress, which signifies disregarding higher-order dissipation due to GND density. However, \cite{fleck2009mathematical} argued that during the experiment, the core energy storage due to dislocation is much less 
	than the dissipation due to dislocation motion. On the other hand, \cite{gurtin2000plasticity,gurtin2002} argued that the plastic slip gradient mainly participates in free energy formation. In order to accurately represent the size effect, it is crucial to take into account both energetic and dissipative parts, as emphasized in \cite{bardella2006deformation}. Moreover, due to experimental limitations, the participation of these slip gradients in energetic or dissipative processes is an open issue. Therefore, it is essential to develop a theory that can incorporate both the energetic and dissipative components of the slip gradient, without any presence of an elastic-gap.   
	
	\cite{panteghini2019potential} pioneered an elastic-gap free model within the distortion gradient plasticity framework by decomposing plastic strain gradient into energetic and dissipative components. 
	\cite{jebahi2023alternative} proposed an elastic-gap free model within the SGCP framework by similar  splitting of the plastic slip gradient into energetic and dissipative parts. Their higher-order stress quantities were derived from a defect energy. Notably, when employing a quadratic defect energy, a two-step plastic behavior was observed. In the case of a non-quadratic defect energy, a reversal in curvature during reverse loading was observed. Recently, \cite{mukherjee2024elastic} introduced an elastic-gap free model within the SGP framework by employing a similar split of the plastic strain gradient.

	In this study, we propose an elastic-gap free strain gradient crystal plasticity model that addresses dissipation caused by each plastic slip gradient and grain boundary Burger tensor. The model involves splitting plastic slip gradient and GB Burger tensor into energetic dissipative quantities. The determination of the higher-order stress within the grain is based on a quadratic potential that takes into account the energetic plastic slip gradient. On the other hand, the energetic stress at the grain boundary is determined by the quadratic potential of the energetic Burger tensor. The higher-order stress for each individual slip systems and GB stresses are obtained by minimizing a dissipative functional using necessary thermodynamic constraints. As a result, the higher order stress and GB stress exhibit nonlinearity and saturation, following a time evolution similar to the Armstrong-Frederick type backstress model \citep{armstrong1966mathematical,chaboche1986time}. The evolution equations consist of a hardening term and a relaxation term. The presence of the relaxation term introduces nonlinearity in the hardening mechanism and gives rise to an extra dissipation, which is consistent with the experimental findings \citep{lim2011simulation}. 
	
	The numerical investigation of an infinite shear layer, conducted through finite element analysis, has confirmed that the suggested model does not exhibit stress jump at the onset of yield or under nonproportional loading conditions. However, the enhanced yield stress can be accurately represented without any abrupt jumps caused by additional higher-order dissipation. In the case of 
	cyclic loading condition, the stress-strain curve maintains its curvature during unloading. The evaluation of the proposed GB model involves an analysis of the shearing of a single-slip periodic bicrystal. The critical assessment focuses on different combinations of bulk and GB relaxation coefficients to highlight the significant roles these coefficients play in controlling the movement of GND between two grains. The numerical simulation demonstrates that the bulk relaxation coefficient influences the movement of GND towards the GB, while the GB relaxation coefficient governs the dissipation and accumulation of GND near the GB. Finally, an example of bicrystal tension is presented, considering a double slip crystal with grain boundary nonlinearity.
	
	Following this introduction, the paper is structured as follows: In Section \ref{gurtin}, a concise overview of the existing Gurtin-type single crystal plasticity model with grain boundary considerations is provided. The derivation of the proposed bulk model is outlined in Section \ref{proposal}, while Section \ref{model_GB} is dedicated to the proposed GB model. The two-dimensional version of the proposed model, along with its finite element implementation, is presented in Section \ref{fem}. To ensure comprehensiveness, a detailed matrix formulation is provided in an \ref{2dimplement}. Section \ref{Result} delves into the discussion of the obtained results and provides accompanying analysis. Lastly, Section \ref{conculsion} serves as the concluding section for this investigation.

	\section{Gurtin type SGCP theory considering grain boundary contribution}
	\label{gurtin}
	In this section, we briefly describe the strain gradient single crystal plasticity theory proposed by \citep{gurtin2007gradient} and the \cite{gurtin2008theory} type grain boundary model to incorporate the grain boundary orientation and grain misorientation, as these theories are building blocks for our proposed model.  
	\subsection{Single  crystal Kinematics} 
	Let $\boldsymbol{u}$ be the displacement of a material point $\boldsymbol{x}$ inside body $\Omega$ with a boundary $\Gamma$. In the case of small deformation consideration, the displacement gradient can be additively split into elastic and plastic distortion as follows: 
	\begin{equation}
		\boldsymbol{\nabla} \boldsymbol{u}=\boldsymbol{H}^e + \boldsymbol{H}^p.
	\end{equation}
	$\boldsymbol{H}^e$ is the elastic distortion due to elastic stretching and rotation, and $\boldsymbol{H}^p$ is the plastic distortion associated with plastic flow. The elastic distortion can be written considering the symmetric and anti-symmetric  part as the sum of elastic strain ($\boldsymbol{\varepsilon}^e$) and elastic rotation ($\boldsymbol{\omega}^e$)  as follows: 
	\begin{equation}
		\boldsymbol{H}^e = \boldsymbol{\varepsilon}^e+\boldsymbol{\omega}^e, \quad \text{with} \quad \boldsymbol{\varepsilon}^e = \frac{1}{2} \left[\boldsymbol{H}^e + {\boldsymbol{H}^e}^T\right] \qquad \text{and} \qquad  \boldsymbol{\omega}^e = \frac{1}{2} \left[\boldsymbol{H}^e - {\boldsymbol{H}^e}^T\right].
	\end{equation}
	In the single crystal plasticity, it is prevalent to consider that the plastic flow is governed by plastic slip $\gamma^\alpha$ of a prescribed slip system $\alpha$ with slip direction $\boldsymbol{s}^\alpha$ and slip normal $\boldsymbol{m}^\alpha$. The plastic distortion and plastic strain tensor can be written considering all the slip systems as follows: 
	\begin{equation}
		\label{hp}
		\boldsymbol{H}^p= \sum_\alpha \gamma^\alpha \boldsymbol{s}^\alpha \otimes \boldsymbol{m}^\alpha, \qquad \boldsymbol{\varepsilon}^p= \frac{1}{2} \left[ \boldsymbol{H}^p + {\boldsymbol{H}^p}^T\right] = \sum_\alpha \frac{1}{2}\gamma^\alpha (\boldsymbol{s}^\alpha \otimes \boldsymbol{m}^\alpha + \boldsymbol{m}^\alpha\otimes \boldsymbol{s}^\alpha) =\sum_\alpha\gamma^\alpha \boldsymbol{T}^\alpha.
	\end{equation} 
	The tensor $ \boldsymbol{s}^\alpha \otimes \boldsymbol{m}^\alpha = \mathcal{S}^\alpha$ is known as the Schmid tensor, and $\boldsymbol{T}^\alpha$ is the symmetrized Schmid tensor to the slip system $\alpha$. Basic rate-like descriptors $\dot{\boldsymbol{u}}$, $\dot{\boldsymbol{\varepsilon}}^e$, $\dot{\boldsymbol{\omega}}^e$, $\dot{\gamma}^\alpha$ are not independent, and they are constrained by the following relation
	\begin{equation}
		\label{const1}
		\boldsymbol{\nabla} \dot{\boldsymbol{u}} = \dot{\boldsymbol{\varepsilon}}^e  + \dot{\boldsymbol{\omega}}^e + \sum_\alpha \dot{\gamma}^\alpha  \mathcal{S}^\alpha.
	\end{equation}
	\subsection{Grain boundary kinematics} 
	\begin{figure}[h!]
		\centering
		\includegraphics[width=0.3\textwidth]{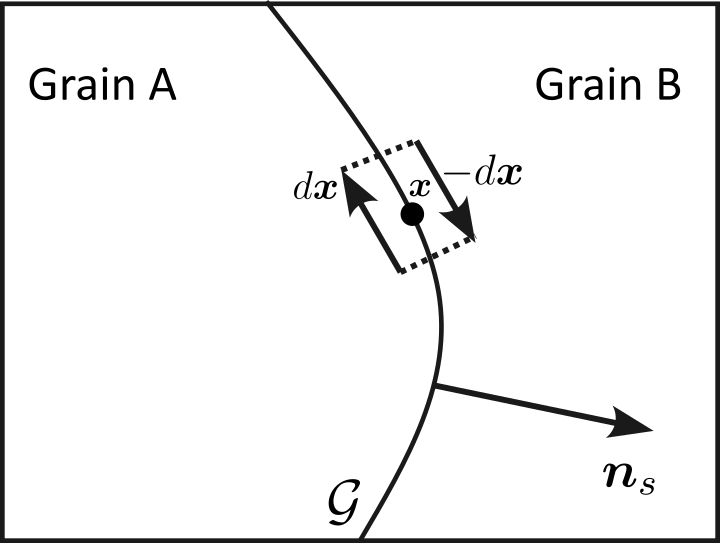}
		\caption{Grain $A$ and $B$ with their  interface $\mathcal{G}$, and the infinitesimal circuit}
		\label{grain}
	\end{figure} 
	Consider two grains $A$ and $B$ separate at the interface $\mathcal{G}$, and the normal to this interface is $\boldsymbol{n}_s$, directed from grain $A$ to grain $B$. Consider a point $\boldsymbol{x}$ at the boundary, and an infinitesimal close circuit $d \Gamma$ is formed around the point. The circuit have $d\boldsymbol{x}$ at the grain $A$ and $-d\boldsymbol{x}$ at grain $B$. Now, denoting jump in any field variable as $\llbracket \left( \right)  \rrbracket = \left( \right) _B- \left( \right) _A$ across the interface, the Burgers vector associated with the close circuit $d\Gamma$ can be written as 
	\begin{equation}
		\boldsymbol{b}(d\Gamma) = \int_{d\Gamma} \boldsymbol{H}^p \cdot d\boldsymbol{x} = \boldsymbol{H}^p_A \cdot d\boldsymbol{x} - \boldsymbol{H}^p_B \cdot d\boldsymbol{x} = - \llbracket \boldsymbol{H}^p \rrbracket \cdot  d\boldsymbol{x}.
	\end{equation}
	The $d\boldsymbol{x}$ is tangential to the $\mathcal{G}$ at point $\boldsymbol{x}$, therefore the Burgers vector can be written as 
	\begin{equation}
		\boldsymbol{b}(d\Gamma) = -\llbracket \boldsymbol{H}^p \rrbracket \cdot[ (\boldsymbol{I}-\boldsymbol{n}_s \otimes \boldsymbol{n}_s) \cdot d\boldsymbol{x}]=(\llbracket\boldsymbol{H}^p \rrbracket (\boldsymbol{n}_s \times))\cdot(\boldsymbol{n}_s\times d\boldsymbol{x}),
	\end{equation}
	where $(\boldsymbol{n}_s \times )$ is a second-order tensor which is defined with permutation symbol $\varepsilon_{ijk}$   as $(\boldsymbol{n}_s \times)_{ij}=\varepsilon_{ikj}n_k$. The $\boldsymbol{n}_s \times d\boldsymbol{x}$ represent an area element within the close circuit, having a direction tangent to the interface and orthogonal to the plane spanned by $\boldsymbol{n}_s$ and $d\boldsymbol{x}$. The tensor $\llbracket\boldsymbol{H}^p \rrbracket (\boldsymbol{n}_s \times)$ can be identified as Burgers vector density. This tensor is defined as grain boundary Burgers tensor  (GB Burgers tensor), and it can be written using equation (\ref{hp}) as follows: 
	\begin{equation}
		\boldsymbol{G}_s = \llbracket \boldsymbol{H}^p \rrbracket (\boldsymbol{n}_s \times) = \boldsymbol{H}^p_B(\boldsymbol{n}_s \times)-\boldsymbol{H}^p_A(\boldsymbol{n}_s \times) = \sum_\alpha[ \gamma^\alpha_B \boldsymbol{s}^\alpha_B  \otimes (\boldsymbol{m}^\alpha_B \times \boldsymbol{n}_s) - \gamma^\alpha_A \boldsymbol{s}^\alpha_A  \otimes (\boldsymbol{m}^\alpha_A \times \boldsymbol{n}_s)].
	\end{equation}
	The GB Burgers tensor and its rate can be written with the Schmid orientation tensor as follows: 
	\begin{equation}
		\label{g_dot}
		\boldsymbol{G}_s= \sum_\alpha[ \gamma^\alpha_B \boldsymbol{N}^\alpha_B -\gamma^\alpha_A \boldsymbol{N}^\alpha_A ], \quad \text{and} \quad \dot{\boldsymbol{G}}_s = \sum_\alpha[ \dot{\gamma}^\alpha_B \boldsymbol{N}^\alpha_B -\dot{\gamma}^\alpha_A \boldsymbol{N}^\alpha_A ],
	\end{equation}
	where $\boldsymbol{N}^\alpha_I = \boldsymbol{s}^\alpha_I  \otimes (\boldsymbol{m}^\alpha_I \times \boldsymbol{n}_s),~ I= A ~\text{or}~B$ is the Schmid orientation tensor that accounts for the grain boundary orientation. In order to find the scalar measure of the $\dot{\boldsymbol{G}}$, we considered the following quantity similar to \cite{gurtin2008theory}
	\begin{equation}
		|\dot{\boldsymbol{G}}_s|^2 = \dot{\boldsymbol{G}}_s : \dot{\boldsymbol{G}}_s = \sum_\alpha[ \dot{\gamma}^\alpha_B \boldsymbol{N}^\alpha_B :  \dot{\boldsymbol{G}}_s -\dot{\gamma}^\alpha_A \boldsymbol{N}^\alpha_A :  \dot{\boldsymbol{G}}_s] = \sum_{\alpha,\beta} C_{BB}^{\alpha\beta} \dot{\gamma}^\alpha_B \dot{\gamma}^\beta_B - 2C_{AB}^{\alpha\beta} \dot{\gamma}^\alpha_A \dot{\gamma}^\beta_B +C_{AA}^{\alpha\beta} \dot{\gamma}^\alpha_A \dot{\gamma}^\beta_A,
	\end{equation}
	where the slip interaction moduli $C^{\alpha\beta}_{IJ} = \boldsymbol{N}_I^\alpha : \boldsymbol{N}_J^\beta$.
	\subsection{Virtual power statement}
	In the SGCP framework, the elastic strain rates ($\dot{\boldsymbol{\varepsilon}}^e$), plastic slip rates ($\dot{\gamma}^\alpha$), and gradient of plastic slip ($\dot{\boldsymbol{\kappa}}^\alpha = \boldsymbol{\nabla} \dot{\gamma}^\alpha$) are involved in the internal virtual power density for the bulk region. The Cauchy stress $\boldsymbol{\sigma}$, power conjugates of $\boldsymbol{\nabla} \dot{\boldsymbol{u}}$, scalar microscopic stress $\pi^\alpha$, power conjugates of $\dot{\gamma}^\alpha$, and vector microscopic stress $\boldsymbol{\xi}^\alpha$,  power conjugates of plastic slip gradient $\dot{\boldsymbol{\kappa}}^\alpha$ form the internal virtual power density. When considering the contribution of grain boundaries, additional internal power is expended due to the presence of grain boundaries, which can also be included in the total internal power for each grain boundary. By taking into account the scalar internal microscopic forces $\Pi_A$ and $\Pi_B$, conjugate to grain boundary slip rates, the internal power can be expressed as follows:
	\begin{equation}
		\label{w_int_gb}
		\delta \dot{W}^{int} = \int_\Omega \boldsymbol{\sigma} : \delta\dot{\boldsymbol{\varepsilon}}^e d\Omega + \sum_\alpha \int_\Omega \left(\pi ^\alpha \delta\dot{\gamma}^\alpha + \boldsymbol{\xi}^\alpha \cdot \delta  \dot{\boldsymbol{\kappa}}^\alpha\right) d\Omega +\sum_\alpha \int_\mathcal{G} [\Pi^\alpha_B \delta \dot{\gamma^\alpha_B} -\Pi^\alpha_A \delta\dot{\gamma}^\alpha_A ] d\mathcal{G}.
	\end{equation}
	The internal virtual power can be written using the divergence theorem as
	\begin{align}
		\nonumber
		\delta \dot{W}^{int} =& \int_\Omega (\text{div}~ \boldsymbol{\sigma})\cdot \delta \dot{\boldsymbol{u}}~ d\Omega + \sum_\alpha \int_\Omega ( \pi^\alpha - \tau^\alpha-\text{div} ~ \boldsymbol{\xi}^\alpha)\delta\dot{\gamma}^\alpha ~ d\Omega + \int_\Gamma (\boldsymbol{\sigma} \cdot  \boldsymbol{n}) \cdot \delta \dot{\boldsymbol{u}} ~d\Gamma    \\
		\label{internal_virt_gb}
		&+ \sum_\alpha \int_\Gamma (\boldsymbol{\xi}^{\alpha} \cdot \boldsymbol{n}) \delta \dot{\gamma}^\alpha ~d\Gamma+\sum_\alpha \int_\mathcal{G} [\Pi^\alpha_B \delta \dot{\gamma^\alpha_B} -\Pi^\alpha_A \delta\dot{\gamma}^\alpha_A ] d\mathcal{G},
	\end{align} 
	where resolved shear stress $\tau^\alpha =\boldsymbol{T}^\alpha : \boldsymbol{\sigma}$, and $\boldsymbol{n}$ is the direction of the normal to the surface $\Gamma$. 
	External virtual power expenditure due to macroscopic traction vector $\boldsymbol{t}$, microscopic traction scalar $\chi^\alpha$ on the surface $\Gamma$, and  microscopic traction on grain boundary $\mathcal{G}$ can be written as 
	\begin{eqnarray}
		\delta \dot{W}^{ext}=\int_\Gamma \boldsymbol{t} \cdot \delta \dot{\boldsymbol{u}} ~d\Gamma + \sum_\alpha \int_\Gamma \chi^{\alpha} \delta \dot{\gamma}^\alpha ~d \Gamma +  \sum_\alpha \int_\mathcal{G}(\boldsymbol{\xi}^\alpha_B \cdot\boldsymbol{n}_s \delta\dot{\gamma}^\alpha_B-\boldsymbol{\xi}^\alpha_A\cdot \boldsymbol{n}_s \delta\dot{\gamma}^\alpha_A) d \mathcal{G}.
	\end{eqnarray}
	Here, it is assumed that displacement is continuous across the grain boundary. From the principle of virtual power, the following equilibrium equation  can be established
	\begin{equation} \label{bal_eq}
		\begin{cases}
			\text{div}~ \boldsymbol{\sigma} = \boldsymbol{0} \qquad\text{in} ~ \Omega, \\  
			\tau^\alpha + \text{div}~ \boldsymbol{\xi}^\alpha -\pi^\alpha = 0 \qquad\text{in}~ \Omega,
		\end{cases}
	\end{equation}
	and the following boundary conditions can be  established
	\begin{equation}
		\begin{cases}
			\boldsymbol{\sigma} \cdot \boldsymbol{n} = \boldsymbol{t} \qquad \text{on} ~\Gamma,\\
			\boldsymbol{\xi}^\alpha \cdot \boldsymbol{n} = \chi^\alpha \qquad \text{on} ~\Gamma,\\
			\boldsymbol{\xi}_A^\alpha \cdot \boldsymbol{n}_s =  \Pi_A^\alpha \qquad \text{on}~ \mathcal{G}, \\
			\boldsymbol{\xi}_B^\alpha \cdot \boldsymbol{n}_s = \Pi_B^\alpha \qquad \text{on}~ \mathcal{G}.
		\end{cases}
	\end{equation}
	\subsection{The dissipation inequality}
	The free energy density in its additive form can be divided into thermal and mechanical components, which are not coupled. This division allows us to focus on a thermomechanical theory that is either uncoupled or weakly coupled, with our discussions limited to mechanical dissipation alone. Therefore, in accordance with the second law of thermodynamics, we solely consider the free energy imbalance \citep{gurtin2010mechanics} for both bulk and grain boundary separately, as defined below 
	\begin{subequations}
		\begin{eqnarray}
			\label{dissip_0}
			\dot{\overline{\int_\Omega \Psi d\Omega}} - \int_\Omega \boldsymbol{\sigma} : \dot{\boldsymbol{\varepsilon}}^e d\Omega - \sum_\alpha \int_\Omega \left(\pi ^\alpha \dot{\gamma}^\alpha - \boldsymbol{\xi}^\alpha \cdot  \dot{\boldsymbol{\kappa}}^\alpha\right) d\Omega \leqslant 0, \\ 
			\dot{\overline{\int_\mathcal{G} \Psi_{GB} d\mathcal{G}}} - \sum_\alpha \int_\mathcal{G} [\Pi^\alpha_B  \dot{\gamma^\alpha_B} -\Pi^\alpha_A \dot{\gamma}^\alpha_A ] d\mathcal{G} \leqslant 0,
		\end{eqnarray}
	\end{subequations} 
	where the free energy density per unit volume is $\Psi$, and the free energy density per unit surface area of GB is $\Psi_{GB}$.  The above inequalities are valid for any arbitrary bulk volume $\Omega$, and arbitrary grain boundary $\mathcal{G}$; therefore, the local dissipation inequality can be written as 
	\begin{subequations}
		\begin{eqnarray}
			\label{dissip}
			D_b=\boldsymbol{\sigma} : \dot{\boldsymbol{\varepsilon}}^e  + \sum_\alpha  \left(\pi ^\alpha \dot{\gamma}^\alpha - \boldsymbol{\xi}^\alpha \cdot  \dot{\boldsymbol{\kappa}}^\alpha\right) -  \dot{ \Psi }  \geqslant 0 \qquad \text{on bulk } \Omega, \\ 
			D_{gb} = \sum_\alpha( \Pi^\alpha_B  \dot{\gamma^\alpha_B} -\Pi^\alpha_A \dot{\gamma}^\alpha_A)  - \dot{\Psi}_{GB} \geqslant0 \qquad \text{on GB } \mathcal{G}.
		\end{eqnarray}
	\end{subequations}
	In the case of the bulk part, the free energy is assumed to be a function of the elastic strain, plastic slip, and its gradient, that is  $\Psi = \Psi (\boldsymbol{\varepsilon}^e, \boldsymbol{\gamma}, \boldsymbol{\kappa})$. $\boldsymbol{\gamma}$ contains plastic slip of all the slip systems; that is $\boldsymbol{\gamma}=(\gamma^1, \gamma^2, ...., \gamma^m)$, and $\boldsymbol{\kappa}$ contain slip gradient  of all the slip systems; that is $\boldsymbol{\kappa}= (\boldsymbol{\kappa}^1, \boldsymbol{\kappa}^2, ..., \boldsymbol{\kappa}^m)$. The grain boundary free energy depends on the GB burgers tensor $\boldsymbol{G}_s$; that is, $\Psi_{GB}  = \Psi_{GB}(\boldsymbol{G}_s)$.  The local dissipation for bulk and grain boundary can be written as 
	\begin{subequations}
		\begin{eqnarray}
			D_b=\left(\boldsymbol{\sigma} - \frac{\partial \Psi}{\partial \boldsymbol{\varepsilon}^e}\right): \boldsymbol{\varepsilon}^e+ \sum_{\alpha} \left(\pi^\alpha - \frac{\partial \Psi}{\partial \gamma^\alpha}\right)\dot{\gamma}^\alpha + \sum_{\alpha}\left(\boldsymbol{\xi}^\alpha- \frac{\partial \Psi}{\partial \boldsymbol{\kappa}^\alpha}\right) \cdot \dot{\boldsymbol{\kappa}}^\alpha \geqslant 0 \label{d_b},\\
			D_{gb} = \sum_\alpha(\Pi^\alpha_B  \dot{\gamma^\alpha_B} -\Pi^\alpha_A \dot{\gamma}^\alpha_A)   - \frac{\partial \Psi_{GB}}{\partial \boldsymbol{G}_s}: \dot{\boldsymbol{G}}_s \geqslant 0 \label{d_gb}.
		\end{eqnarray}
	\end{subequations}
	In the SGCP theory, the Cauchy's stress is assumed to be a purely energetic quantity, and it can be directly derived from the free energy ($\boldsymbol{\sigma} = \frac{\partial \Psi}{\partial \boldsymbol{\varepsilon}^e}$). In the Gurtin type SGCP theory, the scalar and vector microscopic stress quantities are considered to consist of energetic ($(\cdot)_{en}$) and dissipative part ($(\cdot)_{dis}$); that is, 
	\begin{equation}
		\pi^\alpha =\pi^\alpha_{en} + \pi^\alpha_{dis}, \qquad
		\boldsymbol{\xi}^\alpha =\boldsymbol{\xi}^\alpha_{en} + \boldsymbol{\xi}^\alpha_{dis} \quad \text{with} \quad \pi^\alpha_{en} = \frac{\partial \Psi}{\partial \gamma^\alpha}, \qquad
		\boldsymbol{\xi}^\alpha_{en} = \frac{\partial \Psi}{\partial \boldsymbol{\kappa}^\alpha}.
	\end{equation}
	Now,  the dissipation inequality for bulk can be simplified  as follows: 
	\begin{align}
		\label{dissip_1}
		D_b= \sum_\alpha\pi^\alpha _{dis} \dot{\gamma}^\alpha + \boldsymbol{\xi}_{dis}^\alpha \cdot \boldsymbol{\kappa}^\alpha \geqslant 0.
	\end{align}
	The GB dissipation inequality, using equation (\ref{g_dot})$_2$, can be written as the following form
	\begin{equation}
		D_{gb} = \sum_\alpha\left(\Pi^\alpha_B - \frac{\partial \Psi_{GB}}{\partial \boldsymbol{G}_s}:\boldsymbol{N}_B^\alpha\right) \dot{\gamma^\alpha_B} -\sum_\alpha\left(\Pi^\alpha_A-\frac{\partial \Psi_{GB}}{\partial \boldsymbol{G}_s}:\boldsymbol{N}^\alpha_A \right) \dot{\gamma}^\alpha_A  \geqslant 0. 
	\end{equation}
	The scalar internal microscopic force $\Pi_A$ and $\Pi_B$ are also believed to comprise both an energetic and a dissipative component, similar to the bulk part; that is
	\begin{equation}
		\Pi_A^\alpha  = \Pi_{A,en}^{\alpha} + \Pi_{A,dis}^{\alpha}, \quad \Pi_B^\alpha  = \Pi_{B,en}^{\alpha} + \Pi_{B,dis}^{\alpha} \quad \text{with} \quad \Pi_{A,en}^\alpha= \frac{\partial \Psi_{GB}}{\partial \boldsymbol{G}_s}  :\boldsymbol{N}^\alpha_A, \quad \Pi_{B,en}^\alpha= \frac{\partial \Psi_{GB}}{\partial \boldsymbol{G}_s}  :\boldsymbol{N}^\alpha_B.
	\end{equation}
	The GB dissipation inequality can now be written as follows:
	\begin{equation}
		D_{gb} = \sum_\alpha( \Pi^\alpha_{B,dis} \dot{\gamma}_B^\alpha -\Pi^\alpha_{A,dis} \dot{\gamma}_A^\alpha) \geqslant 0.
	\end{equation}
	\subsection{Energetic constitutive relations}
	\subsubsection{Constitutive relation for the bulk} 
	In the Gurtin type SGCP theory \citep{gurtin2007gradient}, the free energy due to elastic strain and plastic slip gradient is considered. The elastic free energy is assumed as a quadratic function of elastic strain, and the defect energy is assumed as a quadratic function of the dislocation densities. The free energy considering quadratic elastic energy and defect energy can be written as 
	\begin{equation}
		\label{defect_eng}
		\Psi= \frac{1}{2} \boldsymbol{\varepsilon}^e : \mathbb{C} : \boldsymbol{\varepsilon}^e + \frac{1}{2} S_0 L_{en}^2\sum_\alpha\left[(\rho^\alpha_\vdash)^2 + (\rho^\alpha_\odot)^2\right],
	\end{equation}  
	where $\mathbb{C}$ is the fourth order linear elastic constitutive tensor,  $S_0$ is the initial slip resistance, $\rho^\alpha_\vdash$ is the edge dislocation density, $\rho^\alpha_\odot$ is the screw dislocation density. The dislocation densities can be related to the plastic slip gradient \citep{arsenlis1999crystallographic} as 
	\begin{equation}
		\label{aresenalis}
		\rho^\alpha_\vdash= -\boldsymbol{s}^\alpha \cdot \boldsymbol{\kappa}^\alpha, \qquad\text{and} \qquad \rho^\alpha_\odot= \boldsymbol{I}^\alpha \cdot \boldsymbol{\kappa}^\alpha, 
	\end{equation}
	where $\boldsymbol{I}^\alpha$ is the line direction, and $\boldsymbol{I}^\alpha = \boldsymbol{m}^\alpha \times \boldsymbol{s}^\alpha$.
	The stress and energetic vector microscopic stress can be derived from the free energy as 
	\begin{subequations}
		\begin{gather} 
			\boldsymbol{\sigma} = \frac{\partial \Psi}{\partial \boldsymbol{\varepsilon}^e} = \mathbb{C} : \boldsymbol{\varepsilon}^e \\ 
			\boldsymbol{\xi}^\alpha_{en}  = \frac{\partial \Psi}{\partial \boldsymbol{\kappa}^\alpha} = \frac{\partial \Psi}{\partial \rho^\alpha_\vdash} \frac{\partial \rho^\alpha_\vdash }{\partial \boldsymbol{\kappa}^\alpha} + 
			\frac{\partial \Psi}{\partial \rho^\alpha_\odot} \frac{\partial \rho^\alpha_\odot }{\partial \boldsymbol{\kappa}^\alpha}
		\end{gather} 
	\end{subequations}
	Using the equation (\ref{aresenalis}), the energetic vector microscopic stress can be written as 
	\begin{equation}
		\boldsymbol{\xi}^\alpha_{en}=  S_0 L_{en}^2[-\rho_\vdash^\alpha \boldsymbol{s}^\alpha + \rho_\odot^\alpha \boldsymbol{I}^\alpha] 
		=  S_0L_{en}^2[(\boldsymbol{s}^\alpha \cdot \boldsymbol{\kappa}^\alpha)  \boldsymbol{s}^\alpha + (\boldsymbol{I}^\alpha \cdot \boldsymbol{\kappa}^\alpha) \boldsymbol{I}^\alpha ] 
		=  S_0 L_{en}^2   \boldsymbol{\kappa}^{ t \alpha },  
	\end{equation}
	where $\boldsymbol{\kappa}^{t\alpha }$ is the tangential gradient of plastic slip,  which is defined as follows: 
	\begin{equation}
		\label{tangent_grad}
		\boldsymbol{\kappa}^{t\alpha } =(\boldsymbol{s}^\alpha \cdot \boldsymbol{\kappa}^\alpha)  \boldsymbol{s}^\alpha + (\boldsymbol{I}^\alpha \cdot \boldsymbol{\kappa}^\alpha) \boldsymbol{I}^\alpha  
		= (\boldsymbol{s}^\alpha \otimes \boldsymbol{s}^\alpha + \boldsymbol{I}^\alpha \otimes \boldsymbol{I}^\alpha)\cdot \boldsymbol{\kappa}^\alpha.
	\end{equation}
	\subsubsection{Constitutive relation for the grain boundary}
	The grain boundary free energy, known as GB defect energy, can be considered to be quadratic with the GB Burgers tensor $\boldsymbol{G}_s$ \citep{gurtin2008theory}, as follows: 
	\begin{eqnarray}
		\Psi_{GB} = \frac{c_0}{2} \boldsymbol{G}_s : \boldsymbol{G}_s,
	\end{eqnarray}
	where $c_0$ is some GB material constant. With this considered free energy, the scalar internal microscopic force quantities can be written as follows: 
	\begin{eqnarray}
		\Pi_{B,en}^\alpha= \frac{\partial \Psi_{GB}}{\partial \boldsymbol{G}_s}  :\boldsymbol{N}^\alpha_B=c_0 \boldsymbol{G}_s :\boldsymbol{N}_B^\alpha, \qquad
		\Pi_{A,en}^\alpha= \frac{\partial \Psi_{GB}}{\partial \boldsymbol{G}_s}  :\boldsymbol{N}^\alpha_A= c_0 \boldsymbol{G}_s :\boldsymbol{N}_A^\alpha. 
	\end{eqnarray}
	\subsection{Dissipative constitutive relation} 
	\subsubsection{Constitutive relation for the bulk}
	The dissipative microscopic stress ($\boldsymbol{\xi}_{dis}^\alpha$) is associated with the dislocation evolution of the $\alpha^{th}$ slip plane, and such dislocation motion is tangential to the slip plane. Therefore, the slip rate gradient $\dot{\boldsymbol{\kappa}}^\alpha$ is replaced by tangential slip gradient $\dot{\boldsymbol{\kappa}}^{t \alpha}$. Accordingly, the dissipation inequality (equation (\ref{dissip_1})) takes the following form 
	\begin{equation}
		\label{dissip_gurtin}
		D_b=\sum_\alpha\pi^\alpha _{dis} \dot{\gamma}^\alpha + \boldsymbol{\xi}_{dis}^\alpha \cdot \dot{\boldsymbol{\kappa}}^{t \alpha} \geqslant 0,
	\end{equation} 
	In order to establish the flow rule, the effective flow rate ($\dot{d}^\alpha$)  containing the plastic slip and its tangential gradient  is considered as follows: 
	\begin{equation}
		\dot{d}^\alpha = \sqrt{(\dot{\gamma}^\alpha)^2+L_d^2 [\dot{\boldsymbol{\kappa}}^{t \alpha}  \cdot \dot{\boldsymbol{\kappa}}^{t \alpha} ]}, \\
	\end{equation}  
	where $L_d$ is the dissipative length scale. Constitutive relation of dissipative stress quantities is derived from the viscoplastic flow rule as follows:
	\begin{equation} \label{ld}
		\pi^\alpha_{dis} = S^\alpha \left(\frac{\dot{d}^\alpha}{\dot{d}_0}\right)^m\frac{\dot{\gamma}^\alpha}{\dot{d}^\alpha}, \qquad
		\boldsymbol{\xi}^\alpha_{dis} =L_d^2 S^\alpha \left(\frac{\dot{d}^\alpha}{\dot{d}_0}\right)^m \frac{\dot{\boldsymbol{\kappa}}^{t \alpha} }{\dot{d}^\alpha},
	\end{equation}   
	where $\dot{d}_0$ is the reference strain rate, $S^\alpha$ is the slip resistance, and $m$ is the rate-sensitive exponent.  
	%
	\subsubsection{Constitutive relation for the GB }
	The constitutive relation for GB dissipative stress can be derived within the viscoplastic framework by considering a cumulative defectiveness ($|\boldsymbol{G}_s|$) as follows: 
	\begin{equation}
		\label{cum_df}
		|\boldsymbol{G}_s| = \int_t |\dot{\boldsymbol{G}}_s| dt.
	\end{equation}
	Following \citep{gurtin2005boundary,gurtin2008theory}, the dissipative part of grain boundary scalar microscopic force $\Pi^\alpha_{A,dis}$ and $\Pi^\alpha_{B,dis}$ can be written 
	\begin{eqnarray}
		\Pi_{B,dis}^\alpha = F(|\boldsymbol{G}_s|) |\dot{\boldsymbol{G}}_s|^m \frac{\boldsymbol{N}^\alpha_B:\dot{\boldsymbol{G}}_s}{|\dot{\boldsymbol{G}}_s|} \qquad \text{and} \qquad 
		\Pi_{A,dis}^\alpha = F(|\boldsymbol{G}_s|) |\dot{\boldsymbol{G}}_s|^m \frac{\boldsymbol{N}^\alpha_A:\dot{\boldsymbol{G}}_s}{|\dot{\boldsymbol{G}}_s|},
	\end{eqnarray}
	where $m > 0$ is rate-sensitive modulus, $F(|\boldsymbol{G}_s|)$ is the isotropic hardening function related to the grain boundary.
	\section{Elastic Gap free SGCP theory}
	\label{proposal}
	\subsection{Decomposition of slip gradient and dislocation density}
	The classical Gurtin type SGCP model \citep{gurtin2007gradient} ensures that dissipation remains non-negative, meeting thermodynamic requirements. However, a key feature of this version of SCGP theory is the existence of a constitutive singularity, specifically nonphysical plastic flow \citep{amouzou2023elastic}, known as the elastic gap due to non-proportional loading \citep{fleck2014strain,fleck2015guidelines}. This is attributed to the non-incremental relationship, equation (\ref{ld})$_2$, between higher-order stress and tangential slip gradient rate, which accounts for the involvement of GND in dissipation.
	
	In the present exploration, our objective is to establish a formulation within the Gurtin-type SGCP framework that eliminates the elastic gap by separating the plastic slip gradient into energetic and dissipative components, in contrast to the original approach by Gurtin where vector microscopic stress quantities are divided. We have decomposed total plastic slip gradient ($\boldsymbol{\kappa}^\alpha$) into energetic ($\boldsymbol{\kappa}^\alpha_{en}$) and dissipative ($\boldsymbol{\kappa}^\alpha_{dis}$) subparts to account for the dissipation related to the plastic slip gradient. Henceforth, the decomposition can be written as follows:
	\begin{equation}
		\boldsymbol{\kappa}^\alpha =\boldsymbol{\kappa}^\alpha_{en} +\boldsymbol{\kappa}^\alpha_{dis}. 
	\end{equation}
	The edge and screw dislocation densities, using equation (\ref{aresenalis}), can be written as 
	\begin{subequations}
		\label{park}
		\begin{equation}
			\label{rho_total}
			\rho^\alpha_\vdash = -\boldsymbol{s}^\alpha \cdot \boldsymbol{\kappa}^\alpha = -\boldsymbol{s}^\alpha \cdot (\boldsymbol{\kappa}^\alpha_{en}+ \boldsymbol{\kappa}^\alpha_{dis}) = (\rho^\alpha_{\vdash})_{en} + (\rho^\alpha_{\vdash})_{dis}, 
		\end{equation}
		\begin{equation}
			\rho^\alpha_\odot=  \boldsymbol{I}^\alpha \cdot \boldsymbol{\kappa}^\alpha = \boldsymbol{I}^\alpha \cdot(\boldsymbol{\kappa}^\alpha_{en}+ \boldsymbol{\kappa}^\alpha_{dis}) =(\rho^\alpha_\odot)_{en} + (\rho^\alpha_\odot)_{dis},
		\end{equation} 
	\end{subequations}
	where $(\rho ^\alpha_\vdash)_{en}$ and  $(\rho ^\alpha_\vdash)_{dis}$  are defined as energetic and dissipative edge dislocation, and $(\rho ^\alpha_\odot)_{en}$ and  $(\rho ^\alpha_\odot)_{dis}$ are the energetic and dissipative screw dislocation. Therefore, the decomposition of plastic slip gradient results in the decomposition of dislocation into the energetic and dissipative parts. Vector microscopic stress, work conjugate to the plastic slip gradient, is considered total quantity, and no subdivision of vector microscopic stress is considered. Similar decomposition of plastic strain gradient-like variable is carried out in single crystal plasticity framework by \citep{jebahi2023alternative}, in distortion gradient plasticity framework by \citep{panteghini2019potential} or phenomenological SGP framework by \citep{mukherjee2024elastic}  to achieve an elastic-gap free formulation.
	
	\subsection{Constitutive relation and free energy imbalance}
	In line with the seminal studies by \cite{gurtin2007gradient}, the free energy ($\Psi$) is considered as a function of elastic strain and plastic slip gradient. The free energy contribution of plastic slip is generally infrequent. Nevertheless, our current proposition is not constrained by this assumption and can be readily expanded to incorporate the influence of plastic slip on free energy. In the current proposition, the classical elastic free energy is augmented with energetic dislocation densities, unlike the Gurtin-type model, where the free energy augmentation is due to total dislocation densities.  Therefore, the free energy is written as the sum of elastic energy $\Psi^e$, and defect energy $\Psi^\rho$ as follows:
	\begin{eqnarray} \label{p_e}
		\Psi= \Psi^e(\boldsymbol{\varepsilon}^e) + \Psi^\rho\left((\rho^\alpha_\vdash)_{en},(\rho^\alpha_\odot)_{en}\right)  = \frac{1}{2} \boldsymbol{\varepsilon}^e : \mathbb{C} : \boldsymbol{\varepsilon}^e+ \frac{1}{2} S_0 L_{*}^2 \sum_{\alpha}\left[(\rho^\alpha_\vdash)_{en}^2 + (\rho^\alpha_\odot)_{en}^2\right],
	\end{eqnarray}
	where $L_*$ is a length scale parameter. The defect energy $\Psi^\rho$ can be simplified with the help of equation (\ref{park}) as follows: 
	\begin{equation}
		\label{phi_rho}
		\Psi^\rho =\frac{1}{2} S_0 L_{*}^2 \sum_{\alpha}\left[(\rho^\alpha_\vdash)_{en}^2 + (\rho^\alpha_\odot)_{en}^2\right] = \frac{1}{2} S_0L_*^2\sum_\alpha[(\boldsymbol{s}^\alpha \cdot \boldsymbol{\kappa}^\alpha_{en})^2 + (\boldsymbol{I}^\alpha \cdot \boldsymbol{\kappa}^\alpha_{en})^2]=\frac{1}{2} S_0 L_*^2 \sum_\alpha(\boldsymbol{\kappa}_{en}^{t\alpha}\cdot \boldsymbol{\kappa}^{t\alpha}_{en}).
	\end{equation}
	By considering the dependence of Cauchy’s stress only on elastic strain ($\boldsymbol{\sigma} = \frac{\partial \Psi}{\partial \boldsymbol{\varepsilon}^e} = \mathbb{C} : \boldsymbol{\varepsilon}^e$), the vector microscopic stress quantities are derived from the free energy as follows:
	\begin{equation}
		\label{defect_en}
		\boldsymbol{\xi}^\alpha = \frac{\partial \Psi}{\partial \boldsymbol{\kappa}_{en}^\alpha}= S_0L_*^2 [(\boldsymbol{s}^\alpha \cdot \boldsymbol{\kappa}^\alpha_{en})  \boldsymbol{s}^\alpha + (\boldsymbol{I}^\alpha \cdot \boldsymbol{\kappa}^\alpha_{en}) \boldsymbol{I}^\alpha] = S_0L_*^2 \boldsymbol{\kappa}_{en}^{t\alpha}.
	\end{equation}
	Now, the dissipation inequality (\ref{d_b}) can be written as follows: 
	\begin{align}
		D_b & =\left(\boldsymbol{\sigma} - \frac{\partial \Psi}{\partial \boldsymbol{\varepsilon}^e}\right): \boldsymbol{\varepsilon}^e+ \sum_{\alpha} \left(\pi^\alpha - \frac{\partial \Psi}{\partial \gamma^\alpha}\right)\dot{\gamma}^\alpha + \sum_{\alpha}\left(\boldsymbol{\xi}^\alpha \cdot \dot{\boldsymbol{\kappa}}^\alpha- \frac{\partial \Psi}{\partial \boldsymbol{\kappa}^\alpha_{en}} \cdot \dot{\boldsymbol{\kappa}}_{en}^\alpha\right) \geqslant 0 \nonumber \\
		& = \sum_\alpha \left(\pi^\alpha \dot{\gamma}^\alpha+\boldsymbol{\xi}^\alpha\cdot\left( \dot{\boldsymbol{\kappa}}^\alpha - \dot{\boldsymbol{\kappa}}_{en}^{\alpha}\right) \right)  \geqslant 0, \quad \text{since,} \quad \boldsymbol{\sigma} = \frac{\partial \Psi}{\partial \boldsymbol{\varepsilon}^e} \quad \text{and} \quad \frac{\partial \Psi}{\partial \gamma^\alpha} = 0 \nonumber \\
		& = \sum_\alpha \left(\pi^\alpha \dot{\gamma}^\alpha+\boldsymbol{\xi}^\alpha\cdot \dot{\boldsymbol{\kappa}}_{dis}^{\alpha} \right)  \geqslant 0 \nonumber \\
		& = \sum_\alpha(\pi^\alpha \dot{\gamma}^\alpha+ \boldsymbol{\xi}^\alpha \cdot \dot{\boldsymbol{\kappa}}^{t\alpha}_{dis}) \geqslant 0 \label{dissip_2}
	\end{align}
	In the above equation, $\dot{\boldsymbol{\kappa}}^\alpha_{dis}$ is replaced by the tangential slip gradient $\dot{\boldsymbol{\kappa}}^{t \alpha}_{dis}$, due to assumption of the dislocation motion along the tangential slip plane direction.
	\subsection{Evolution of vector microscopic stress by minimizing plastic dissipating potential}
	In order to construct the evolution of the $\boldsymbol{\kappa}^{t \alpha}_{en}$, we have considered an effective microscopic stress $\bar{\pi}^\alpha$ and effective slip rate ($\dot{\bar{d}}^\alpha$), associated to $\alpha$ slip plane as follows:
	\begin{equation}
		\bar{\pi}^\alpha = \sqrt{(\pi^\alpha)^2} = |\pi^\alpha|, \qquad\text{and } \qquad \dot{\bar{d}}^\alpha = \sqrt{(\dot{\gamma}^\alpha)^2} = |\dot{\gamma}^\alpha|.
	\end{equation}
	We have considered that a flow surface exists associated with each slip plane. The plastic flow occurs along the slip plane when the effective microscopic stress equals the flow resistance $\pi^\alpha_f(\bar{d}^1, \bar{d}^2, ...)$, which depends on the cumulative effective slip ($\bar{d}^\alpha = \int \dot{\bar{d}}^\alpha~ dt$) of all the slip planes. The flow surface for $\alpha$ slip plane can be written as 
	\begin{equation}
		\label{flow_surf}
		f^\alpha = \bar{\pi}^\alpha - \pi_f^\alpha(\bar{d}^1, \bar{d}^2, ...)=0.
	\end{equation} 
	In classical plasticity, nonlinear backstress models such as the Armstrong-Frederick model or Chaboche model can be defined with the help of a dual dissipating potential ($\Phi^\alpha$) and a plastic potential (defect energy, $\Psi^\alpha$) \citep{chaboche1986time,dettmer2004theoretical}.  
	Here, we consider those two potentials for each slip plane  as follows:  
	\begin{equation}
		\Psi ^\alpha = \frac{S_0 L_*^2}{2} \boldsymbol{\kappa}_{en}^{t\alpha} \cdot \boldsymbol{\kappa}_{en}^{t\alpha}\qquad \text{and} \qquad
		\Phi^\alpha = f^\alpha + \frac{\zeta}{2S_0 L_*^2}  \boldsymbol{\xi}^\alpha \cdot \boldsymbol{\xi}^\alpha,
	\end{equation}
	where $\zeta$ is a positive coefficient. The dual dissipation potential is defined in terms of driving forces derived from the Legendre-Fenschel transformation of the dissipation potential $\Phi^{*\alpha}$, defined in terms of strain-like quantities.
	Evolution law of $\dot{\bar{d}}^\alpha$, and $\dot{\boldsymbol{\kappa}}_{en}^{t\alpha}$ can be obtained by minimizing the dual dissipating potential with positive dissipation as a constraint, similar to \cite{voyiadjis2003thermodynamic,aygun2021coupling,mukherjee2023strain,mukherjee2024elastic}. The Lagrangian corresponding to this minimization problem can be written as
	\begin{equation}
		\mathcal{L} =D^\alpha_b -  \pi^\alpha_f\dot{\bar{d}}^\alpha -\dot{\lambda}\Phi^\alpha,  
	\end{equation}
	where $\dot{\lambda}$ is the Lagrange multiplier and $D^\alpha_b$ is the dissipation associated to slip plane $\alpha$, defined by equation (\ref{dissip_2}). The minimization of the objective function $\mathcal{L}$ is ensured by considering the first order optimality criteria of $\mathcal{L}$ about the driving force $\pi^\alpha$, $\pi^\alpha_f$ and $\boldsymbol{\xi}^\alpha$. The first-order optimality criteria can be written as 
	\begin{subequations}
		\begin{equation}
			\frac{\partial \mathcal{L}}{\partial \pi^\alpha} = 0 = \dot{\gamma}^\alpha- {\dot{\lambda}}\frac{\partial \bar{\pi}^\alpha}{\partial \pi^\alpha}  \implies \frac{\pi^\alpha}{\bar{\pi}^\alpha} = \frac{\dot{\gamma}^\alpha}{\dot{\lambda}} \implies \pi^\alpha = \pi^\alpha_f \frac{\dot{\gamma}^\alpha}{\dot{\bar{d}}^\alpha} ,
		\end{equation}
		\begin{equation}
			\frac{\partial \mathcal{L}}{\partial \boldsymbol{\xi}^\alpha} = \boldsymbol{0} = \dot{\boldsymbol{\kappa}}^{t\alpha}-\dot{\boldsymbol{\kappa}}^{t\alpha}_{en} -\dot{\lambda}\frac{\zeta}{S_0 L_*^2} \boldsymbol{\xi}^\alpha \label{evo_k},
		\end{equation}
		\begin{equation}
			\frac{\partial \mathcal{L}}{\partial \pi^\alpha_f} = - \dot{\bar{d}}^\alpha + \dot{\lambda} \implies \dot{\lambda} = \dot{\bar{d}}^\alpha.
		\end{equation}
	\end{subequations}
	The flow resistance $\pi^\alpha_f$ of slip plane $\alpha$ (with the help of slip resistance $S^\alpha$ and viscoplastic rate function $R(\dot{d}^\alpha)$) and time derivative of the equations (\ref{defect_en}) are used to establish the constitutive relations as follows: 
	\begin{subequations}
		\begin{gather}
			\label{microstr}
			\pi^\alpha = S^\alpha R(\dot{\bar{d}}^\alpha) \frac{\dot{\gamma}^{\alpha}}{\dot{\bar{d}}^\alpha}, \\
			\label{xi_evol}
			\dot{\boldsymbol{\xi}}^\alpha = S_0 L_*^2 \dot{\boldsymbol{\kappa}}^{t\alpha} -\dot{\bar{d}}^\alpha \zeta \boldsymbol{\xi}^\alpha.
		\end{gather}
	\end{subequations}
	The last evolution equation looks similar to the Armstrong-Frederick type \citep{armstrong1966mathematical} backstress evolution equation, where each vector microscopic stress (i.e., for each slip system $\alpha$) evolves independently.  Each term's contribution will depend on the positive coefficients $S_0L_*^2$ and $\zeta$. When the $\zeta=0$, the second term becomes zero, and the evolution equation becomes similar to the conventional Gurtin-type energetic model. The first term provides linear hardening similar to the Gurtin-type model. However, the second term relaxes some of the energy and brings the nonlinearity to the hardening. In every condition, the hardening should be more than the relaxation from material stability criteria \cite{chaboche1986time}.
	The evolution equation (\ref{evo_k}) can be rearranged as follows:
	\begin{equation}
		\label{kappa_evol}
		\dot{\boldsymbol{\kappa}}^{t\alpha}_{en}=\dot{\boldsymbol{\kappa}}^{t\alpha} - \dot{\bar{d}}^\alpha \zeta 
		\boldsymbol{\kappa}_{en}^{t\alpha}  \qquad \text{or} \qquad \dot{\boldsymbol{\kappa}}^{t\alpha}_{dis}=\dot{\bar{d}}^\alpha \zeta 
		\boldsymbol{\kappa}_{en}^{t\alpha} \qquad \text{or} \qquad \boldsymbol{\xi}^\alpha = \frac{S_0L_*^2}{\zeta} \frac{\dot{\boldsymbol{\kappa}}^{t\alpha}_{dis}}{\dot{\bar{d}}^\alpha}.
	\end{equation}
	Above equations can be linked to the GND densities with help of equation (\ref{park}) as follows: 
	\begin{subequations}
		\begin{eqnarray}
			-\boldsymbol{s}^\alpha \cdot \dot{\boldsymbol{\kappa}}^{t\alpha}_{en}=-\boldsymbol{s}^\alpha \cdot\dot{\boldsymbol{\kappa}}^{t\alpha} + \dot{\bar{d}}^\alpha \zeta \boldsymbol{s}^\alpha \cdot  
			\boldsymbol{\kappa}_{en}^{t\alpha} \implies \dot{(\rho_\vdash^\alpha)}_{en} = \dot{(\rho_\vdash^\alpha)} - \zeta \dot{\bar{d}}^\alpha (\rho_\vdash^\alpha)_{en} \quad \text{or} \quad \dot{(\rho_\vdash^\alpha)}_{dis} = \zeta \dot{\bar{d}}^\alpha (\rho_\vdash^\alpha)_{en}, \\
			\boldsymbol{I}^\alpha \cdot \dot{\boldsymbol{\kappa}}^{t\alpha}_{en}=\boldsymbol{I}^\alpha \cdot\dot{\boldsymbol{\kappa}}^{t\alpha} - \dot{\bar{d}}^\alpha \zeta \boldsymbol{I}^\alpha \cdot  
			\boldsymbol{\kappa}_{en}^{t\alpha} \implies \dot{(\rho_\odot^\alpha)}_{en} = \dot{(\rho_\odot^\alpha)} - \zeta \dot{\bar{d}}^\alpha (\rho_\odot^\alpha)_{en} \quad \text{or} \quad \dot{(\rho_\odot^\alpha)}_{dis} = \zeta \dot{\bar{d}}^\alpha (\rho_\odot^\alpha)_{en}.
		\end{eqnarray}
	\end{subequations}
	The last two equations signify that the proposed evolution equation (\ref{xi_evol}) leads to saturation to the edge and screw dislocation density as the plastic flow occurs. The coefficient $\zeta$ controls how much GND will participate in the energetic and dissipative counterpart. 
	
	Referring back to equation (\ref{dissip_2}), the dissipation inequality can be rearranged with the help of equation (\ref{kappa_evol}) as follows:
	\begin{eqnarray}
		D_b= \sum_\alpha(\pi^\alpha \dot{\gamma}^\alpha+ \boldsymbol{\xi}^\alpha \cdot \dot{\boldsymbol{\kappa}}^{t\alpha}_{dis}) 
		=\sum_\alpha(\pi^\alpha \dot{\gamma}^\alpha+\zeta \dot{\bar{d}}^\alpha \boldsymbol{\xi}^\alpha \cdot {\boldsymbol{\kappa}}^{t\alpha}_{en})
		=\sum_\alpha(\pi^\alpha \dot{\gamma}^\alpha+ \frac{\zeta \dot{\bar{d}}^\alpha}{S_0L_*^2} \boldsymbol{\xi}^\alpha\cdot \boldsymbol{\xi}^\alpha). \label{db22}
	\end{eqnarray}
	The last equation contains an additional term compared to classical crystal plasticity. The second term, quadratic in nature, is always positive (as the coefficient is positive). $\boldsymbol{\xi}^\alpha$ evolves nonlinearly, which saturates over time; therefore, the additional dissipation also follows similar pattern. The additional dissipation increases linearly at the beginning of the plastic flow, and it eventually saturates over time, similar to $\boldsymbol{\xi}^\alpha$. 
	
	It should be emphasized that the constitutive proposal mentioned earlier does not result in any stress jump at the onset of yield \citep{fleck2015guidelines}. This can be observed from equation (\ref{kappa_evol})$_3$. At the onset of yield, the incremental quantities for each slip system ($\dot{\bar{d}}^\alpha, \dot{\boldsymbol{\kappa}}^{t\alpha} $) are nonzero, but the non-incremental quantities ($\bar{d}^\alpha,\boldsymbol{\kappa}^{t\alpha}$) are zero. This condition leads to the energetic component of the slip gradient ($\boldsymbol{\kappa}_{en}^{t\alpha}$), and consequently, $\boldsymbol{\xi}^\alpha$ is zero. As a result, the incremental quantity $\dot{\boldsymbol{\kappa}}_{dis}^{t\alpha}$ in equation (\ref{kappa_evol})$_3$ is also zero. Finally, according to equation (\ref{db22}), at the onset of yield, $D_b = \pi_f^{\alpha}\dot{\bar{d}}^\alpha$. In the case of Gurtin type theory,  higher-order dissipation is possessed by higher-order dissipative stress $\boldsymbol{\xi}^\alpha_{dis}$. However, this dissipative vector microscopic stress becomes nonzero (due to nonzero $\dot{\boldsymbol{\kappa}}^{t\alpha}$) at the onset of yield resulting in an elastic-gap \citep{fleck2014strain}. This is further evident by the micro force balance equation (\ref{bal_eq})$_2$ as follows:
	\begin{subequations}
		\begin{align}
			& \tau^\alpha + \underbrace{  S_0 L_{*}^2~\text{div}~\boldsymbol{\kappa}^{t\alpha}_{en} }_{\text{Energetic}}  = \underbrace{ S^\alpha R(\dot{\bar{d}}^\alpha) \frac{\dot{\gamma}^{\alpha}}{\dot{\bar{d}}^\alpha}}_{\text {Dissipative}},\\
			& \tau^\alpha  - \underbrace{ \frac{S_0 L_{*}^2}{\zeta}~\text{div}\left( \frac{\dot{\boldsymbol{\kappa}}^{t\alpha}_{en} - \dot{\boldsymbol{\kappa}}^{t\alpha}}{\dot{\bar{d}}^\alpha} \right)  }_{\text{Energetic}}  = \underbrace{ S^\alpha R(\dot{\bar{d}}^\alpha) \frac{\dot{\gamma}^{\alpha}}{\dot{\bar{d}}^\alpha}}_{\text {Dissipative}}. \label{saturation_micro2}
		\end{align}
	\end{subequations}
	It can be noted here that the above equation cannot capture grain size-dependent yield stress or higher order dissipative hardening \citep{jebahi2023alternative,mukherjee2024elastic}.
	
	\subsection{Vector microscopic stress update and material tangent calculation}\label{mat}
	In order to implement the evolutionary vector microscopic stress model in a finite element framework, an accurate stress update and material tangent calculation are required. This section details the derivation of stress update routine and material tangent calculation.  
	An implicit time integration rule is followed. Time increment of field variables is denoted as $\Delta (\cdot)$, which can be calculated as $\Delta (\cdot) = \dot{(\cdot)} \Delta t$. Increment in $\boldsymbol{\xi}^\alpha$ between $n$ to $n+1$ time step can be found from the time discretization of the evolution equation (\ref{xi_evol}) as follows:
	\begin{subequations}
		\begin{eqnarray}
			\Delta \boldsymbol{\xi}^\alpha = S_0 L_*^2 \Delta \boldsymbol{\kappa}^{t\alpha} - \Delta \bar{d}^\alpha \zeta \boldsymbol{\xi}^\alpha_{n+1}, \\ 
			\implies	\boldsymbol{\xi}^\alpha_{n+1}- \boldsymbol{\xi}^\alpha_n= S_0 L_*^2\Delta\boldsymbol{\kappa}^{t\alpha} - \Delta \bar{d}^\alpha\zeta \boldsymbol{\xi}^\alpha_{n+1},\\ 
			\label{xi_inc}
			\implies \boldsymbol{\xi}^\alpha_{n+1}(1+\zeta\Delta\bar{d}^\alpha) = S_0 L_*^2 \Delta\boldsymbol{\kappa}^{t\alpha} +\boldsymbol{\xi}^\alpha_n,\\
			\label{str_update}
			\implies \boldsymbol{\xi}^\alpha_{n+1} = \frac{1}{1+ \zeta \Delta\bar{d}^\alpha} \left(S_0 L_*^2 \Delta\boldsymbol{\kappa}^{t\alpha} +\boldsymbol{\xi}^\alpha_n\right).
		\end{eqnarray}
	\end{subequations} 
	
	In SGCP formulation, $\Delta \gamma^\alpha$ is treated as an unknown variable along with displacement $\boldsymbol{u}$. Finite element implementation of the evolutionary $\boldsymbol{\xi}^\alpha$ requires the derivatives of $\boldsymbol{\xi}^\alpha_{n+1}$ with respect to increment of nodal plastic slip $\hat{\gamma}^\beta$; that is $\partial\boldsymbol{\xi}^\alpha_{n+1}/\partial\Delta\hat {\gamma}^\beta $  need to be calculated at each material point. Calculation of the derivative is done as follows: 
	\begin{eqnarray}
		\frac{\partial \boldsymbol{\xi}^\alpha_{n+1}(\Delta \gamma^\alpha, \Delta \boldsymbol{\kappa}^{\alpha})}{\partial \Delta\hat{\gamma}^\beta} = \frac{\partial \boldsymbol{\xi}^\alpha_{n+1}}{\partial \Delta {\gamma}^\beta} \frac{\partial \Delta \gamma^\beta }{\partial \Delta \hat{\gamma}^\beta} + \frac{\partial \boldsymbol{\xi}^\alpha_{n+1}}{\partial \Delta\boldsymbol{\kappa}^{\beta}}  \frac{\partial \Delta\boldsymbol{\kappa}^{\beta}}{\partial \Delta \hat{\boldsymbol{\kappa}}^\beta}.
	\end{eqnarray}  
	Calculation of $\frac{\partial \boldsymbol{\xi}^\alpha_{n+1}}{\partial \Delta \gamma^\beta}$ using equation  (\ref{xi_inc}):
	\begin{gather}
		\nonumber
		\frac{\partial }{\partial \Delta \gamma^\beta} \left[\boldsymbol{\xi}^\alpha_{n+1}(1+\zeta\Delta\bar{d}^\alpha)\right] = 	\frac{\partial }{\partial \Delta \gamma^\beta}\left[ S_0 L_*^2 \Delta\boldsymbol{\kappa}^{t\alpha} +\boldsymbol{\xi}^\alpha_n\right], \\ 
		\implies	(1+\zeta \Delta \bar{d}^\alpha) \frac{\partial \boldsymbol{\xi}^\alpha_{n+1}}{\partial \Delta\gamma^\beta} +  \zeta \boldsymbol{\xi}^\alpha_{n+1 }  \frac{\partial \Delta \bar{d}^\alpha}{\partial \Delta \gamma^\beta} = S_0 L_*^2 \frac{\partial \Delta \boldsymbol{\kappa}^{t\alpha}}{\partial \Delta \gamma^\beta} + \frac{\partial \boldsymbol{\xi}^\alpha_n}{\partial \Delta \gamma^\beta}.
	\end{gather}
	The right-hand side of the last equation can be identified as zero. Therefore, it can be rearranged as
	\begin{equation}
		\label{dxi_dgamma}
		\frac{\partial \boldsymbol{\xi}^\alpha_{n+1}}{\partial \Delta\gamma^\beta} = -\frac{\zeta}{1+ \zeta \Delta\bar{d}^\alpha} \frac{\partial \Delta \bar{d}^\alpha}{\partial \Delta\gamma^\beta} \boldsymbol{\xi}^\alpha_{n+1}.
	\end{equation}
	Calculation of $\frac{\partial \boldsymbol{\xi}^\alpha_{n+1}}{\partial \Delta\boldsymbol{\kappa}^{\beta}}$ using equation (\ref{xi_inc}):
	\begin{gather}
		\nonumber
		\frac{\partial }{\partial \Delta \boldsymbol{\kappa}^{\beta}} \left[\boldsymbol{\xi}^\alpha_{n+1}(1+\zeta\Delta\bar{d}^\alpha)\right] = 	\frac{\partial }{\partial \Delta \boldsymbol{\kappa}^{\beta}}\left[ S_0 L_*^2 \Delta\boldsymbol{\kappa}^{t\alpha} +\boldsymbol{\xi}^\alpha_n\right],\\
		\implies (1+\zeta \Delta \bar{d}^\alpha) \frac{\partial \boldsymbol{\xi}^\alpha_{n+1}}{\partial \Delta\boldsymbol{\kappa}^{\beta}} +  \underbrace{\zeta \boldsymbol{\xi}^\alpha_{n+1 }  \frac{\partial \Delta \bar{d}^\alpha}{\partial \Delta \boldsymbol{\kappa}^{\beta}}}_{\text{I}} = S_0 L_*^2 \frac{\partial \Delta \boldsymbol{\kappa}^{t\alpha}}{\partial \Delta \boldsymbol{\kappa}^{\beta}} + \underbrace{\frac{\partial \boldsymbol{\xi}^\alpha_n}{\partial \Delta \boldsymbol{\kappa}^{\beta}}}_{\text{II}}.
	\end{gather}
	The term $\text{I}$ and $\text{II}$ of the last equation can be identified as zero; therefore the last equation can be arranged as 
	\begin{equation}
		\label{dxi_dkappa}
		\frac{\partial \boldsymbol{\xi}^\alpha_{n+1}}{\partial \Delta \boldsymbol{\kappa}^{\beta}} = \frac{S_0L_*^2}{1+ \zeta \Delta \bar{d}^\alpha} \frac{\partial\Delta \boldsymbol{\kappa}^{t\alpha}}{\partial \Delta \boldsymbol{\kappa}^{\beta}}.
	\end{equation}
	
	\section{An effective modification of GB constitutive relation to account for GB dissipation} \label{model_GB}
	In the Gurtin type theory \citep{gurtin2008theory}, the GB stress variable is also decomposed into an energetic and a dissipative stress variable. This decomposition allows for the consideration of GB hardening and GB dissipation, respectively. The GB energetic stress accounts for the free energy related to GB, and the GB dissipative stress takes care of the dissipation that occurs at the grain boundary (for example, grain boundary burst \citep{soer2005detection,wang2004indentation,britton2009nanoindentation}). Experimental findings reveal a nonlinear  Hall-Petch hardening \citep{lim2011simulation}, which is predominantly governed by the grain boundary. In this proposition, we aim to consider the nonlinear hardening along with GB dissipation. To accomplish this, we have opted to divide the GB-related kinematic variable instead of decomposing the GB stress variable. The GB Burgers tensor has been separated into two components: an energetic component referred to as $\boldsymbol{G}_{s,en}$ and a dissipative component referred to as $\boldsymbol{G}_{s,dis}$. The expression for this decomposition is as follows:
	\begin{equation}
		\label{gs_split}
		\boldsymbol{G}_s = \boldsymbol{G}_{s,en} + \boldsymbol{G}_{s,dis}.
	\end{equation}
	\subsection{Proposed grain boundary defect energy and the corresponding dissipation inequity}
	We have considered free energy dependence with the energetic part of the GB burger tensor, unlike the Gurtin type model \citep{gurtin2008theory}, where free energy depends on the total GB Burgers tensor.  The free energy of the grain boundary, considering energetic GB Burgers tensor can be written as
	\begin{equation}
		\label{gb_free}
		\Psi_{gb} = \frac{c_s}{2} \boldsymbol{G}_{s,en} : \boldsymbol{G}_{s,en},
	\end{equation}
	where $c_s$ is a positive material constant. The energetic stress $\boldsymbol{M}$, power conjugate to the rate of GB Burgers tensor, is considered as a total quantity. The energetic stress quantity can be derived from the free energy as 
	\begin{equation}
		\label{energetic_str}
		\boldsymbol{M} = \frac{\partial \Psi_{gb}}{\partial \boldsymbol{G}_{s,en}} = c_s \boldsymbol{G}_{s,en}.
	\end{equation}
	Considering the grain boundary internal power as $\boldsymbol{M}:\dot{\boldsymbol{G}}_s$, the grain boundary dissipation at any point (equation (\ref{d_gb})) can be written as
	\begin{equation}
		\label{dissip_gb}
		D_{gb} = \boldsymbol{M} : \dot{\boldsymbol{G}}_s - \dot{\Psi}_{gb}=  \boldsymbol{M} : \dot{\boldsymbol{G}}_s - \frac{\partial \Psi_{gb}}{\partial \boldsymbol{G}_{s,en}} : \dot{\boldsymbol{G}}_{s,en} = \boldsymbol{M} : \dot{\boldsymbol{G}}_s - \boldsymbol{M} : \dot{\boldsymbol{G}}_{s,en} = \boldsymbol{M} : \dot{\boldsymbol{G}}_{s,dis} \geqslant 0.
	\end{equation}
	It can be noted here that the dissipation at the grain boundary is associated with the dissipative part of the GB Burgers tensor.
	
	\subsection{Evolution of GB stress by minimizing plastic dissipating potential}
	This section suggests an evolution of the GB Burger tensor (or stress $\boldsymbol{M}$) by taking into account the decomposition (\ref{gs_split}) and the GB free energy (\ref{gb_free}) through the minimization of a dissipating potential ($\Phi^*_{GB}$) within a viscoplastic framework. In order to derive such a relation we first define a slip resistance or a hardening/softening function (see equation 6.8 in \cite{gurtin2008theory}) $\mathcal{M}_R:= \mathcal{M}_R\left(|\boldsymbol{G}_s|\right)$ conjugate to $\dot{|\boldsymbol{G}_s|}$. $|\boldsymbol{G}_s|$ is cumulative defectiveness defined in equation (\ref{cum_df}). The GB dissipation inequality, equation (\ref{dissip_gb}), can be written as
	\begin{equation}
		D_{gb} = \boldsymbol{M} : \dot{\boldsymbol{G}}_{s,dis} = \mathcal{M}_R |\dot{\boldsymbol{G}}_s| .
	\end{equation}  
	Similar to the bulk, we consider the following dual dissipating potential in order to formulate an evolution principle for $\boldsymbol{M}$, 
	\begin{equation}
		\Phi_{GB} =  \frac{\zeta_{s}}{2c_s} \boldsymbol{M} : \boldsymbol{M} - \mathcal{M}_R.
	\end{equation}
	where $\zeta_s$ is a positive coefficient. The evolution in $\boldsymbol{M}$ can be derived by minimizing the dual dissipating potential with plastic dissipation as a constraint. The Lagrangian to this minimization problem can be written as follows:
	\begin{equation}
		\mathcal{L}_{gb} = D_{gb} - \mathcal{M}_R \dot{|\boldsymbol{G}_s|} - \dot{\lambda}_s \Phi_{GB},
	\end{equation}
	$ \dot{\lambda}_s$ is a Lagrange multiplier for this optimization problem. Now, the evolution equations can be derived by minimizing the Lagrangian with respect to  the driving force $\boldsymbol{M}$, $\mathcal{M}_R$. Using the first-order optimality criteria, the following equation can be formed 
	\begin{subequations}
		\begin{gather}
			\label{partiail_1}
			\frac{\partial \mathcal{L}_{gb}}{\partial \boldsymbol{M}} = \boldsymbol{0}= \dot{\boldsymbol{G}}_s - \dot{\boldsymbol{G}}_{s,en} -  \dot{\lambda}_s \frac{\zeta_{s}}{c_s} \boldsymbol{M}, \\ 
			\frac{\partial \mathcal{L}_{gb}}{\partial \mathcal{M}_R} = 0 = - |\dot{\boldsymbol{G}}_s| +  \dot{\lambda}_s \implies  \dot{\lambda}_s = |\dot{\boldsymbol{G}}_s|.
		\end{gather}
	\end{subequations}
	Equation (\ref{energetic_str}) can be used to rewrite the equation (\ref{partiail_1}) as follows: 
	\begin{equation}
		\label{evol}
		\frac{\dot{\boldsymbol{M}}}{c_s}= 	\dot{\boldsymbol{G}}_s - |\dot{\boldsymbol{G}}_s| \frac{\zeta_{s}}{c_s} \boldsymbol{M} \implies \dot{\boldsymbol{M}} = c_s \dot{\boldsymbol{G}}_s -  \zeta_{s} |\dot{\boldsymbol{G}}_s| \boldsymbol{M}.
	\end{equation}
	Similar to bulk (equation (\ref{xi_evol})), the above equation is analogous to the nonlinear kinematic hardening model by Armstrong-Frederick \citep{armstrong1966mathematical}. The coefficient $c_s$ causes linear hardening due to GB Burgers tensor, and the second term is the relaxation term. $\zeta_s$ controls the percentage of relaxation from the grain boundary. Nonlinearity in hardening is incorporated due to the relaxation term. In the absence of a second term, the hardening model becomes equivalent to the Gurtin-type model \citep{gurtin2008theory}. The GB energetic stress evolution equation (\ref{evol}) can be rearranged as follows: 
	\begin{eqnarray}
		\dot{\boldsymbol{G}}_s -\dot{\boldsymbol{G}}_{s,en} = \dot{\boldsymbol{G}}_{s,dis} = \frac{\zeta_s}{c_s} |\dot{\boldsymbol{G}}_s| \boldsymbol{M}.
	\end{eqnarray} 
	Now, the dissipation inequality (\ref{dissip_gb}) can be written as follows: 
	\begin{eqnarray}
		D_{gb} = \frac{\zeta_s}{c_s} |\dot{\boldsymbol{G}}_s| \left({\boldsymbol{M}} : {\boldsymbol{M}}\right) \geqslant 0.
	\end{eqnarray}
	Due to the fact that $D_{gb}$ varies quadratically with $\boldsymbol{M}$, and the coefficients are all positive, dissipation will consistently remain positive. Consequently, when $\zeta_s \ne 0$, a fraction of the accumulated slips at GB will dissipate, leading to the eventual saturation of GB stress.
	\subsection{Stress update and material tangent calculation of GB energetic stress evolution equation} 
	Similar to section \ref{mat}, a backward time integration scheme is used between two consecutive time steps $t_n$ and $t_{n+1}$. The increment in energetic stress can be calculated from the time discretization of the evolution equation (\ref{evol}) as follows:
	\begin{eqnarray}
		\nonumber
		\Delta \boldsymbol{M} = \boldsymbol{M}_{n+1} -\boldsymbol{M}_n = c_s \Delta \boldsymbol{G}_s - \zeta_{s} |\Delta \boldsymbol{G}_s| \boldsymbol{M}_{n+1}, \\ 
		\nonumber
		\implies(1+ \zeta_{s} |\Delta \boldsymbol{G}_s|)	\boldsymbol{M}_{n+1}= c_s \Delta \boldsymbol{G}_s + \boldsymbol{M}_{n}, \\
		\implies \boldsymbol{M}_{n+1} = \frac{1}{1+ \zeta_{s} |\Delta \boldsymbol{G}_s|}( c_s \Delta \boldsymbol{G}_s + \boldsymbol{M}_{n}).
	\end{eqnarray}
	The above equation is nonlinear; therefore, material tangent at step $n$ must be known to proceed to the next step. The material tangent can be calculated as follows:
	\begin{gather}
		\nonumber
		(1+ \zeta_{s}|\Delta \boldsymbol{G}_s|) \frac{\partial \boldsymbol{M}_{n+1}}{\partial \Delta {\gamma}_I^\alpha} + \zeta_{s} \boldsymbol{M}_{n+1}\frac{\partial |\Delta\boldsymbol{G}_s|}{\partial \Delta{\gamma}^\alpha_I} = c_s \frac{\partial(\Delta\boldsymbol{G}_s)}{\partial \Delta \gamma_I^\alpha}, \qquad \forall ~I=A,B \\ 
		\label{moment_deriv}
		\implies \frac{\partial\boldsymbol{M}_{n+1}}{\partial \Delta\gamma_I^\alpha} = \frac{1}{1+\zeta_{s}|\Delta\boldsymbol{G}_s|} \left( c_s \frac{\partial(\Delta\boldsymbol{G}_s)}{\partial \Delta \gamma_I^\alpha} - \zeta_{s} \boldsymbol{M}_{n+1}\frac{\partial |\Delta\boldsymbol{G}_s|}{\partial \Delta{\gamma}^\alpha_I}\right).
	\end{gather} 
	\section{Two-dimensional version of the proposed model and its finite element implementation}
	\label{fem}
	This section explains a simple two-dimensional version of the proposed model for bulk and grain boundary and its finite element implementation. 
	\subsection{Simplified 2D version for bulk} 
	A plane section of the single crystal spanned by direction $x_1$, $x_2$, and $x_3$ direction is perpendicular to the plane with plane strain condition considered. Here, planer slip planes are considered; that is, slip plane $\boldsymbol{s}^\alpha$ and its normal $\boldsymbol{m}^\alpha$ lies on the same plane as the considered section.   Out-of-plane contribution to the slip system is zero; that is
	\begin{eqnarray}
		\boldsymbol{s}^\alpha \cdot \boldsymbol{e}_3=0, \qquad\boldsymbol{m}^\alpha \cdot \boldsymbol{e}_3=0, \qquad\boldsymbol{s}^\alpha \times \boldsymbol{m}^\alpha = \boldsymbol{e}_3=-\boldsymbol{I}^\alpha,
	\end{eqnarray}
	where $\boldsymbol{e}_3$ is the unit vector perpendicular to the plane section. The slip $\gamma^\alpha$ is independent of direction $x_3$. This condition makes the screw dislocation to vanish; that is 
	\begin{eqnarray}
		\rho^\alpha_\odot = \boldsymbol{I}^\alpha \cdot \boldsymbol{\kappa}^\alpha=0.
	\end{eqnarray}
	The tangential gradient of the plastic slip can be written as 
	\begin{eqnarray}
		\boldsymbol{\kappa}^{t\alpha} = (\boldsymbol{s}^\alpha \cdot \boldsymbol{\kappa}^\alpha) \boldsymbol{s}^\alpha .
	\end{eqnarray}
	If a slip system $\alpha$ makes an $\theta$ angle to the positive $x_1$ axis, then the slip plane and its normal can be written as follows:
	\begin{eqnarray}
		\boldsymbol{s}^\alpha = \cos(\theta) \boldsymbol{e}_1 + \sin (\theta) \boldsymbol{e}_2, \qquad \text{and} \qquad\boldsymbol{m}^\alpha = -\sin(\theta) \boldsymbol{e}_1+ \cos(\theta)\boldsymbol{e}_2, 
	\end{eqnarray}
	where $\boldsymbol{e}_1$ and $\boldsymbol{e}_2$ are the unit vector along $x_1$ and $x_2$ direction respectively. Considering the isotropic material property and the proposed defect energy (\ref{p_e}), the constitutive relations to the proposed theory can be summarized as 
	\begin{eqnarray}
		\begin{cases}
			\boldsymbol{\sigma} = \lambda(\text{tr}~ \boldsymbol{\varepsilon}^e) \boldsymbol{I} + 2\mu \boldsymbol{\varepsilon}^e, \\ 
			\pi^\alpha  = S^\alpha R(\dot{\bar{d}}^\alpha) \frac{\dot{\gamma}}{\dot{\bar{d}}^\alpha}, \qquad \dot{\bar{d}}^\alpha = |\dot{\gamma}^\alpha|,\\
			\dot{\boldsymbol{\xi}}^\alpha = S_0L_*^2 (\boldsymbol{s}^\alpha \cdot \dot{\boldsymbol{\kappa}}^\alpha)\boldsymbol{s}^\alpha - \dot{\bar{d}}^\alpha \zeta \boldsymbol{\xi}^\alpha.
		\end{cases}
	\end{eqnarray}
	The virtual power statement of the SGCP theory consists of displacement and plastic slip as variables. Therefore, in order to implement the SGCP theory, one needs a mixed element with plastic slip and displacement as nodal variables. For the present numerical investigation, we have considered eight-node serendipity elements with plastic slip and displacement as primary variables \citep{mukherjee2023strain,mukherjee2024elastic}. Full numerical integration is carried out by considering the $3 \times 3$ quadrature rule. A backward Euler scheme is used for the time integration of the vector microscopic stress. The element stiffness matrix and force vector are prepared with the help of a user subroutine (UEL), which is assembled and solved with the help of the commercial finite element package ABAQUS/Standard \citep{abaqus}. Detailed finite element implementation is explained in  \ref{2dimplement}. 
	\subsection{2D Finite element implementation of the grain boundary model} 
	In order to show the efficacy of the proposed model, a simple two-dimensional version of the grain boundary model is implemented here. The grain boundary and its normal lie in the same plane as the two grains.
	
	The grain boundary model is implemented with the help of a three-node zero-thickness mixed element similar to \cite{ozdemir2014modeling}. The displacement and plastic strain of the nodes are considered the primary variables. The zero-thickness element behaves as a one-dimensional element, where the variable values vary quadratically along the direction of the element. Numerical integration is carried out using the full three-point quadrature rule. Time integration of the GB energetic stress evolution equation (\ref{evol}) is carried out considering the backward Euler time integration scheme. The element stiffness matrix and force vector are prepared with the help of the user subroutine UEL, and they are augmented to the global stiffness matrix of the bulk. Finite element solution is carried out in commercial package ABAQUS/Standard \citep{abaqus}. Detailed finite element implementation and stiffness matrix calculation are provided in  \ref{gb_model}. 
	\section{Result and discussion}
	\label{Result}
	\subsection{Infinite shear layer with double slip system}
	\begin{figure}[h!]
		\centering
		\includegraphics[width=0.35\textwidth]{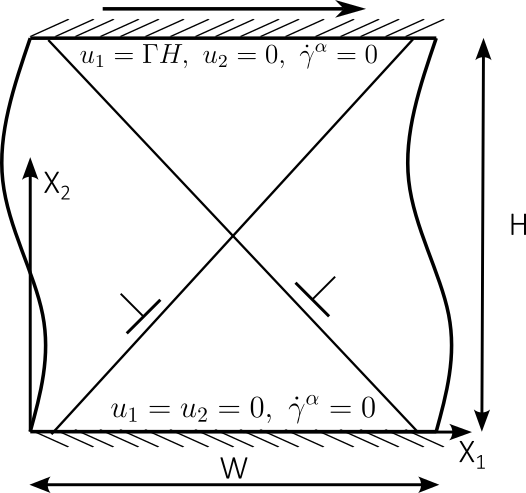}
		\caption{The infinite shear layer with boundary conditions}
		\label{inf_shear_layer}
	\end{figure}  
	In order to show the effectiveness of the proposed model, a simple two-dimensional version of the proposed theory is applied to analyze one infinite shear layer. The layer spans infinite along $x_1$ direction and is of finite thickness along $x_2$ direction, as shown in figure \ref{inf_shear_layer}.   Two symmetric slip systems with $\theta=\pm60^0$ are considered. Response of the shear layer will only depend on $x_2$, making the problem one-dimensional. The $x_2$ direction is divided into 100 elements, while only single layer of element is used along $x_1$, which is satisfactory. 
	
	The layer is fixed at the top and bottom, with displacement being implemented solely at the top. The boundary conditions can be written as  
	\begin{eqnarray}
		\nonumber
		u_1 = u_2  = 0	\qquad \text{at} \qquad x_2=0, \qquad \forall x_1\\
		u_1 = H \Gamma(t), u_2=0 \qquad \text{at}  \qquad x_2=H, \qquad \forall x_1 \nonumber\\
		\gamma^\alpha(x_1,0,t)=\gamma^\alpha(x_1,H,t)=0, \qquad \text{for}~ \alpha=1, ~2 \qquad \forall x_1,t 
	\end{eqnarray}
	where $\Gamma$ is the value of the applied shear strain. 
	The infinite length of the layer is implemented by imposing periodic boundary conditions along the left and right sides of the layer and can be written as 
	\begin{eqnarray}
		\nonumber
		u_i (0,x_2,t) = u_i(W,x_2,t), 	\qquad \text{for}~ i=1,2 	\qquad \forall ~x_2,t \\
		\gamma^\alpha (0,x_2,t) = \gamma^\alpha(W,x_2,t), 	\qquad \text{for}~ \alpha=1,2 	\qquad \forall~ x_2,t
	\end{eqnarray}
	The material properties used in finite element implementation are taken from \cite{gurtin2007gradient}, listed in table \ref{mat_prop}. 
	\begin{table}[h!]
		\centering
		\caption{Material properties used in the infinite shear layer problem}
		\label{mat_prop}
		\begin{tabular}{ccccl}
			\hline
			Material parameter name  & Symbol & Value & Unit  \\
			\hline
			Young's Modulus  & E      &   260 & GPa          \\
			Poisson's ratio      &  $\nu$     &   0.3 &    \\
			Reference slip rate	& $\dot{d_0}$     &    0.02 & $s^{-1}$         \\
			Rate-sensitive exponent 	&  $m$      & 0.05  &          \\
			Initial slip resistance	&   $S_0$     &   50    &  MPa        \\  
			\hline   
		\end{tabular}
	\end{table}
	\subsubsection{Validation of the model implementation}
\begin{figure}[h!]
		\centering
		\begin{subfigure}{0.49\textwidth}
			\includegraphics[width=\textwidth]{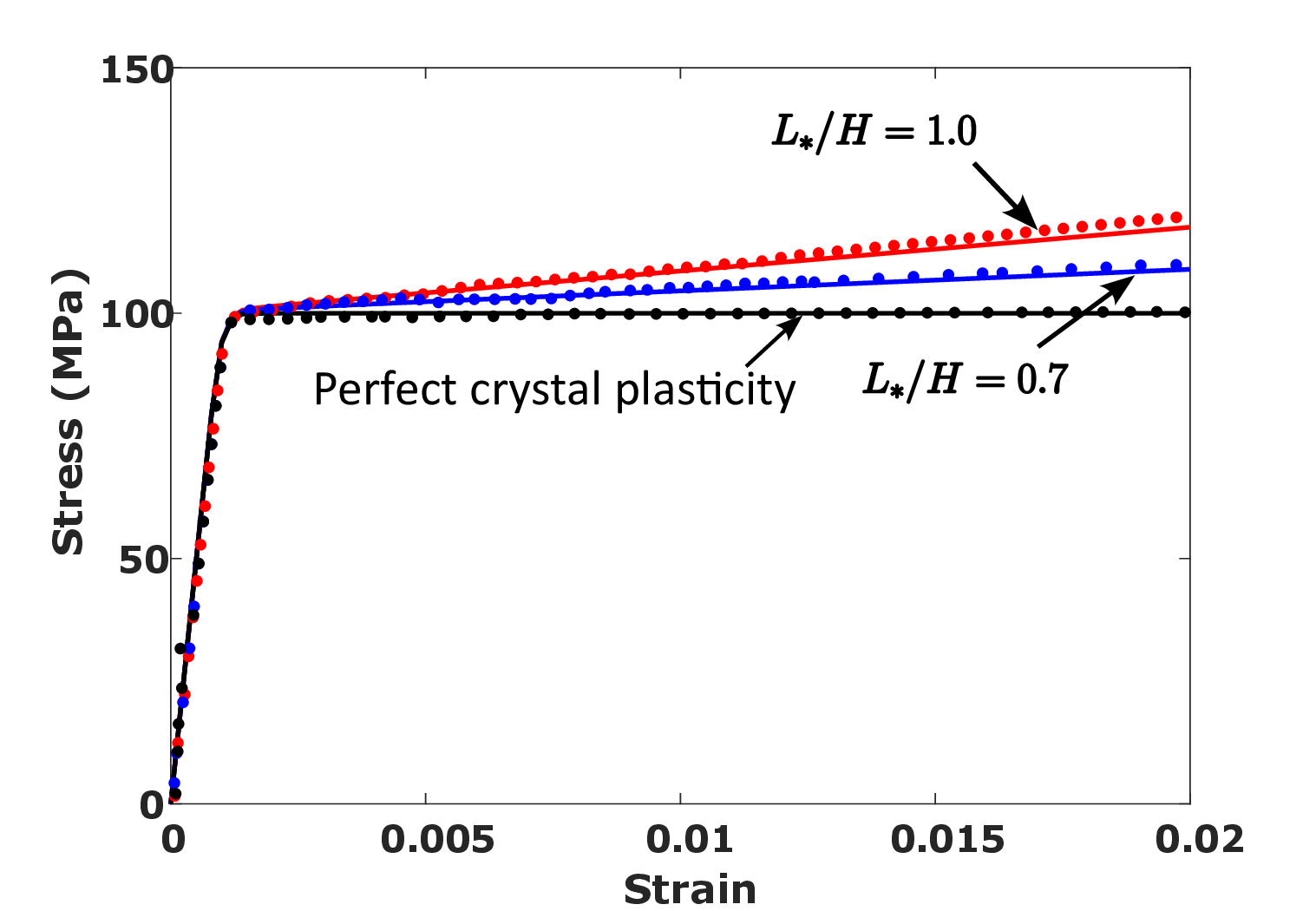}
			\caption{Stress strain curve}
			\label{validation_raction}
		\end{subfigure} 
		\begin{subfigure}{0.49\textwidth}
			\includegraphics[width=\textwidth]{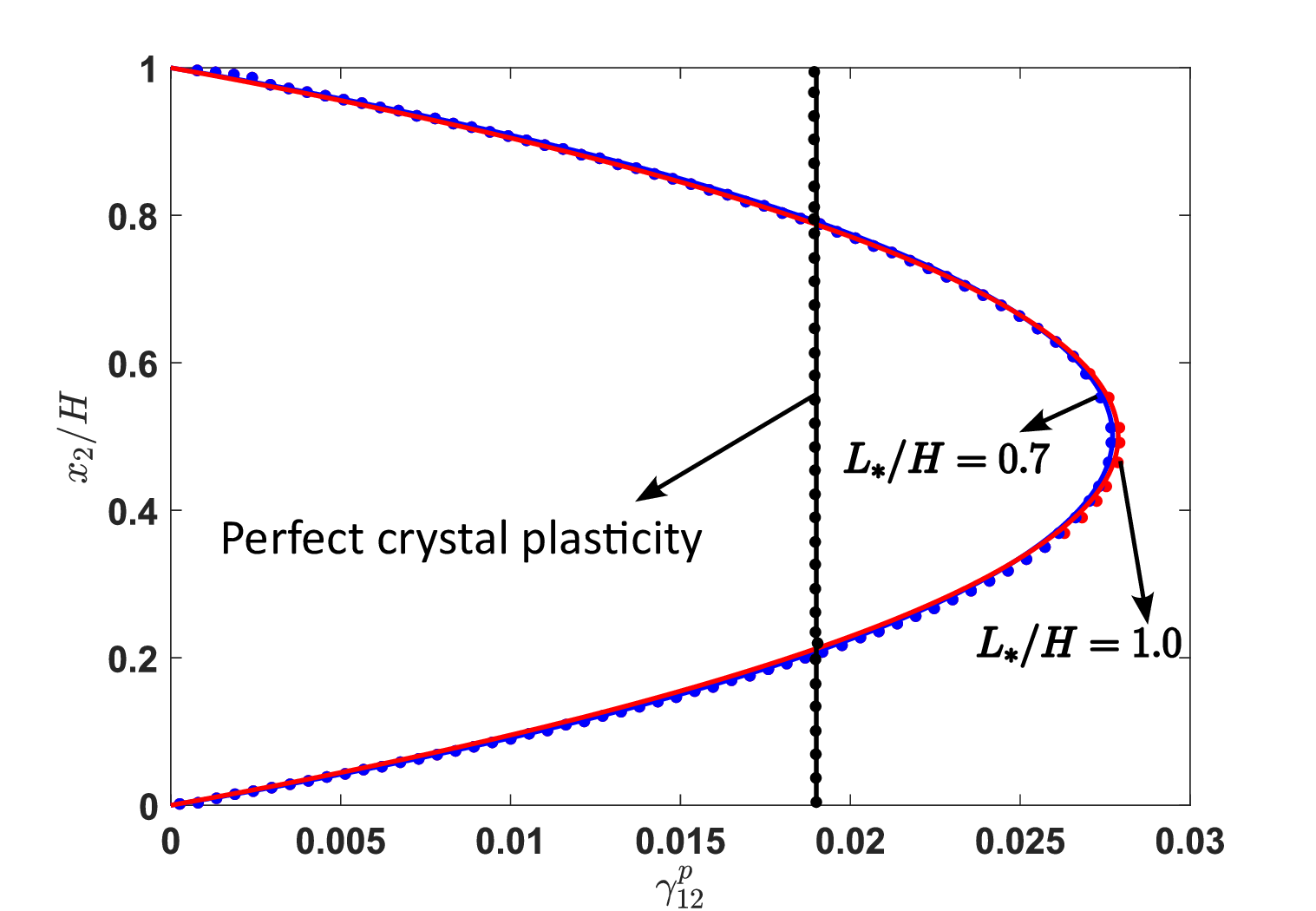}
			\caption{Plastic strain distribution}
			\label{validation_strain}
		\end{subfigure} 
		\caption{Validation of the model implementation with results from \cite{gurtin2007gradient} (Firm lines are FEM solution and dots are taken from literature)}
		\label{validation}
	\end{figure}

	In order to validate our model implementation, we consider the \cite{gurtin2007gradient} theory. In the absence of the $\zeta$, the proposed model becomes equivalent to the Gurtin-type energetic model. Considering $\zeta=0$ and $L_e = L_*$, the shear layer is analyzed. The stress-strain curve obtained with different $L_e$ is compared in figure  \ref{validation_raction}, and the corresponding plastic strain distribution is compared in figure \ref{validation_strain}. The implemented model perfectly matches the result obtained by \cite{gurtin2007gradient}, which validates our model implementation.   
	\subsubsection{Effect of $\zeta$ on monotonic loading}
	\begin{figure}[h!]
		\centering
		\begin{subfigure}{0.45\textwidth}
			\includegraphics[width=\textwidth]{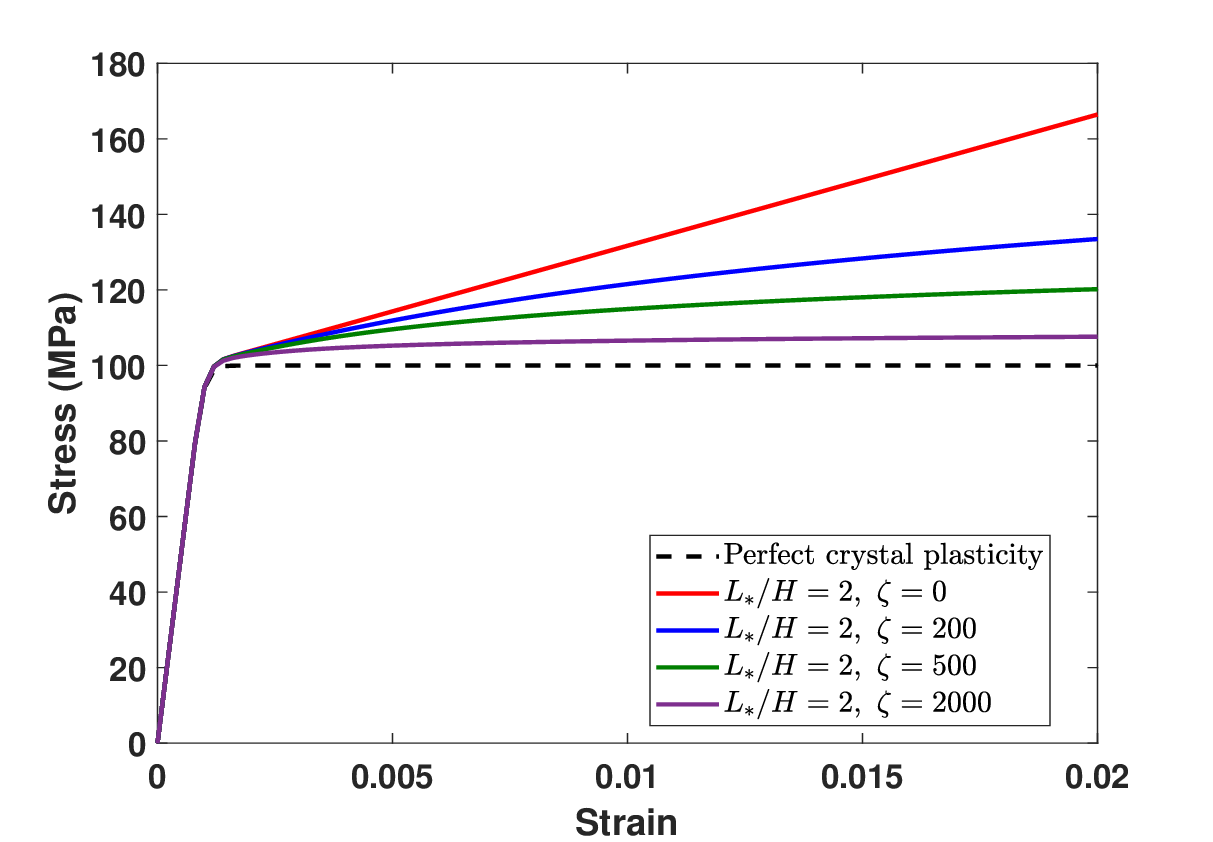}
			\caption{Stress-strain curve}
			\label{zeta_effect_fd_bulk}
		\end{subfigure} 
		\begin{subfigure}{0.45\textwidth}
			\includegraphics[width=\textwidth]{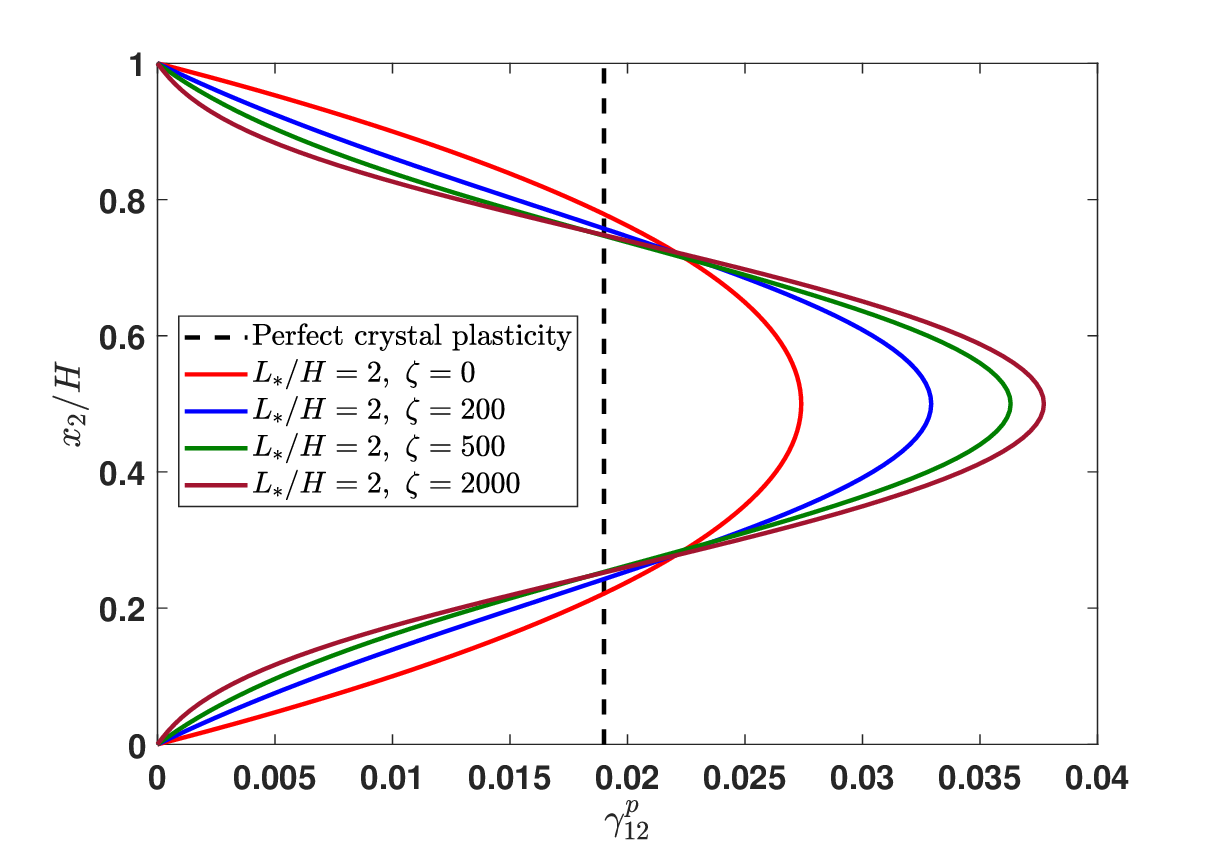}
			\caption{Plastic strain distribution}
			\label{zeta_effect_strain_bulk}
		\end{subfigure} 
		\begin{subfigure}{0.45\textwidth}
			\includegraphics[width=\textwidth]{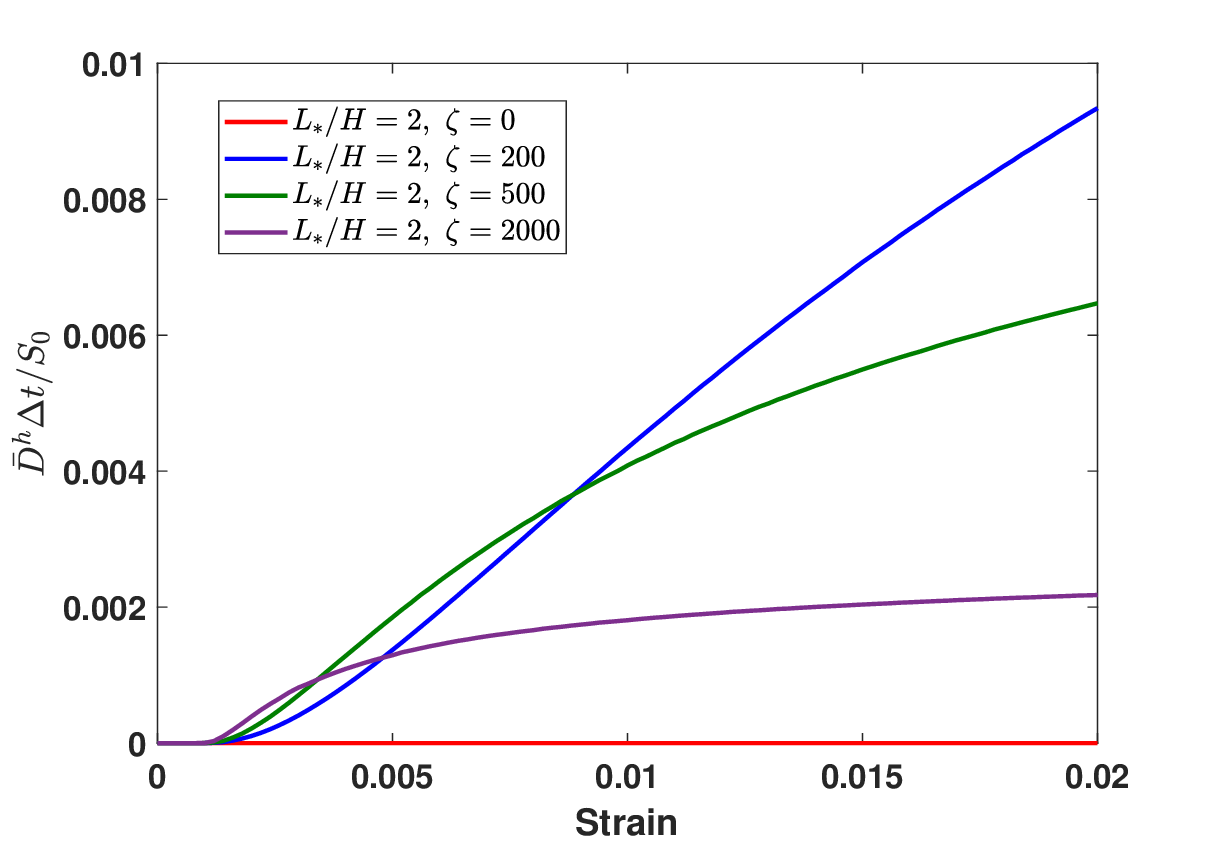}
			\caption{Average higher-order dissipation}
			\label{zeta_effect_dissip_bulk}
		\end{subfigure} 
		\begin{subfigure}{0.45\textwidth}
			\includegraphics[width=\textwidth]{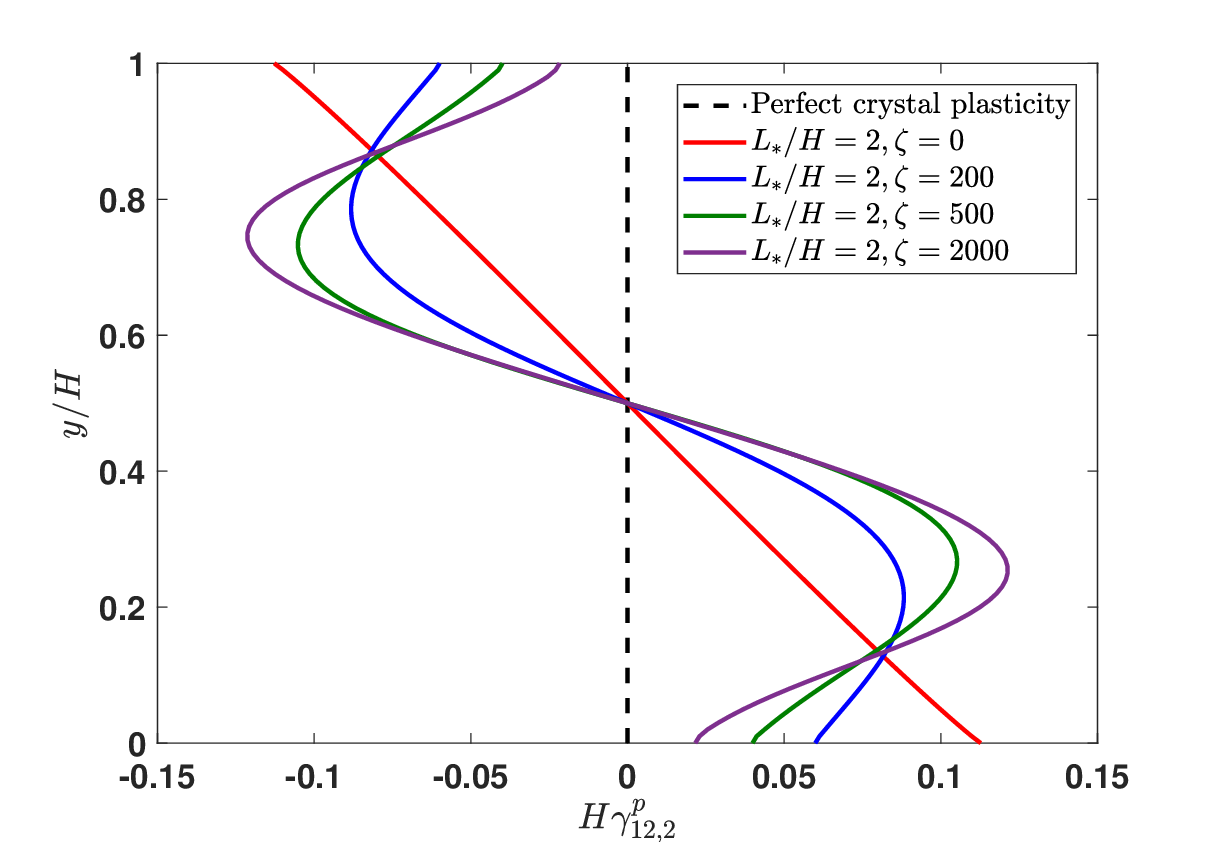}
			\caption{Plastic strain gradient distribution ($\partial\gamma^p_{12}/\partial x_2$)}
			\label{zeta_effect_str_grad_bulk}
		\end{subfigure} 
		\caption{Effect of $\zeta$ on monotonic loading}
	\end{figure}
	This section explains the effect of considering $\zeta$ to the infinite shear layer problem. The shear layer is  considered with $L_*/H=2$, and $\zeta$ value is increased slowly. The shear stress-strain curve for different $\zeta$ values is presented in figure \ref{zeta_effect_fd_bulk}. In the case of $\zeta=0$, the proposed model coincides with the Gurtin-type energetic model, possessing linear hardening. $\zeta$ brings the nonlinearity to the hardening as observed from the figure \ref{zeta_effect_fd_bulk}. At a higher $\zeta$ value, a saturation to the hardening is observed possessing a pseudo strengthening.
	
	We introduce average higher order dissipation ($\bar{D}^h$) as
	\begin{equation}
		\bar{D}^h=\frac{1}{H} \int_0^H D^h(x_2) dx_2, \qquad D^h =\sum_\alpha \boldsymbol{\xi}^\alpha \cdot \dot{\boldsymbol{\kappa}}^{t\alpha}_{dis},
	\end{equation} 
	where $D^h$ is the higher-order dissipation related to the plastic slip gradient. The average higher-order dissipation is plotted in figure \ref{zeta_effect_dissip_bulk}. In the case of $\zeta = 0$, the model possesses no dissipation (The Gurtin-type model has no dissipation for pure energetic conditions). However, with the incorporation of $\zeta$, additional dissipation is observed, which is generally borne by $L_d$ in the Gurtin type model. With the increase in $\zeta$, the $\bar{D}^h$ increases and then saturates. At a higher value of $\zeta$, the hardening almost equals the recovery term, and consequently, dissipation becomes almost constant.
	
	The plastic strain profile is presented in figure  \ref{zeta_effect_strain_bulk}. In case $\zeta=0$, the plastic strain takes a parabolic form. With the increase in $\zeta$, the plastic strain at the middle increases; however, the symmetricity of the plastic strain is maintained for all values of $\zeta$. 
	Plastic strain gradient $\partial \gamma^p_{12}/\partial x_2$ is presented in figure  \ref{zeta_effect_str_grad_bulk}. In the case of linear hardening, the plastic strain gradient is linearly varying. The nonlinearity in plastic strain gradient is introduced with $\zeta$.
	With the increase in   $\zeta$, the slope of the plastic strain decreases near the top and bottom, where the plastic slip is restricted. However, the plastic strain gradient at the middle remains zero for all $\zeta$ values.  
	\subsubsection{Effect of $\zeta $ on the cyclic loading}
	\begin{figure}[h!]
		\centering
		\begin{subfigure}{0.49\textwidth}
			\includegraphics[width=\textwidth]{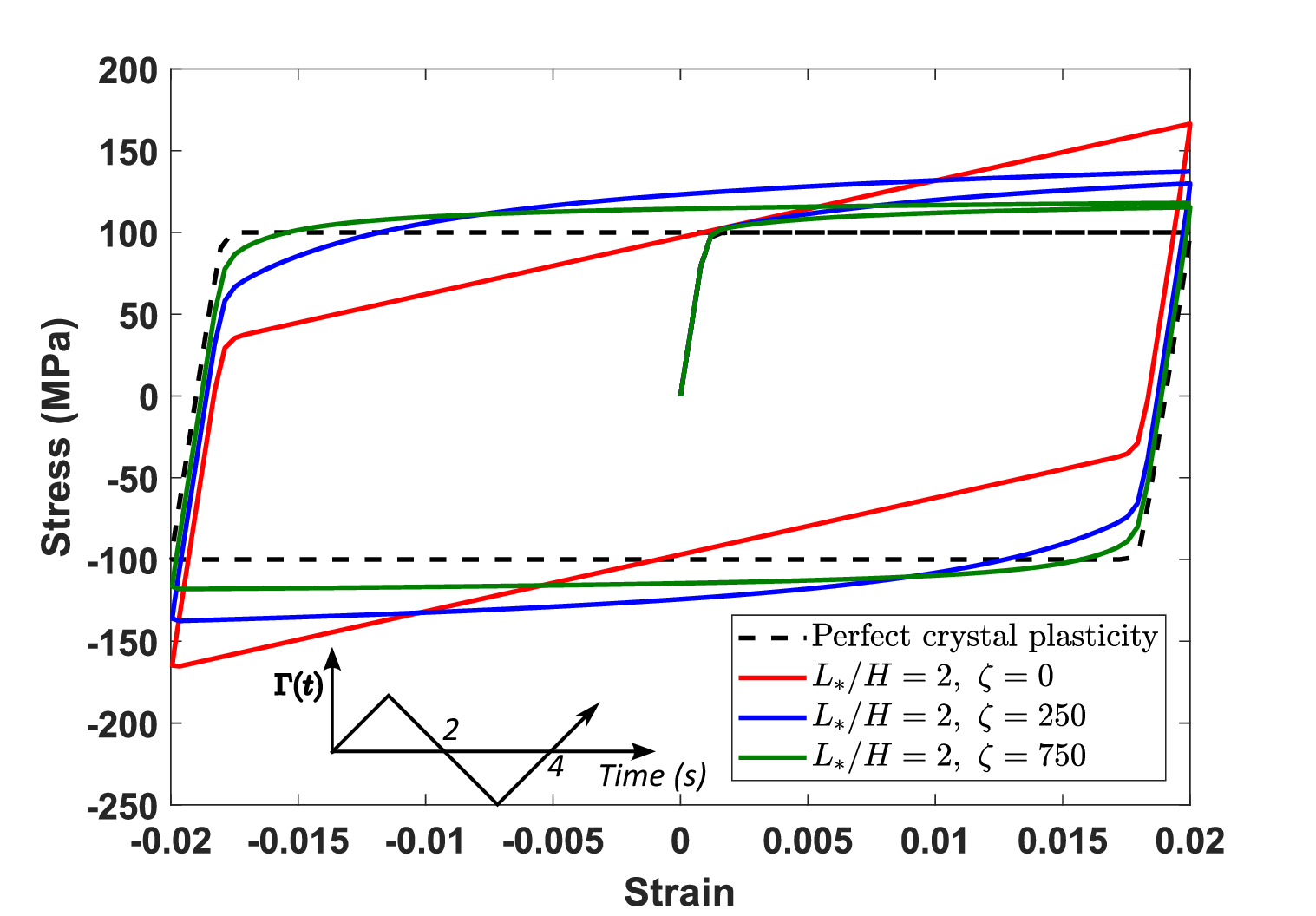}
			\caption{Stress-strain curve}
			\label{zeta_effect_cyclic_fd}
		\end{subfigure} 
		\begin{subfigure}{0.49\textwidth}
			\includegraphics[width=\textwidth]{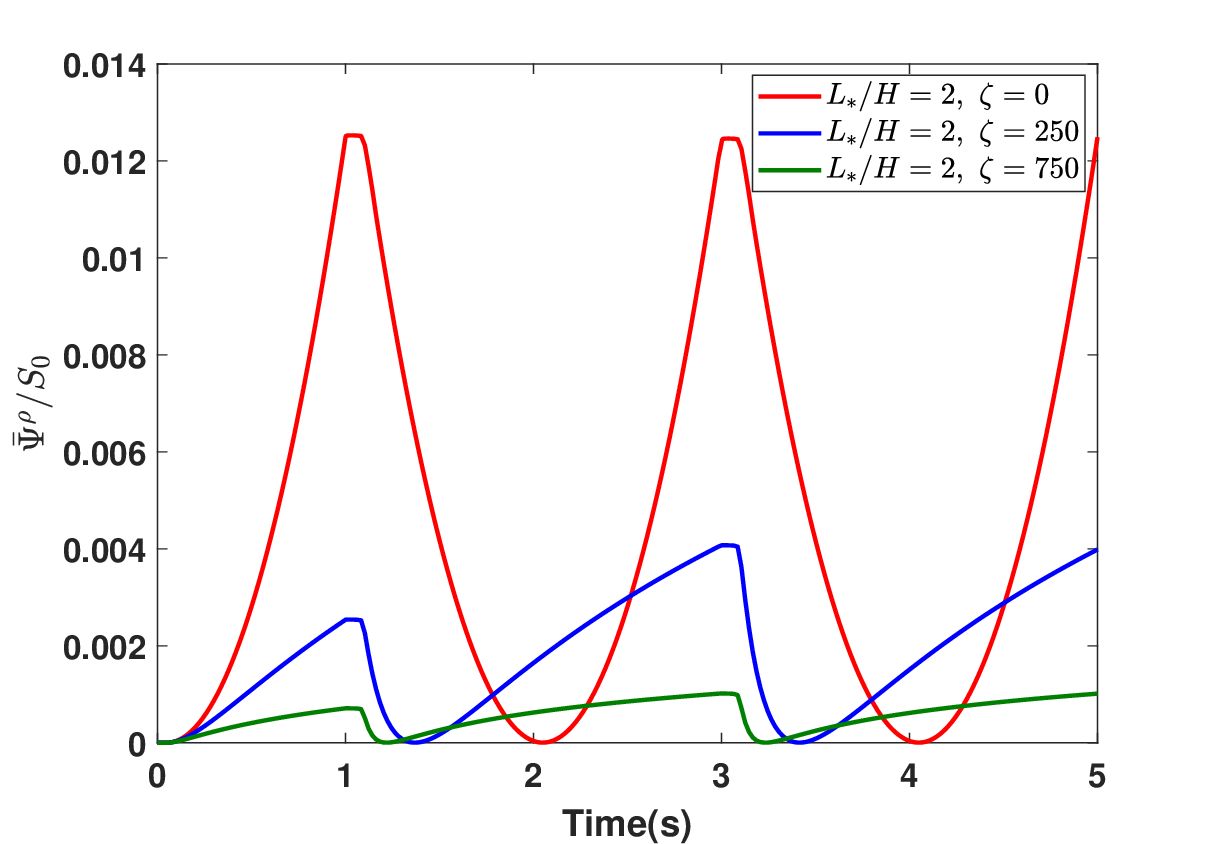}
			\caption{Average defect energy}
			\label{zeta_effect_cyclic_energy}
		\end{subfigure} 
		\caption{Effect of $\zeta$ on cyclic loading}
	\end{figure}
	The previous example is now considered to show the effectiveness of the proposed model during cyclic loading. The shear layer is subjected to linearly varying shear strain to its top. One complete cyclic loading is applied within 4 s time. Stress-strain curve during cyclic loading is presented in figure \ref{zeta_effect_cyclic_fd}, and the corresponding average defect energy variation ($\bar{\Psi}^\rho$) is depicted in figure \ref{zeta_effect_cyclic_energy}. The average defect energy can be written as follows:
	\begin{equation}
		\bar{\Psi}^\rho=\frac{1}{H}\int_0^H \Psi^\rho(x_2) dx_2 , \qquad \Psi^\rho(x_2)= \frac{S_0L_*^2}{2 }\sum_\alpha \boldsymbol{\kappa}^{t\alpha}_{en}(x_2) \cdot \boldsymbol{\kappa}^{t\alpha}_{en}(x_2). \label{avg_en}
	\end{equation}
	
	In the absence of $\zeta$, the shear post-yield stress-strain curve is observed to be linear. However, as $\zeta$ increases, a nonlinearity in the stress-strain curve becomes apparent. As $\zeta$ values increase, the stress-strain curve reaches a saturation point, and the hardening effect becomes negligible. It may noted here that we did not observe any unusual kinematic hardening during reverse loading (concavity change during reverse plasticity), unlike other nonlinear kinematic hardening models based on non-quadratic defect energy \citep{WULFINGHOFF20151,bardella2015modelling,jebahi2020strain,jebahi2023alternative}.
	
	In the absence of $\zeta$ with no dissipation, the defect energy is a quadratic functional with total plastic slip gradient as demonstrated in  figure \ref{zeta_effect_cyclic_energy}. During the loading stage, the average defect energy increases and remains constant during the elastic unloading phase, after which it decreases. However, when $\zeta$ is present, the defect energy deviates from its quadratic behavior over time. This deviation is due to the fact that $\zeta$ introduces dissipation into the system, thereby influencing both the energetic and dissipative components of the slip gradient. As the strain values increase, the defect energy reaches a saturation point, indicating a saturation in higher-order stress. The recovery coefficient $\zeta$ is associated with energy relaxation. Consequently, for a fixed $L_*$, an increase in $\zeta$ decreases the defect energy.
	
	\subsubsection{Response of the shear layer during non-proportional loading condition} 
	\begin{figure}[h!]
		\centering
		\begin{subfigure}{0.49\textwidth}
			\includegraphics[width=\textwidth]{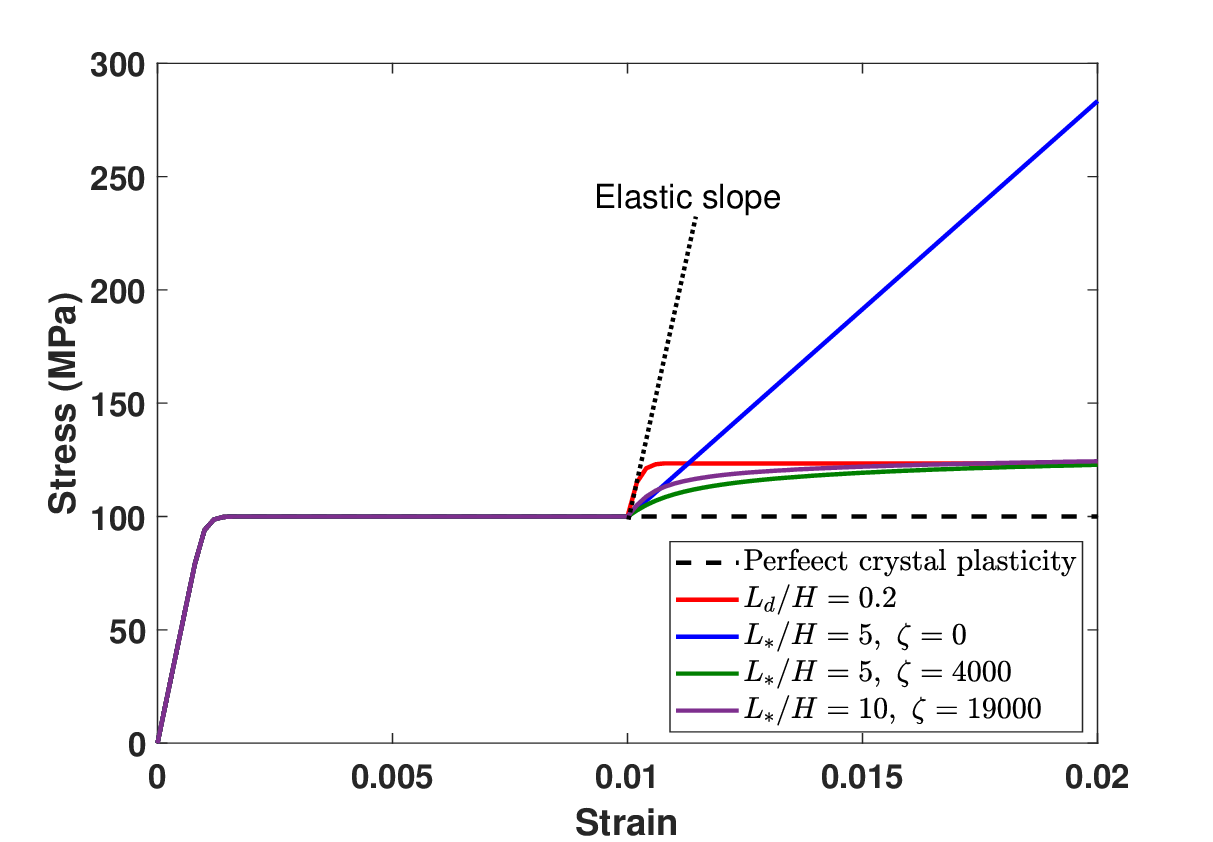}
			\caption{Stress-strain curve}
			\label{elastic_gap_fd}
		\end{subfigure} 
		\begin{subfigure}{0.49\textwidth}
			\includegraphics[width=\textwidth]{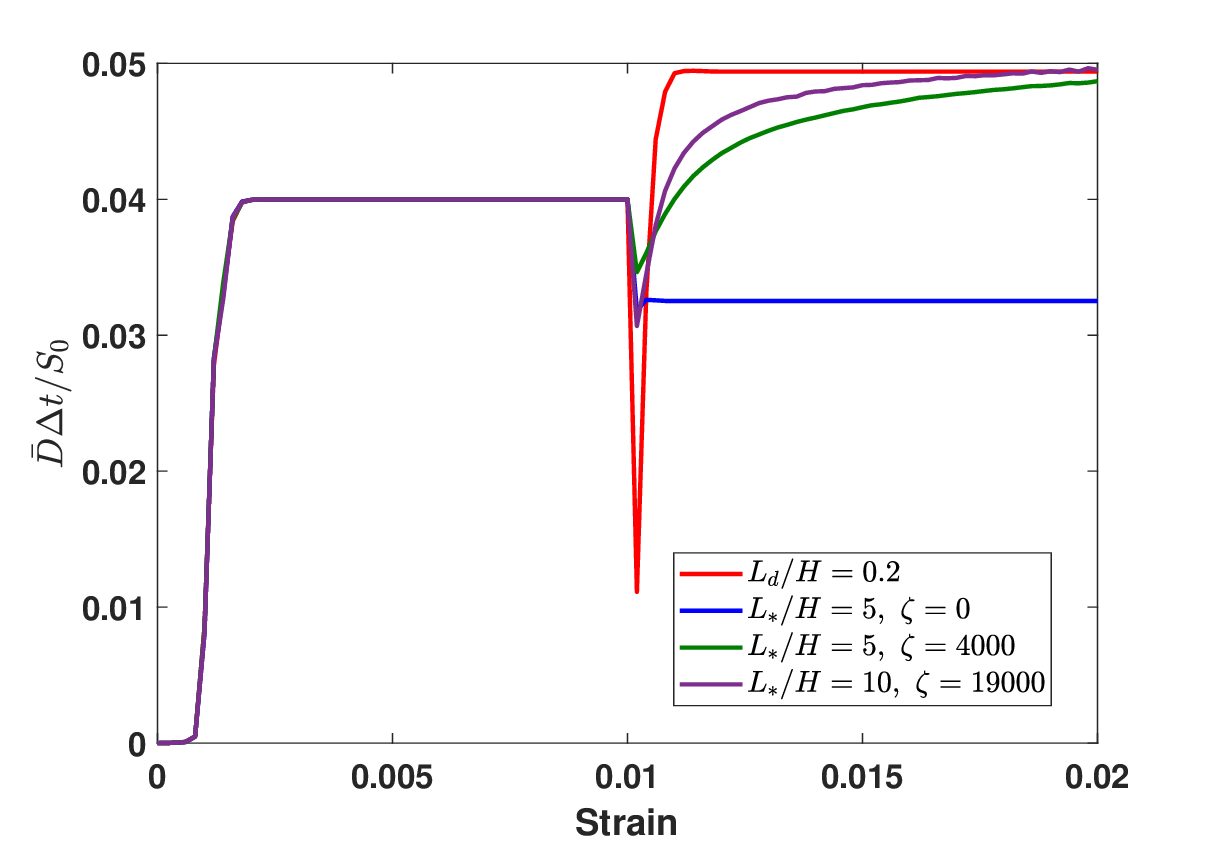}
			\caption{Average dissipation $\bar{D}$}
			\label{elastic_gap_dissip}
		\end{subfigure} 
		\begin{subfigure}{0.49\textwidth}
			\includegraphics[width=\textwidth]{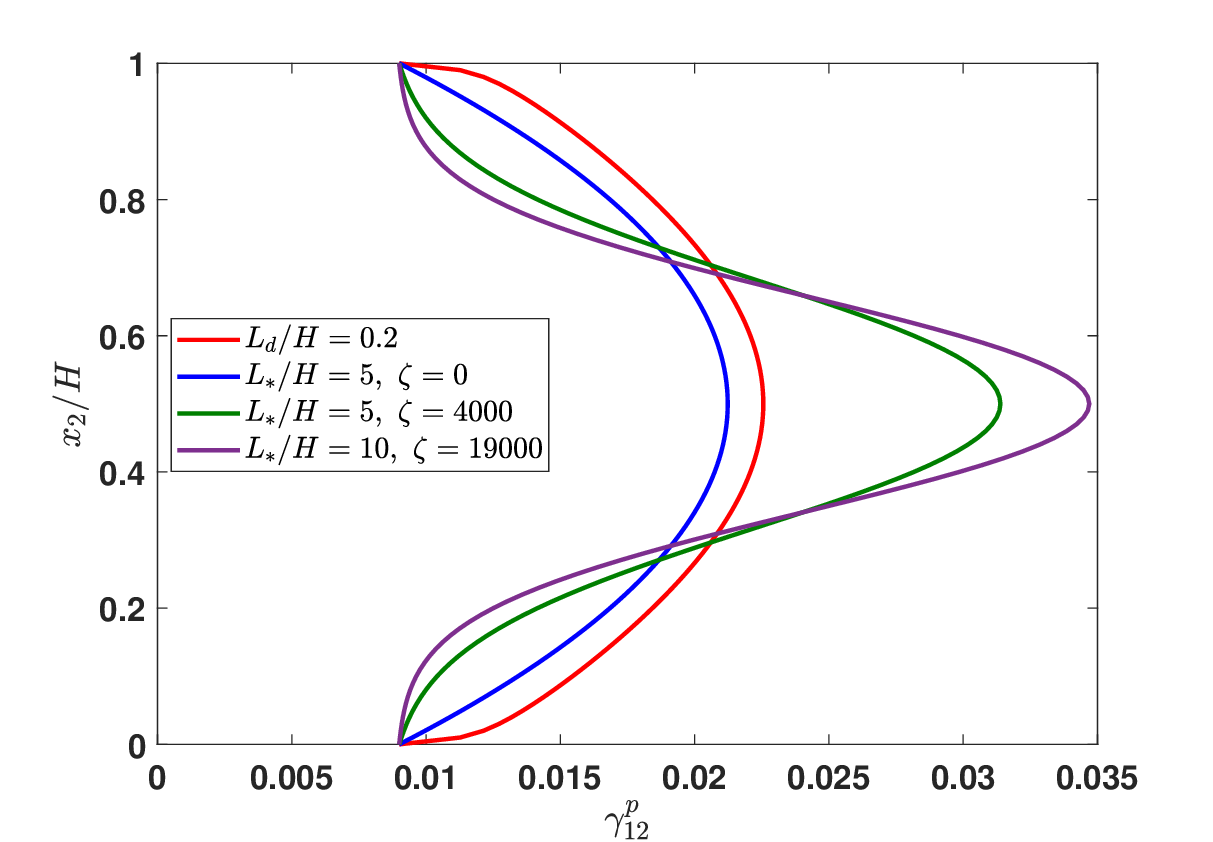}
			\caption{Plastic strain distribution}
			\label{elastic_gap_strain}
		\end{subfigure} 
		\begin{subfigure}{0.49\textwidth}
			\includegraphics[width=\textwidth]{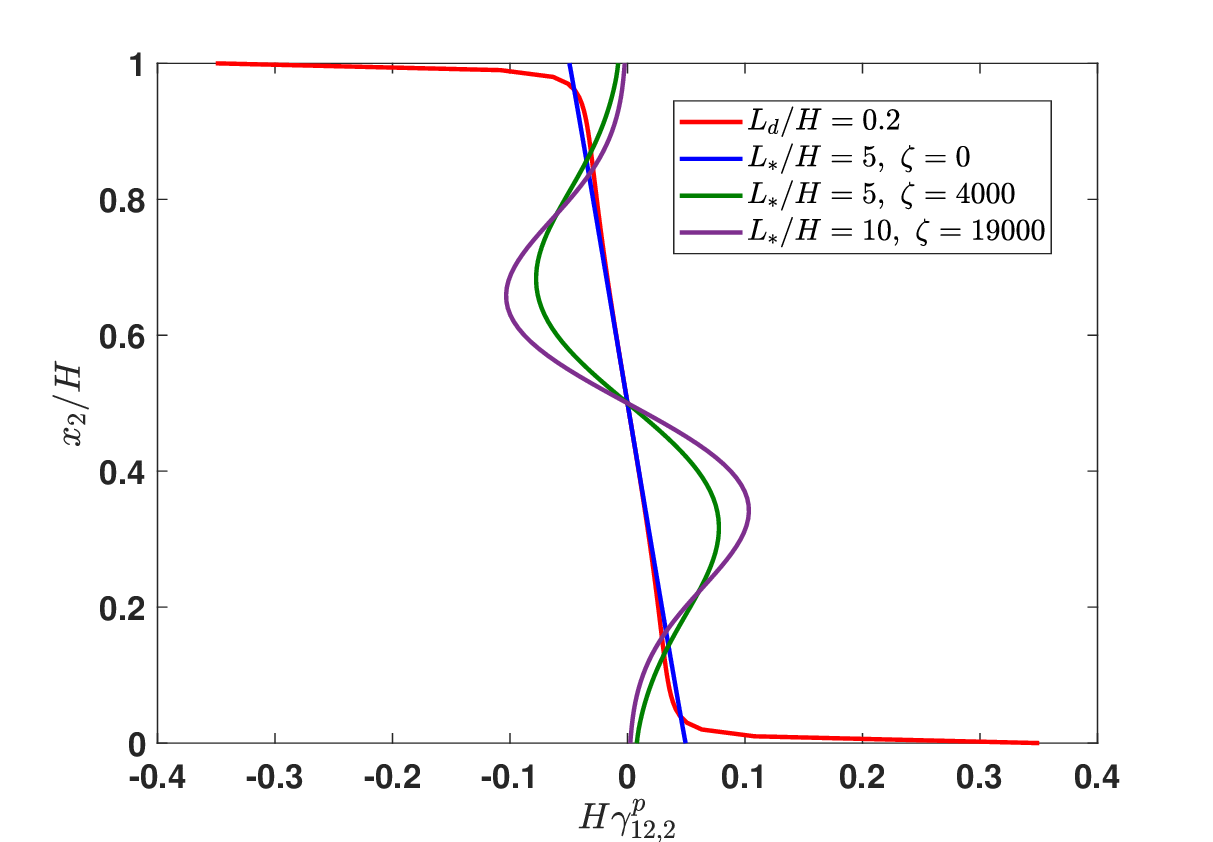}
			\caption{Plastic strain gradient ($\partial \gamma_{12}^p/\partial x_2$) distribution}
			\label{elastic_gap_str_grad}
		\end{subfigure} 
		\caption{Response of the shear layer during non-proportional loading condition}
		\label{elastic_gap}
	\end{figure}
	In this problem, the performance of the proposed model is assessed via the non-proportional loading condition of the infinite shear layer. The previous infinite shear layer is considered, and it is sheared up to 2 \% strain. During the initial 1\% strain, the top and bottom surfaces are micro-free. Next, the top and bottom boundary conditions are abruptly changed into micro hard at a 1\% strain value. 
	
	In the non-proportional loading condition, the micro hard boundary condition (plastic slip rate equals zero) is applied at time $t_0$. The boundary conditions of plastic slip at the top and bottom are initially unrestricted; then, the restriction is applied at time $t_0$. The boundary condition of plastic slip can be written as 
	\begin{equation}
		\label{gap_bc}
		\dot{\gamma}^\alpha(x_1,0,t > t_0)=\dot{\gamma}^\alpha(x_1,H,t>t_0)=0, \qquad \text{for}~ \alpha=1, ~2 \qquad \forall x_1, t > t_0.
	\end{equation}  
	The shear response is shown in figure \ref{elastic_gap_fd}. Gurtin-type theory with $L_d=0$ does not follow the elastic slope onset of application of micro hard boundary condition. However, $L_d\ne0$ shows a jump in shear response, causing an elastic-gap. It is observed that the proposed model can asymptotically follow the shear response for $L_d \ne 0$. Different $L_*, ~\zeta$ combinations can be used to smoothly maneuver the $L_d$ curve. With the increase in $L_*$ values, the curve saturates earlier; however, the requirement of $\zeta$ also increases, bringing more nonlinearity. A similar observation is available in \cite{mukherjee2024elastic} concerning isotropic SGP theory. Now we introduce an average dissipation $\bar{D}$ as follows: 
	\begin{align}
		\bar{D} = \frac{1}{H} \int_0^H D_b(x_2) dx_2 
	\end{align}
	Variation of average dissipation over time is represented in figure \ref{elastic_gap_dissip}. It is observed that up to 1\% strain, the dissipation is due to plastic slip only. Due to the sudden application of the micro hard boundary, the microscopic force is reduced by incorporating dissipative vector microscopic stress. Therefore, an instantaneous reduction in dissipation is observed at 1\% strain; however, the presence of $L_d$ quickly saturates the dissipation. Our proposed model asymptotically follows the dissipation caused by the nonzero $L_d$ curve. 
	
	The average dissipation at saturation with the proposed model equals the Gurtin type model with nonzero $L_d$ as evident from figure \ref{elastic_gap_dissip}. Plastic strain and its gradient at 2\% strain value are presented in figure \ref{elastic_gap_strain} and \ref{elastic_gap_str_grad} respectively. It is observed that the proposed model may possess equivalent shear response or dissipation to the Gurtin-type dissipative model ($L_d=0.2$); however, the plastic strain or its gradient profile is different. In addition, unlike the Gurtin theory, the present theory utilizes the relaxation term of the plastic strain gradient hardening evolution equation (thus energetic moment stress) to match the dissipative hardening caused by dissipative moment stress in Gurtin-type theory.    
	\subsection{Shearing of a periodic bicrystal}
	\begin{figure}[h!]
		\centering
		\includegraphics[width=0.4\textwidth]{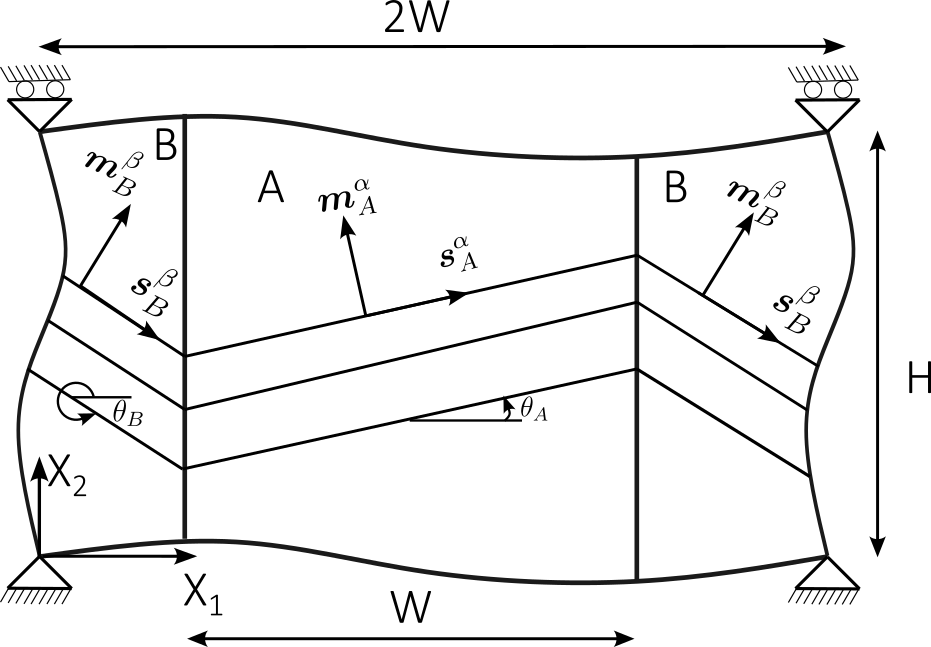}
		\caption{The periodic bicrystal}
		\label{bicrystal}
	\end{figure}
	In order to show the efficacy of the proposed grain boundary model,  one periodic bicrystal \citep{van2013grain} is considered. The Periodic bicrystal consists of two types of columnar grains of width $W$, alternating, as shown in figure \ref{bicrystal}. The bicrystal is modeled by considering one full-grain (A) and two half-grain (B) placed alternately. The bicrystal has infinite length along the $x_2$ direction, modeled by considering H thickness with periodic boundary at the top and bottom. The infinite numbers of alternating grains are modeled by applying periodic boundary conditions at left and right. Both grains are assumed to have a single slip system with slip angle $\theta_A$ and $\theta_B$. 
	
	It is shown in \cite{van2013grain} that the considered problem is essentially one dimension in nature; that is, the response depends only on $x_1$. Therefore, a single layer of mesh is considered along the $x_2$ direction. Each grain is discretized with 100 elements along the $x_1$ direction. The bicrystal is subjected to shear strain $\Gamma $, which is applied considering the periodicity.  The periodic boundary condition of this shearing problem can be written as 
	\begin{eqnarray}
		\nonumber
		\text{Left-right:}~~	\boldsymbol{u}(0,x_2,t) = \boldsymbol{u}(2W,x_2,t), \qquad \gamma^\alpha(0,x_2,t) = \gamma^\alpha (2W,x_2,t)\qquad \forall~ x_2,\\
		\text{Top-bottom:}~~	\boldsymbol{u}(x_1,H,t) - \boldsymbol{u}(x_1,0,t)=\Gamma H, \qquad \gamma^\alpha(x_1,H,t) = \gamma^\alpha (x_1,0,t)\qquad \forall~ x_1.
	\end{eqnarray}
	The material properties \citep{van2013grain,lei_cai} used in the computation are listed in table \ref{bicrystal_shear}.
	\begin{table}[h!]
		\centering
		\caption{Material properties used in periodic bicrystal shear}
		\label{bicrystal_shear}
		\begin{tabular}{ccccl}
			\hline
			Material parameter name  & Symbol  & Values & Unit  \\
			\hline
			Young's Modulus  & E      &   60.84  & GPa          \\
			Poisson's ratio      &  $\nu$     &    0.3 &  -  \\
			Reference slip rate	& $\dot{d_0}$     &     0.001 & $s^{-1}$         \\
			Rate-sensitive exponent 	&  $m$       & 0.05  &   -       \\
			Initial slip resistance	&   $S_0$     &    60.84   &  MPa        \\  
			Inclination of first slip of  grain A & $\theta_A$ & $10$  & Degree \\
			Inclination of first slip of  grain B & $\theta_B$ & $-10$  & Degree \\
			Isotropic hardening  & $h^{\alpha \beta}$ &0   & - \\ 
			Length scale of bulk & $L_*/W$ & 2  & -\\
			Recovery coefficient  of bulk & $\zeta$ &Case study & -\\
			GB hardening coefficient & $c_s$ & Case study & N/m \\
			GB recovery coefficient & $\zeta_s$  & Case study & - \\
			\hline   
		\end{tabular}
	\end{table}
	\subsubsection{Influence of GB hardening  and recovery coefficient with energetic bulk plastic potential $\left( \zeta=0\right)$} 
	This part examines the impact of the GB hardening coefficient $c_s$ and GB relaxation coefficient $\zeta_s$ in the absence of the bulk recovery coefficient $\zeta$, specifically when the bulk defect energy $\Psi$ is regarded as a purely quadratic functional of the total plastic slip gradient.  
	\subsubsubsection{Influence of GB hardening coefficient $c_s$ ($\zeta_s=0$)}
	\begin{figure}[h!]
		\centering
		\begin{subfigure}{0.45\textwidth}
			\includegraphics[width=\textwidth]{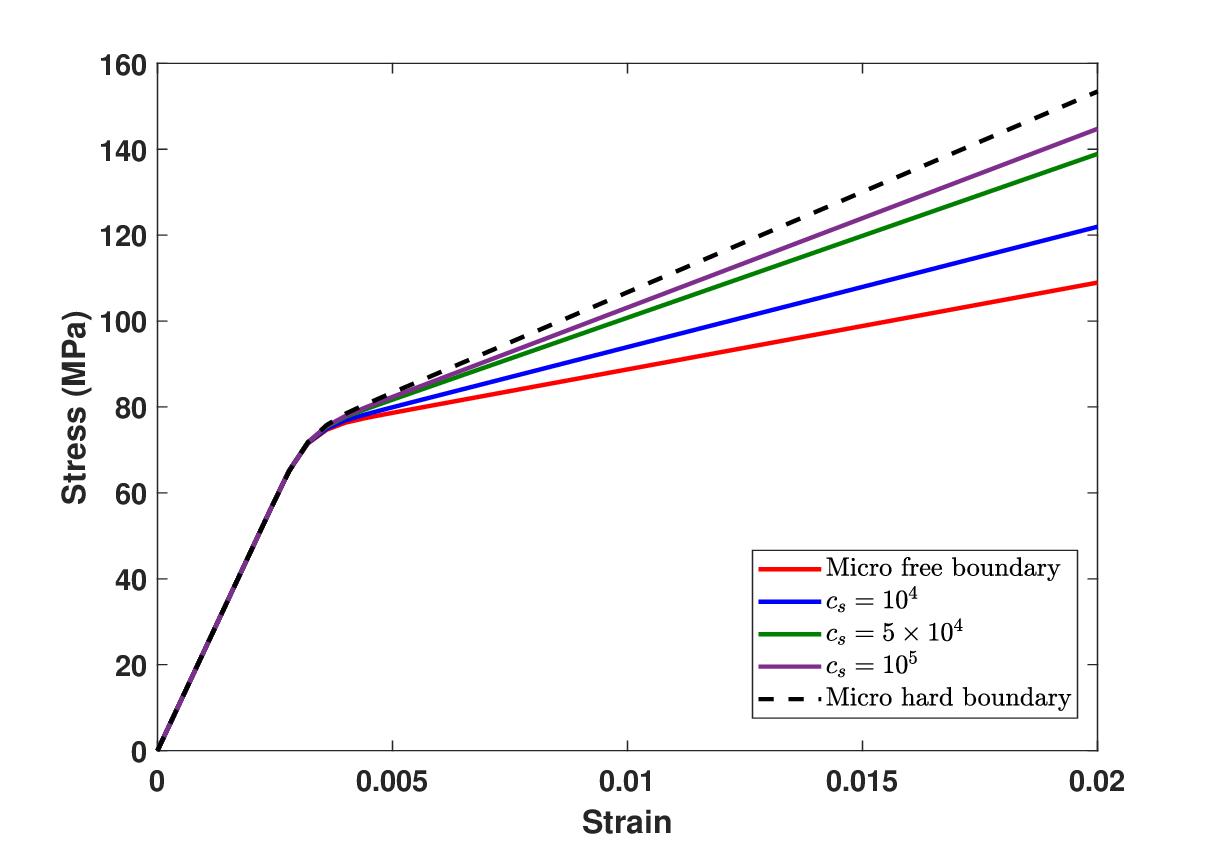}
			\caption{Shear response }
			\label{cs_eff_fd}
		\end{subfigure} 
		\begin{subfigure}{0.45\textwidth}
			\includegraphics[width=\textwidth]{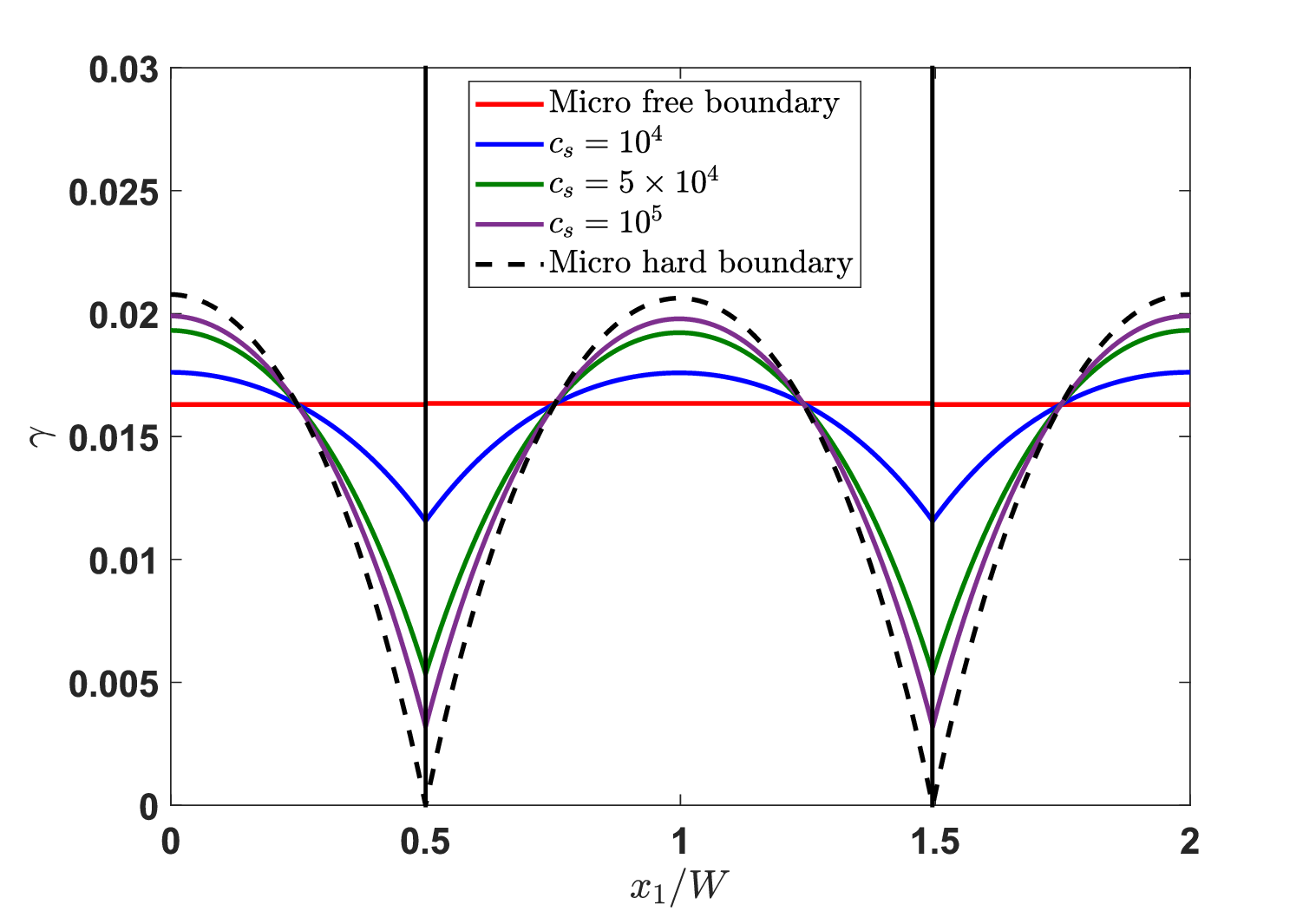}
			\caption{Plastic slip distribution}
			\label{cs_eff_slip}
		\end{subfigure}
		\begin{subfigure}{0.45\textwidth}
			\includegraphics[width=\textwidth]{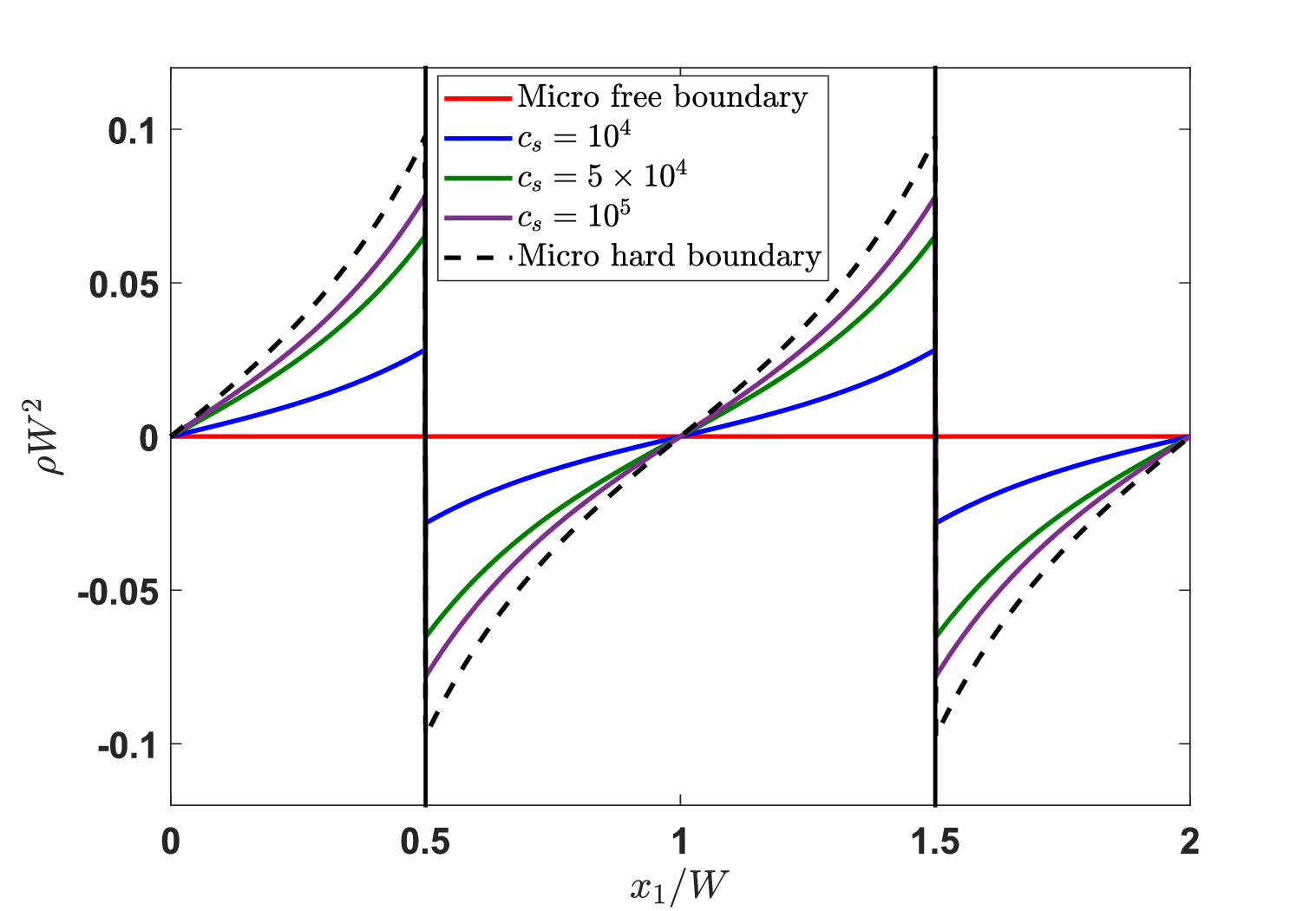}
			\caption{GND distribution}
			\label{cs_eff_gnd}
		\end{subfigure} 
		\caption{Influence of $c_s$ on the shearing of periodic bicrystal}
	\end{figure}
	In order to assess the effect of $c_s$ on the shear response, we first consider GB relaxation coefficient $\zeta_s = 0$. The resulting formulation is equivalent to the \cite{gurtin2008theory} type grain boundary model with only an energetic effect.
	
	The stress-strain curve due to the different values of $c_s$ is plotted in figure \ref{cs_eff_fd}. Micro-free boundary conditions are implied when the grain boundary potential is zero, resulting in zero grain boundary contribution to hardening (i.e., $c_s = 0$ ). With an increase in $c_s$, the grain boundary contribution to hardening increases, increasing the stress-strain curve's post-yield hardening. The micro hard boundary condition is achieved when $c_s \to \infty$. In micro hard conditions, the GB hardening attains the maximum values. Therefore, as expected, the curve prepared by varying $c_s$  lies between micro hard and micro free curves. 
	
	Plastic slip variation over the crystals due to different $c_s$ is presented in figure \ref{cs_eff_slip}. During the micro-free boundary condition, the plastic slip distribution is uniform as no energy from the grain boundary is participating. With an increase in $c_s$ values, the plastic slip at the grain boundary reduces, increasing a plastic slip gradient over the crystal. At the micro hard boundary condition, the plastic slip at the grain boundary becomes zero, and the plastic slip gradient over the crystal becomes maximum. 
	
	The GND variation over the crystal is presented in figure \ref{cs_eff_gnd}. In the figure, $\rho$ implies the total edge GND density, calculated according to equation (\ref{rho_total}). There is no plastic slip gradient during the micro-free boundary condition, and GND density is zero everywhere in the crystals. With an increase in $c_s$, the GND density increases, which takes maximum value at the micro hard boundary condition. Due to the application of $c_s$, a GND pileup near the grain boundary is observed. This observation is consistent with the experiment by \cite{sun1998mesoscale,sun2000observations}, where GND pileup was observed near the grain boundary during the low compressive strain of an aluminum bicrystal.
	
	\subsubsubsection{Influence of GB recovery coefficient $\zeta_s$}
	\begin{figure}[h!]
		\centering
		\begin{subfigure}{0.45\textwidth}
			\includegraphics[width=\textwidth]{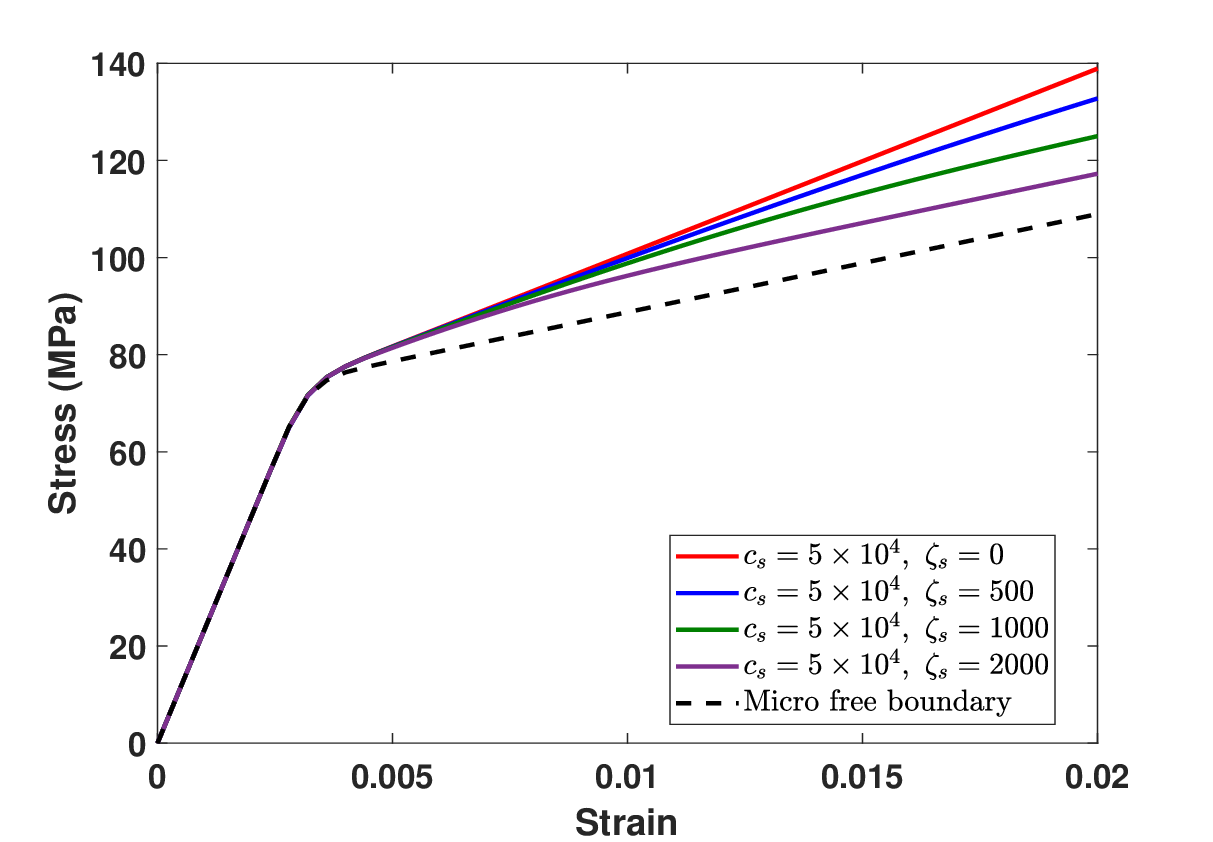}
			\caption{Shear response }
			\label{zeta_eff_fd}
		\end{subfigure} 
		\begin{subfigure}{0.44\textwidth}
			\includegraphics[width=\textwidth]{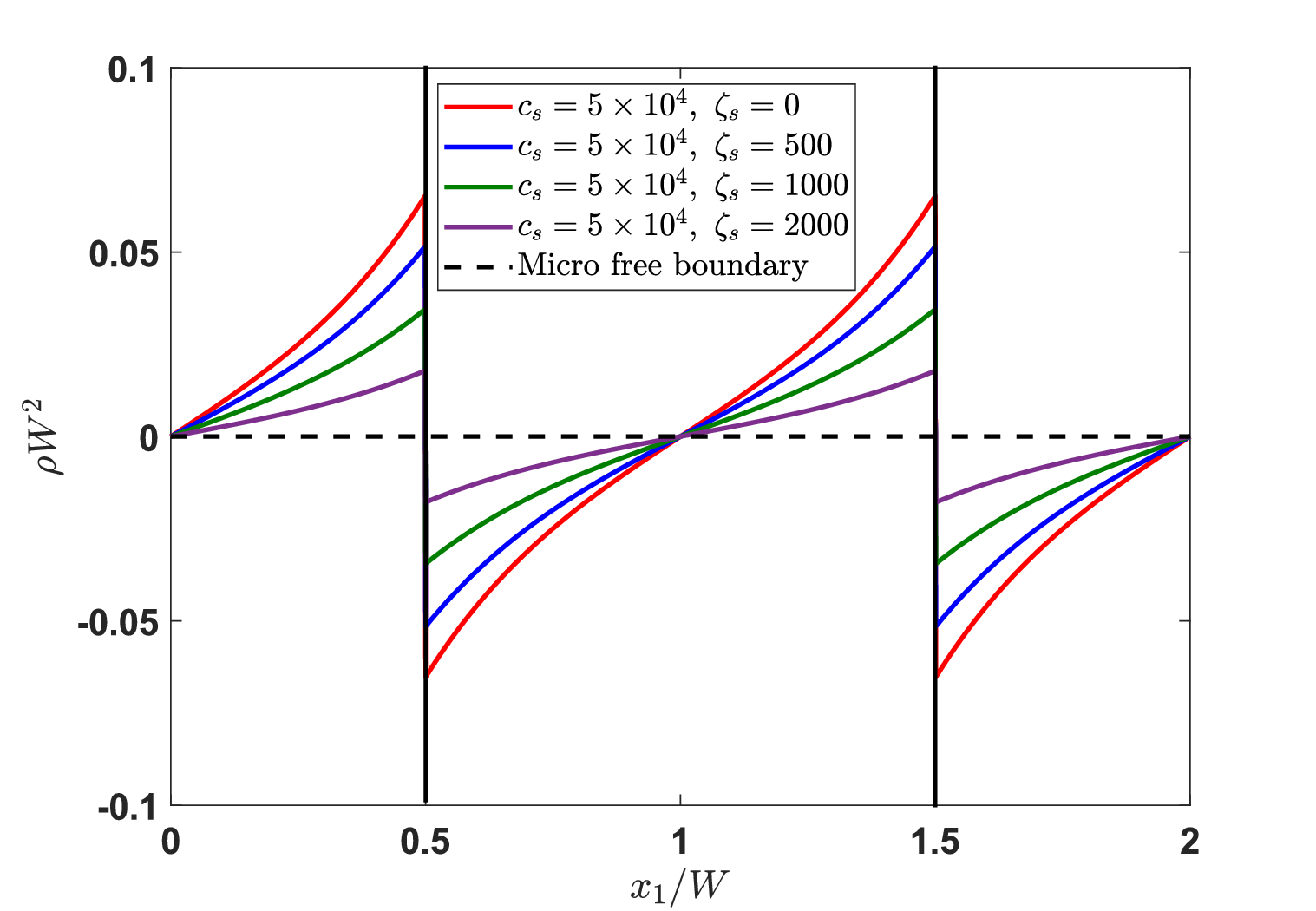}
			\caption{GND distribution}
			\label{zeta_eff_gnd}
		\end{subfigure}
		\begin{subfigure}{0.45\textwidth}
			\includegraphics[width=\textwidth]{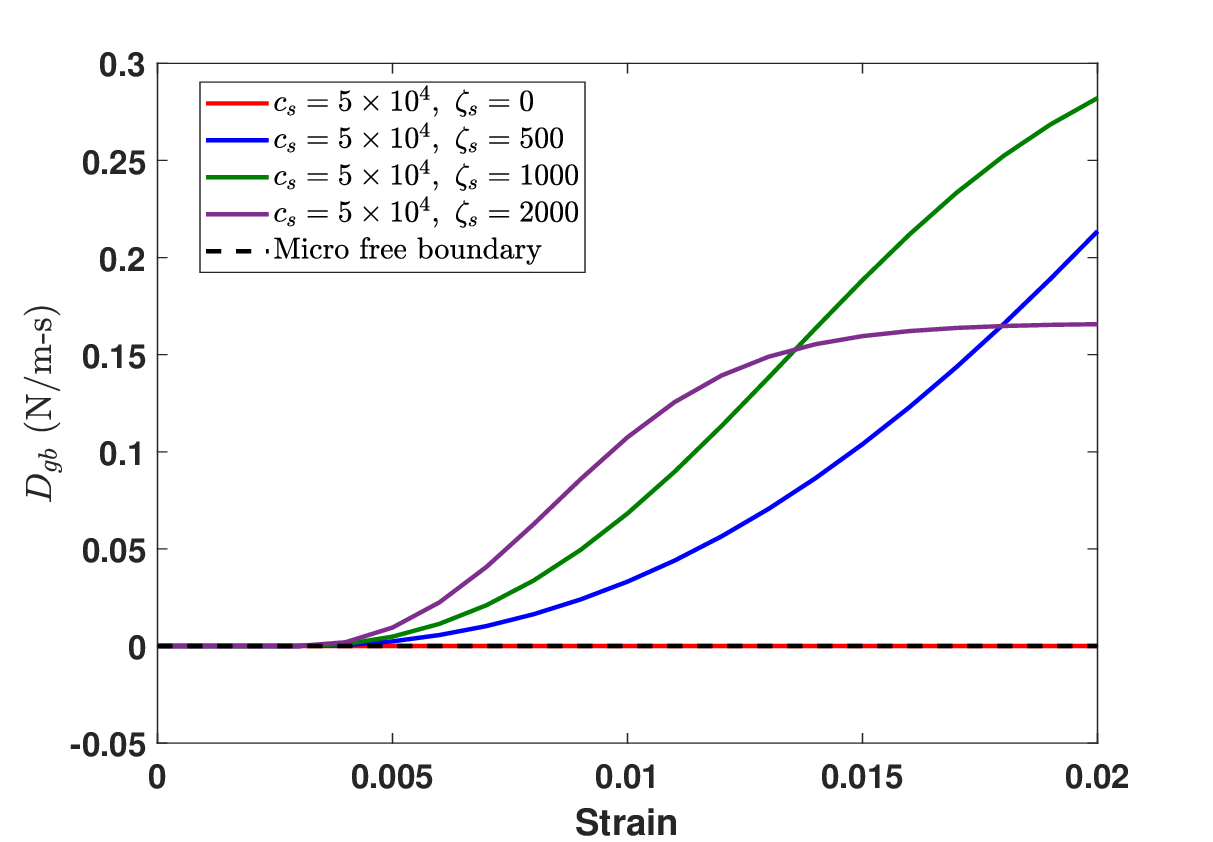}
			\caption{$D_{gb}$ variation with time}
			\label{zeta_eff_dissipation}
		\end{subfigure} 
		\begin{subfigure}{0.45\textwidth}
			\includegraphics[width=\textwidth]{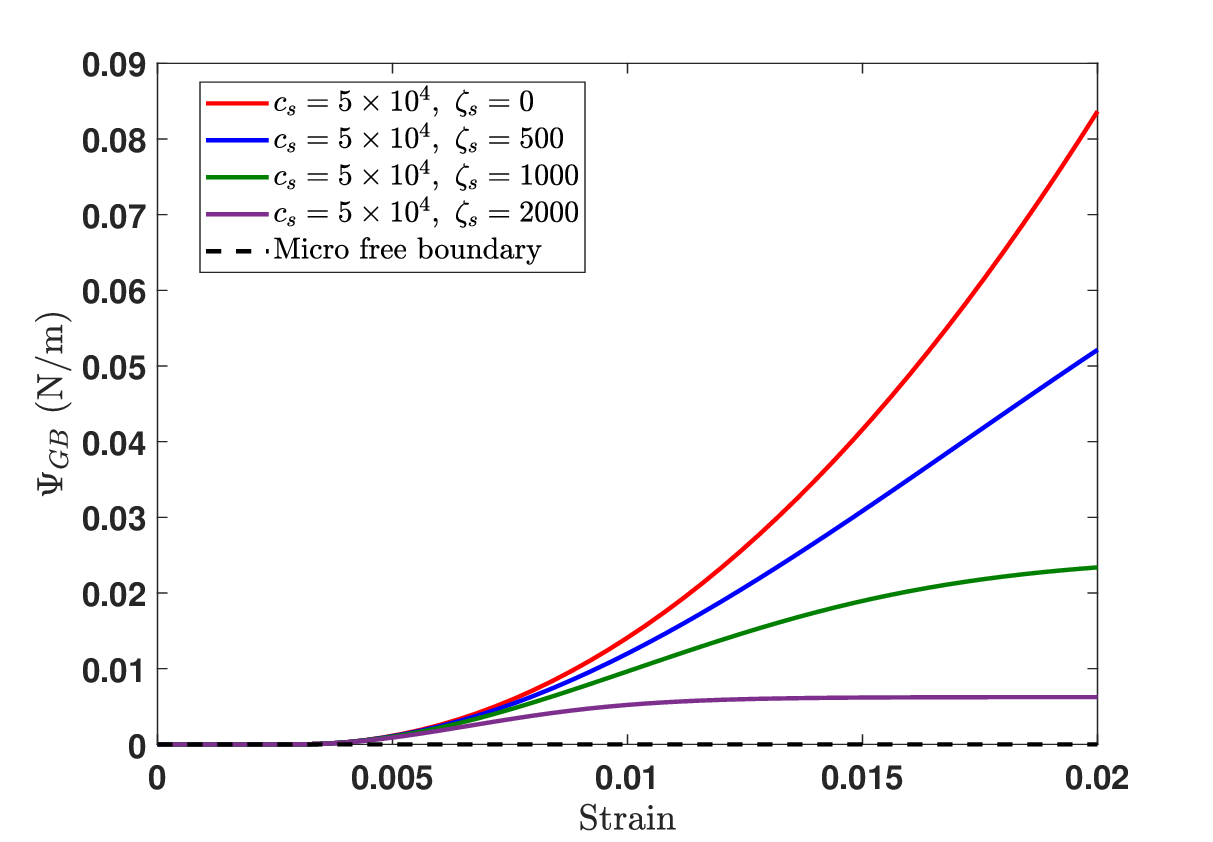}
			\caption{$\Psi_{gb}$ variation with time}
			\label{zeta_eff_energy}
		\end{subfigure}
		\caption{Effect of $\zeta_s$ on the shearing of periodic bicrystal}
	\end{figure}
	\begin{figure}[h!]
		\centering
		\begin{subfigure}{0.45\textwidth}
			\includegraphics[width=\textwidth]{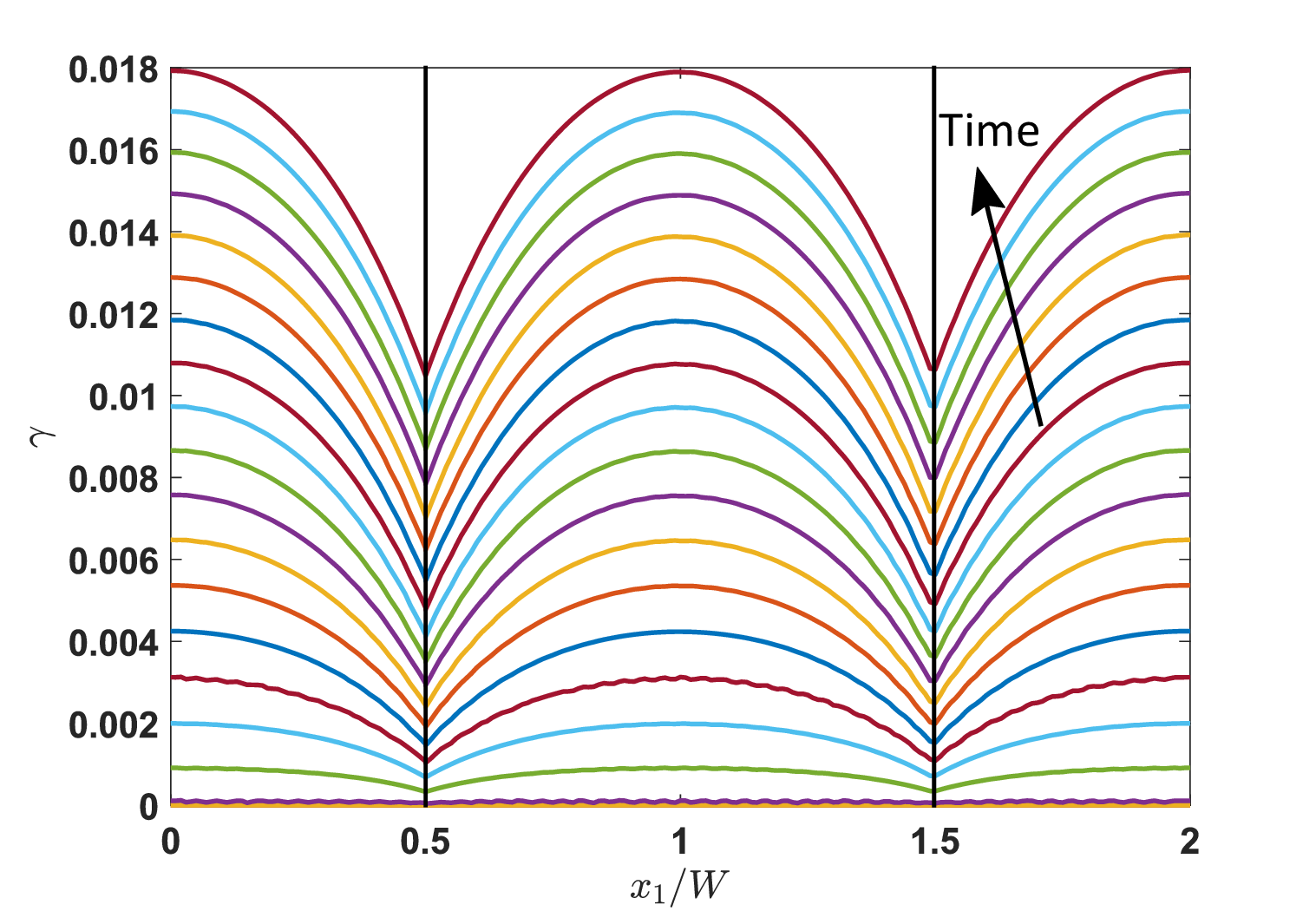}
			\caption{Slip evolution}
			\label{slip_evol_cs_0_05_zeta_s_1000}
		\end{subfigure} 
		\begin{subfigure}{0.45\textwidth}
			\includegraphics[width=\textwidth]{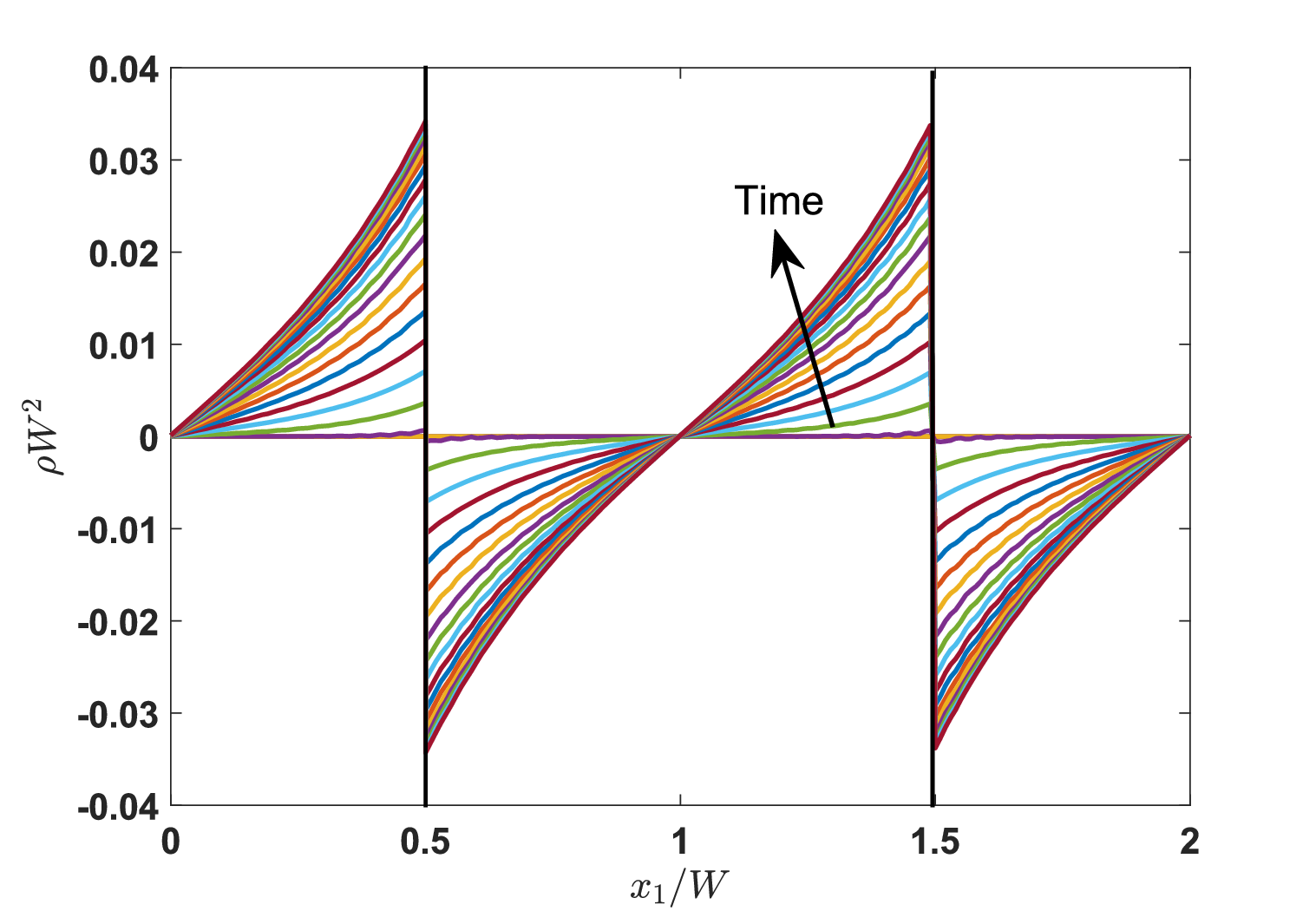}
			\caption{GND evolution}
			\label{gnd_evol_cs_0_05_zeta_s_1000}
		\end{subfigure} 
		\caption{Evolution of plastic slip and GND over time ($c_s=5 \times 10^4, ~\zeta_s=1000$)}
		\label{zetas_evol}
	\end{figure}
	To investigate the dissipation related to grain boundaries, we revisit the aforementioned scenario with a $c_s=5 \times 10^4$ value. Subsequently, the parameter $\zeta_s$ is raised from zero to analyze its impact on the deformation of the periodic bicrystal. The shear response is presented in figure \ref{zeta_eff_fd}. The shear response curve initially follows the $\zeta_s=0$ curve; then, it deviates, showing some nonlinearity. With an increase of $\zeta_s$, more deviation from the $\zeta_s=0$ curve indicates dissipation.  
	
	The GND distribution with varying $\zeta_s$ and corresponding dissipation is presented in figure \ref{zeta_eff_gnd} and \ref{zeta_eff_dissipation} respectively. When $\zeta_s=0$, the grain boundary contributes only free energy formation; therefore, grain boundary dissipation $D_{gb}$ is zero. However, with an increase in $\zeta_s$, some additional dissipation related to the grain boundary is observed. Therefore, the grain boundary GND density reduces with an increase in $\zeta_s$. The additional dissipation $D_{gb}$ increases with increase in $\zeta_s$. However, at the higher value of $\zeta_s$, it saturates to some low peak value. 
	
	The grain boundary defect energy $\Psi_{gb}$ is shown in figure \ref{zeta_eff_energy}.    Due to the absence of $\zeta_s$, it follows the quadratic path. However, with an increase in  $\zeta_s$, the defect energy deviates from a quadratic nature due to additional dissipation. Consequently, GB energy saturates at higher values of $\zeta_s$.  
	
	In order to show the capability of the proposed formulation to capture GND saturation, we have presented the GND and corresponding plastic slip evolution for $c_s=5 \times 10^4,$ and $\zeta_s=1000$ in figure \ref{zetas_evol}. The plastic slip increases with time, and the difference between two consecutive time steps is approximately constant. This indicates that the plastic slip increment does not saturate over time. On the contrary, the GND shows saturation at a higher time (higher applied strain values). This saturation in GND value is consistent with recent experiment \citep{zhang2023geometrically}.
	\subsubsection{Influence of GB hardening and recovery coefficient in the presence of bulk recovery coefficient $\zeta$}
	This illustration will elucidate the impact of the GB hardening and recovery coefficient on the shear response when the bulk recovery coefficient $\zeta$ is taken into account. 
	\subsubsubsection{Influence of bulk recovery coefficient ($\zeta \ne0$) in absence of GB recovery coefficient ($\zeta_s=0$)}
	\begin{figure}[h!]
		\centering
		\begin{subfigure}{0.45\textwidth}
			\includegraphics[width=\textwidth]{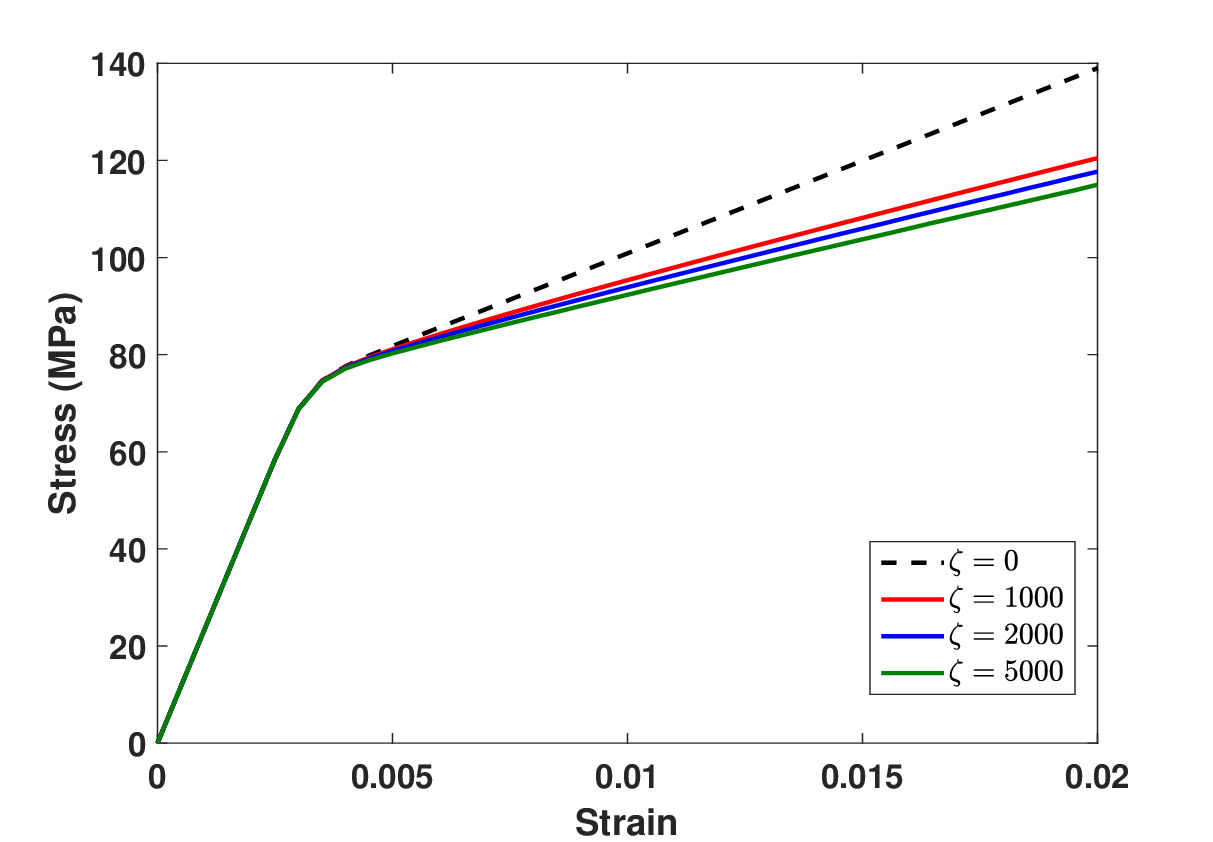}
			\caption{Shear response }
			\label{zeta_bulk_fd}
		\end{subfigure} 
		\begin{subfigure}{0.45\textwidth}
			\includegraphics[width=\textwidth]{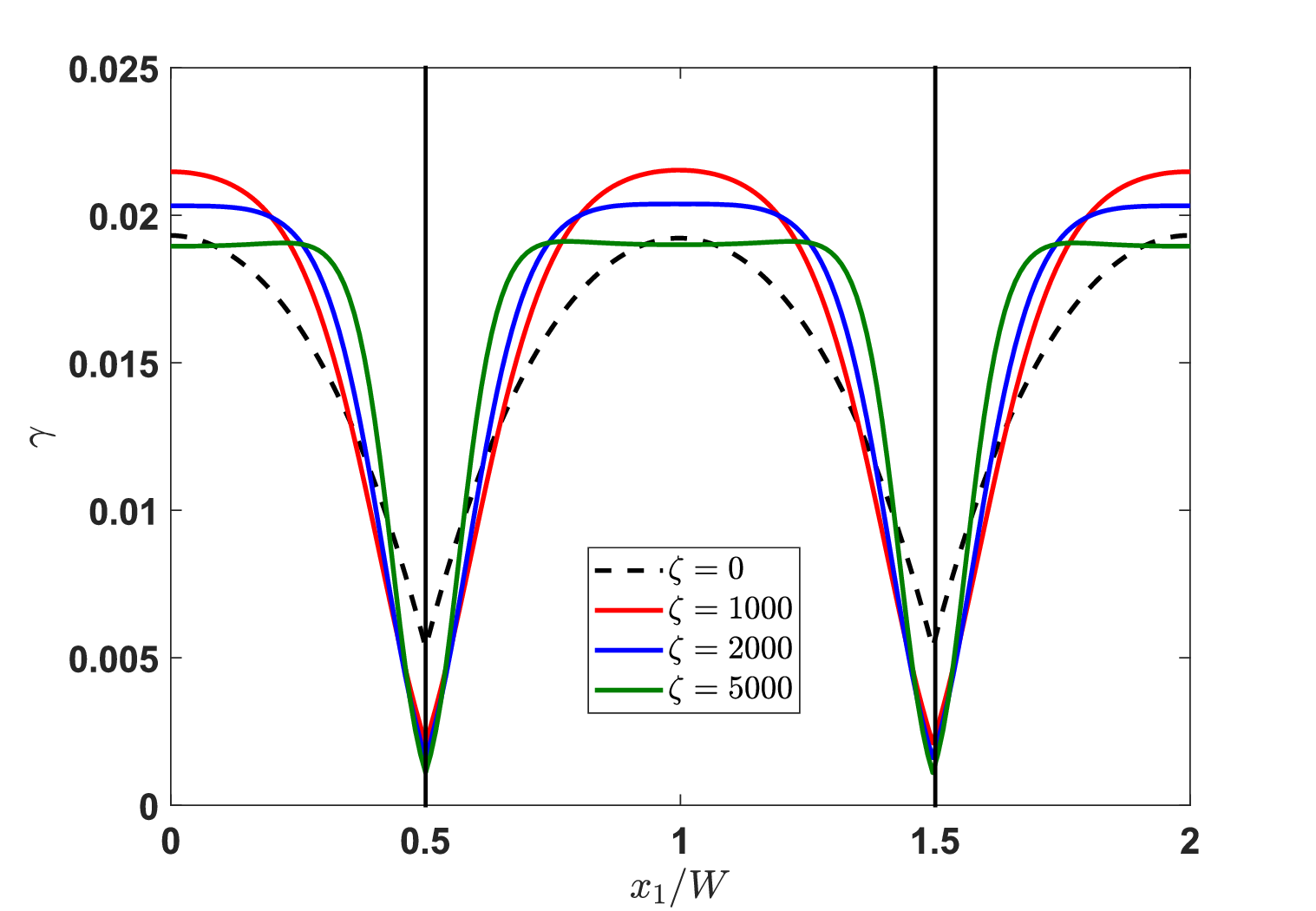}
			\caption{Plastic slip distribution}
			\label{zeta_bulk_slip}
		\end{subfigure}
		\begin{subfigure}{0.45\textwidth}
			\includegraphics[width=\textwidth]{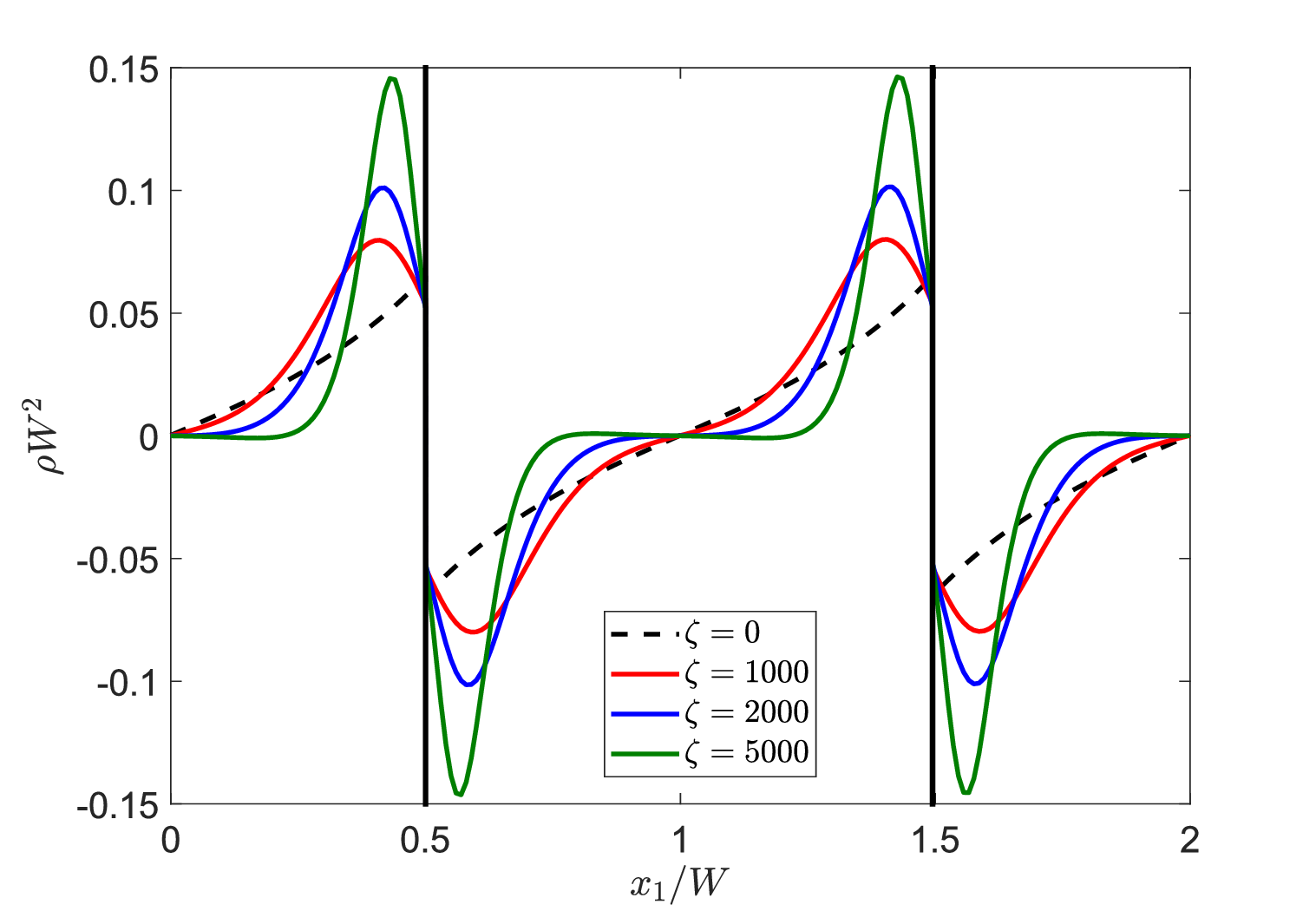}
			\caption{GND distribution}
			\label{zeta_bulk_gnd}
		\end{subfigure} 
		\caption{Effect of recovery coefficient $\zeta$ ($ c_s=5 \times 10^4,~\zeta_s=0$)}
	\end{figure}
	\begin{figure}[h!]
		\centering
		\begin{subfigure}{0.45\textwidth}
			\includegraphics[width=\textwidth]{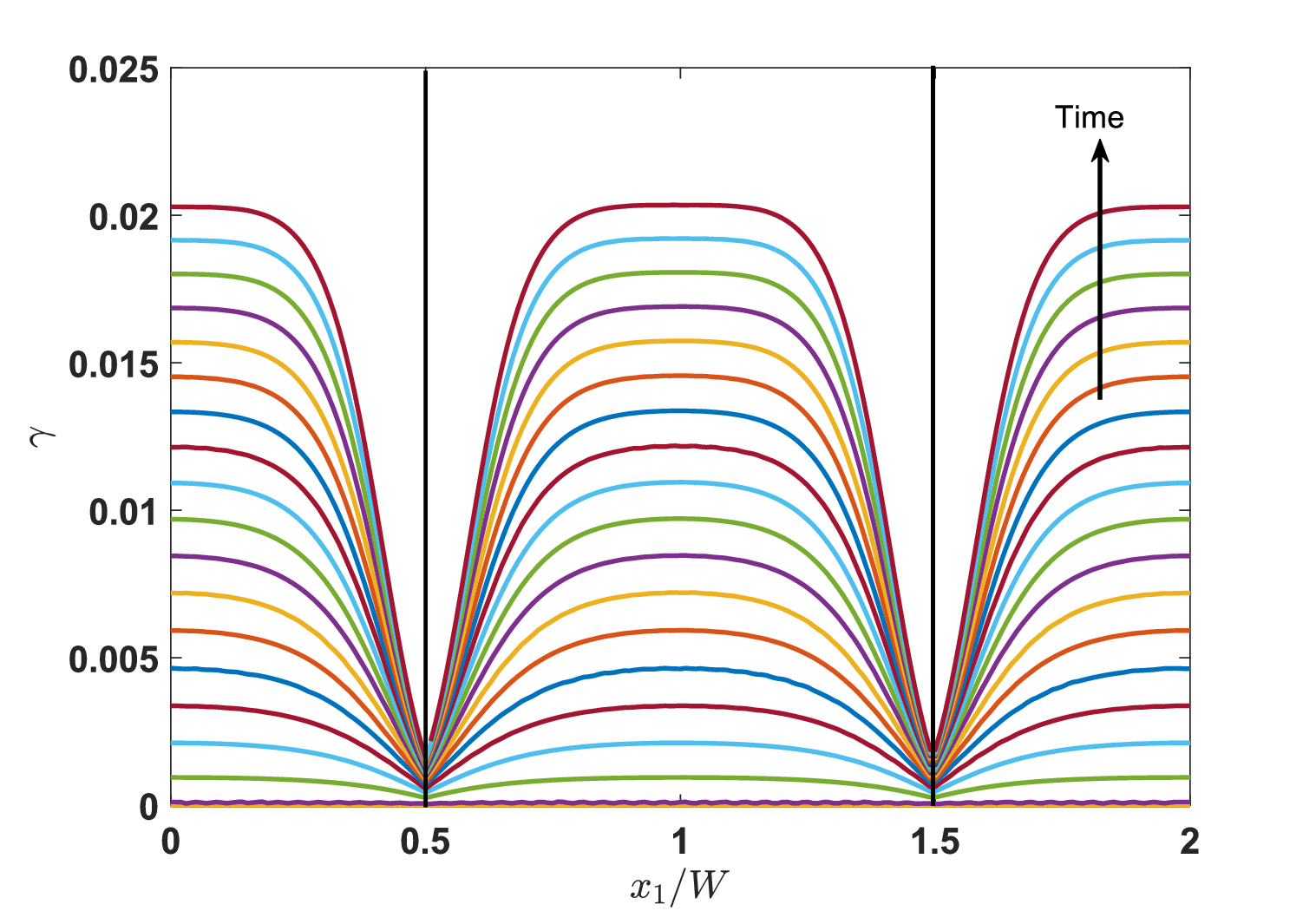}
			\caption{Slip evolution}
			\label{slip_evol_L_2_zeta_2000_cs_0_05}
		\end{subfigure} 
		\begin{subfigure}{0.45\textwidth}
			\includegraphics[width=\textwidth]{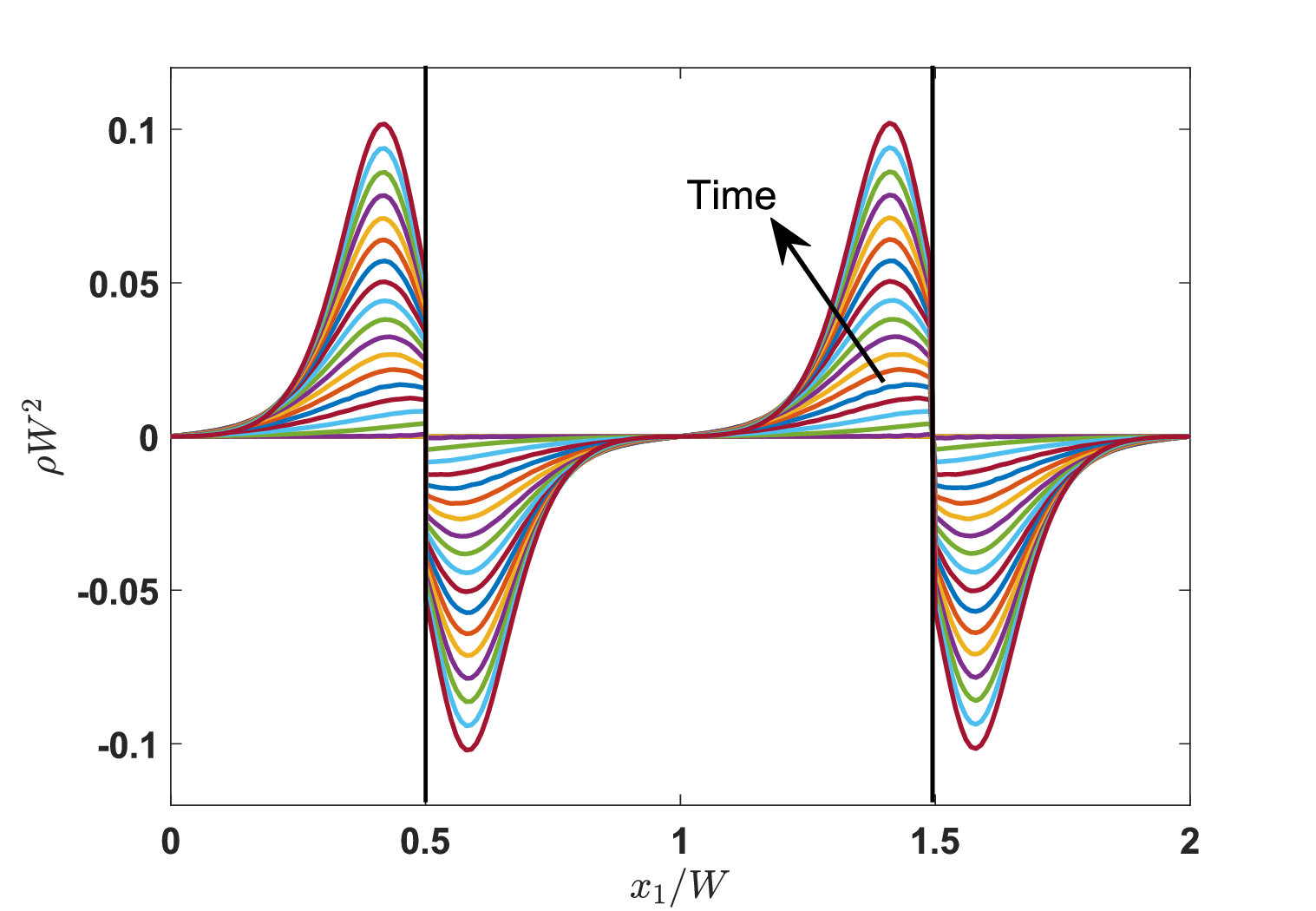}
			\caption{GND evolution}
			\label{gnd_evol_L_2_zeta_2000_cs_0_05}
		\end{subfigure} 
		\caption{Evolution of plastic slip and GND over time ($\zeta=2000, ~c_s=5 \times 10^4$, $\zeta_s=0$)}
		\label{zeta_evol}
	\end{figure}
	In this example, we investigate the influence of bulk recovery coefficient $\zeta$ on the GB hardening in the absence of GB recovery coefficient $\zeta_s$. The previous example is considered with quadratic $\Psi_{gb}$ ($\zeta_s=0$); however, bulk recovery coefficient $\zeta \ne 0$.   $L_*/W=2$, $c_s=5 \times 10^4$ is considered and bulk recovery coefficient $\zeta$ is increased from zero.  The shear response is shown in figure \ref{zeta_bulk_fd}, and corresponding plastic slip and GND distribution are presented in figure \ref{zeta_bulk_slip} and \ref{zeta_bulk_gnd}, respectively. 
	
	In the event that $\zeta$ is not present, the quadratic defect energy is evaluated within the bulk, the hardening occurs linearly, and the distribution of plastic slip is roughly parabolic. However, the inclusion of $\zeta$ introduces nonlinearity to the hardening curve. Increasing $\zeta$ decreases the curvature of the plastic slip variation within the grains, but a sharp variation near the grain boundary is observed. This slip distribution leads to an approximate zero GND density inside the grain, with more GND pileup near the boundary. Essentially, the presence of $\zeta$ causes the GND to be pushed towards the boundary, resulting in increased pileup near the boundary. In this particular example, no dissipation related to grain boundaries is expected since $\zeta_s=0$ is considered. The evolution of plastic slip and  GND for $\zeta=2000$ is portrayed in figure  \ref{zeta_evol}. As expected, in contrast to the previous example, there is no saturation in GND observed in this case as $\zeta_s = 0$.
	\subsubsubsection{Interaction between bulk and GB recovery coefficient ($\zeta\ne0$ and  $\zeta_s\ne0$)}
	\begin{figure}[h!]
		\centering
		\begin{subfigure}{0.45\textwidth}
			\includegraphics[width=\textwidth]{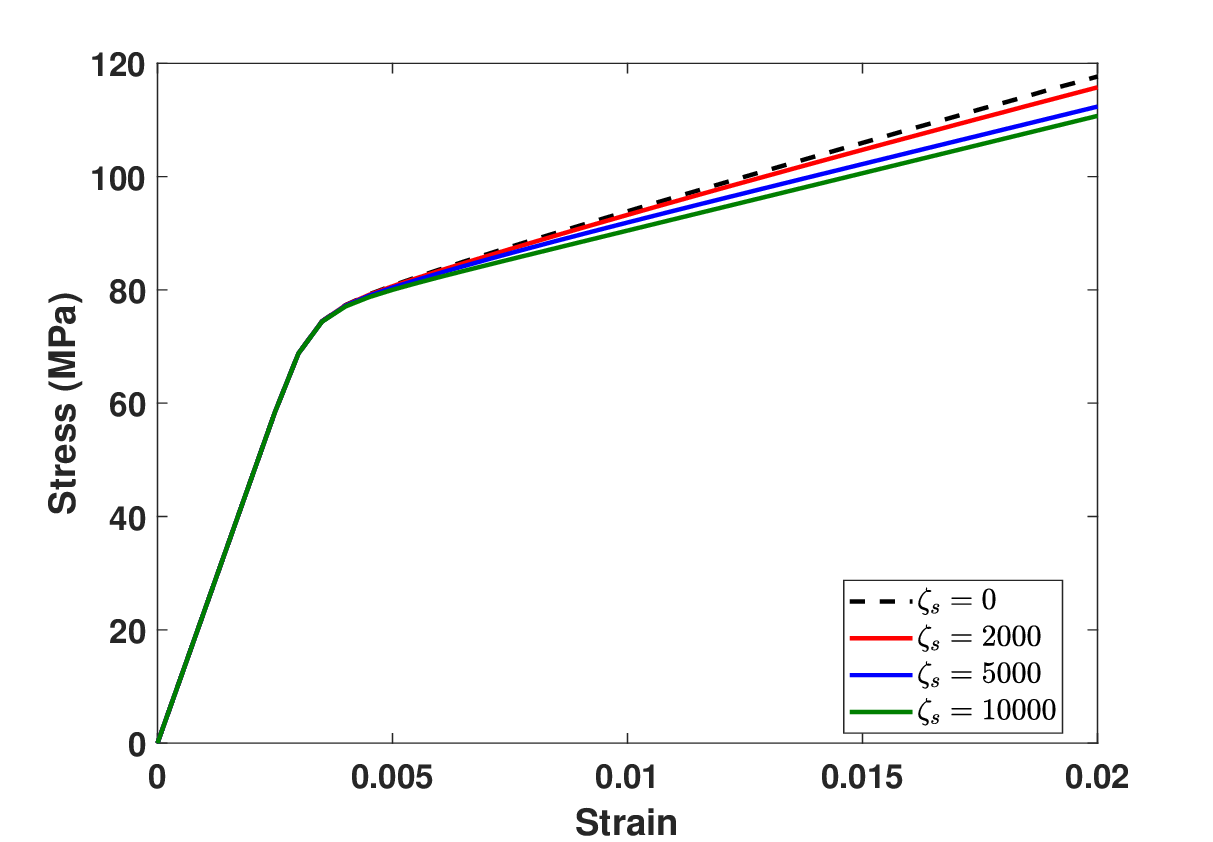}
			\caption{Shear response }
			\label{zeta_zetas_fd}
		\end{subfigure} 
		\begin{subfigure}{0.45\textwidth}
			\includegraphics[width=\textwidth]{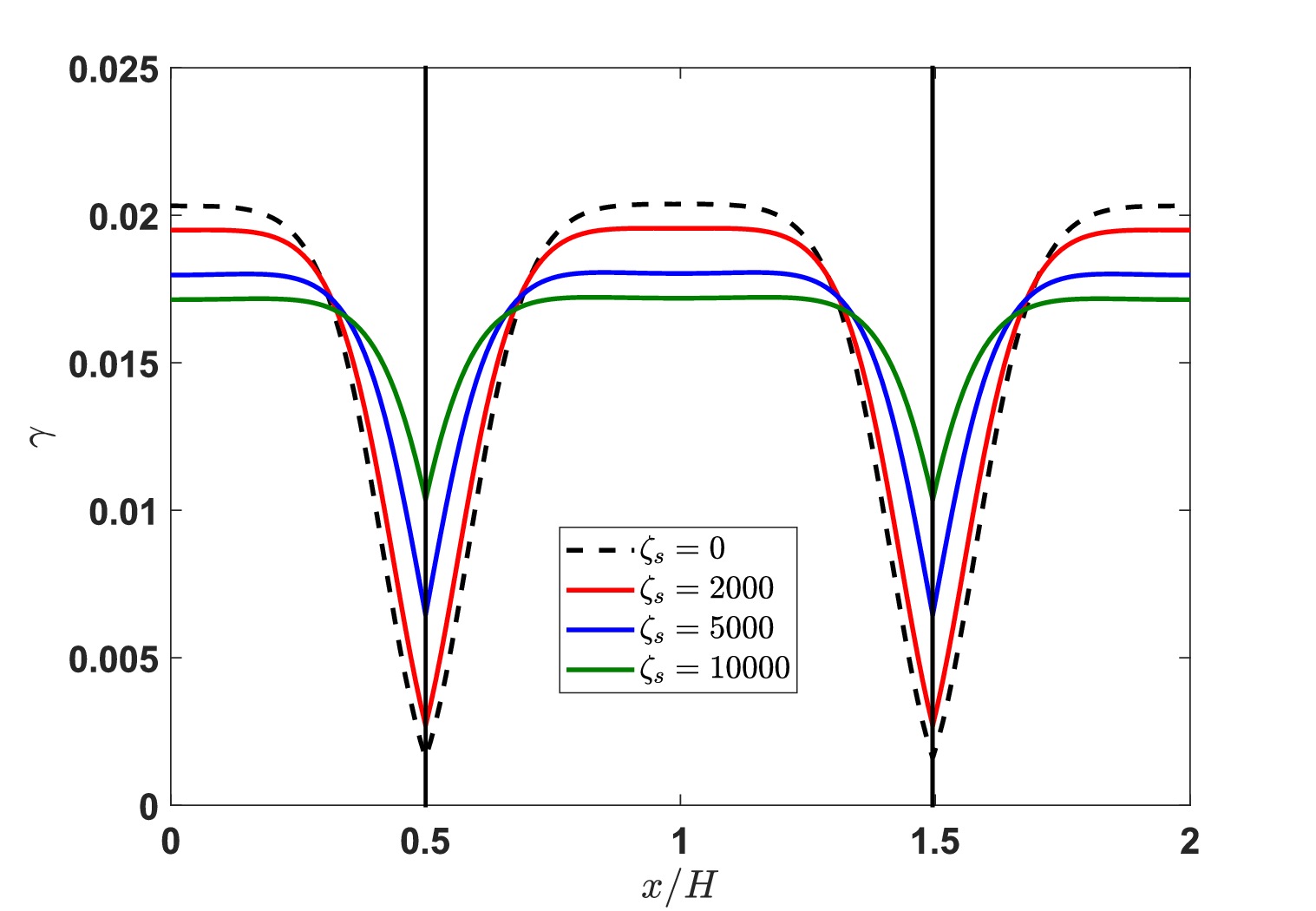}
			\caption{Plastic slip distribution}
			\label{zeta_zetas_slip}
		\end{subfigure}
		\begin{subfigure}{0.45\textwidth}
			\includegraphics[width=\textwidth]{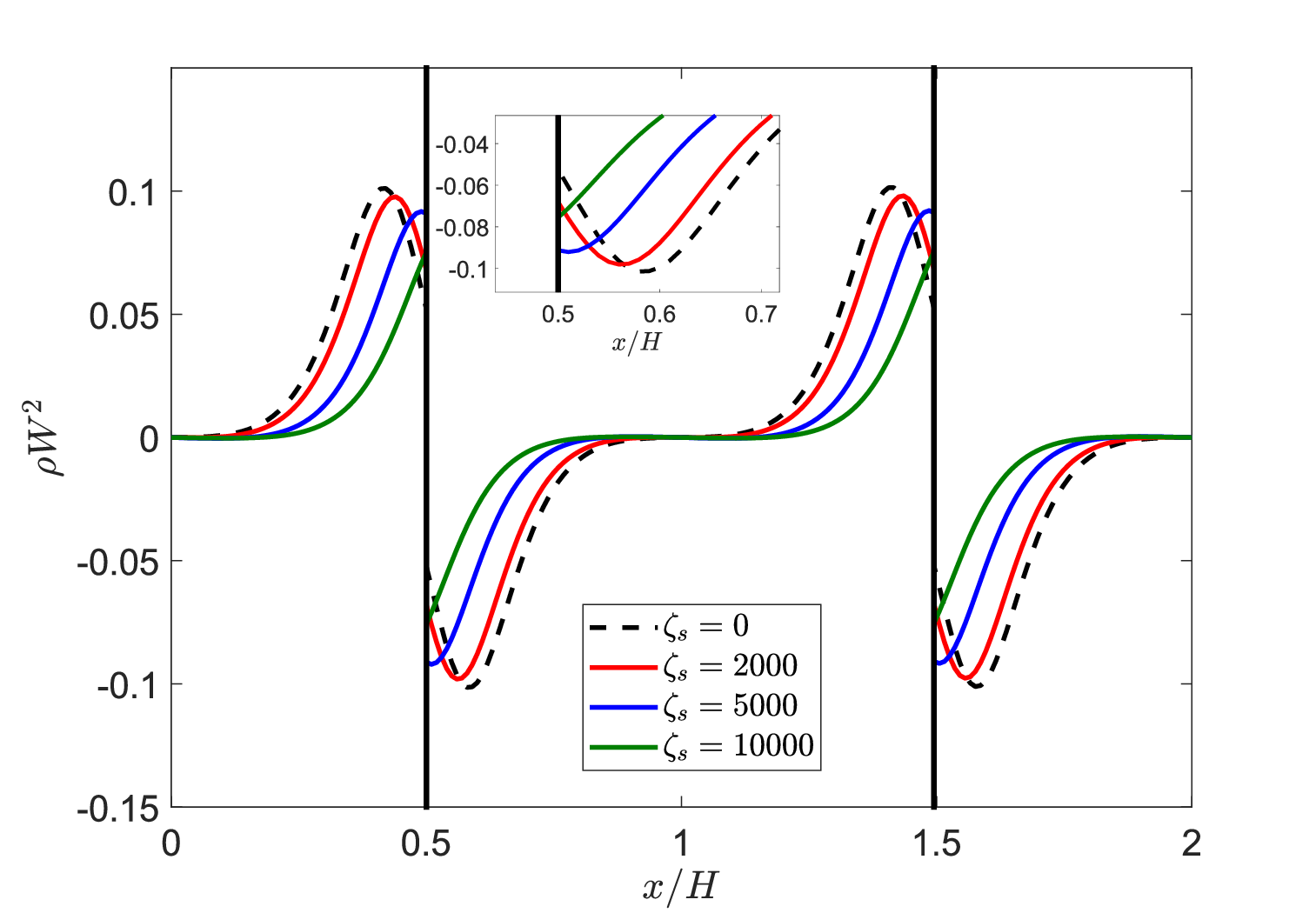}
			\caption{GND distribution}
			\label{zeta_zetas_gnd}
		\end{subfigure} 
		\begin{subfigure}{0.45\textwidth}
			\includegraphics[width=\textwidth]{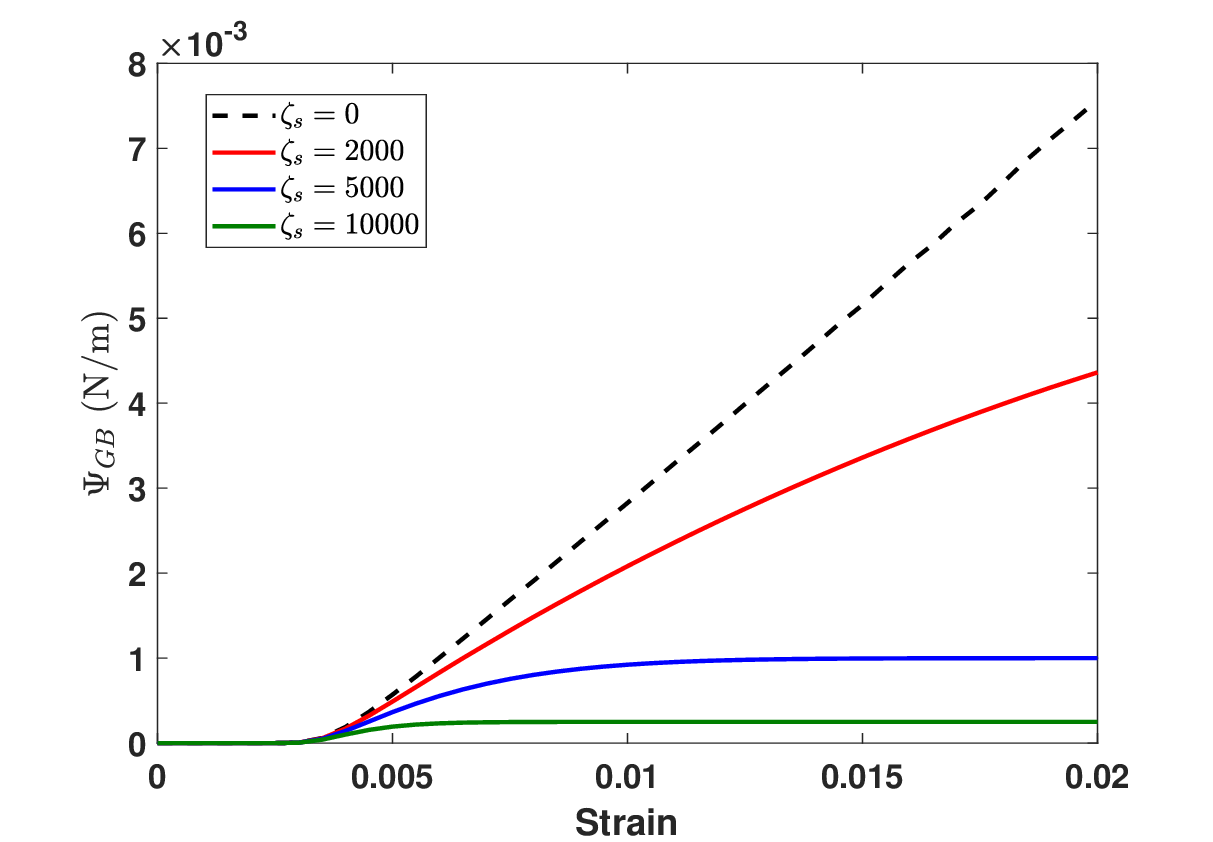}
			\caption{$\Psi_{gb}$ variation with time}
			\label{zeta_zetas_energy}
		\end{subfigure}
		\caption{Effect of recovery coefficient $\zeta_s$ ($c_s=5 \times 10^4,~\zeta=2000$)}
	\end{figure}
	
	Finally, the influence of the GB recovery coefficient in the presence of the bulk recovery coefficient is considered. The shear layer is now considered with $\zeta=2000,~c_s=5\times 10^4$, and  $\zeta_s$ increases from zero. The shear response is presented in figure \ref{zeta_zetas_fd}. Corresponding plastic slip, GND and $\Psi_{gb}$ variation is presented in figure \ref{zeta_zetas_slip}, \ref{zeta_zetas_gnd}, \ref{zeta_zetas_energy} respectively.  
	
	In the previous case, the bulk recovery coefficient was observed to influence the movement of the geometrically necessary dislocations (GND) towards the grain boundary. However, when we consider the nonzero grain boundary recovery coefficient $\zeta_s$, it leads to dissipation at the boundary, reducing the GND density near the grain boundary. Moreover, as the value of $\zeta_s$ increases, the dissipation increases, leading to decrease in the GND density near the grain boundary. Interestingly, increasing $\zeta_s$ causes more regions within the crystal to experience a uniform plastic slip, resulting in zero GND density in these areas. In a series of distinct microscale experiments, it has been noted that the GND pile-up is confined to the vicinity of the grain boundary and does not spread to the crystal itself \citep{shen1988dislocation,feaugas1999origin,feaugas2003grain} at low strain. The stacking of GND densities at the boundary, facilitated by $\zeta$ and $\zeta_s$, aligns with the findings of these experiments. In conclusion, the bulk recovery coefficient $\zeta$ influences the movement of GND towards the grain boundary, while the grain boundary recovery coefficient $\zeta_s$ controls the dissipation and the accumulation of GND near the boundary.
	\subsection{Bicrystal tension} 
	\begin{figure}[h!]
		\centering
		\includegraphics[width=0.7\textwidth]{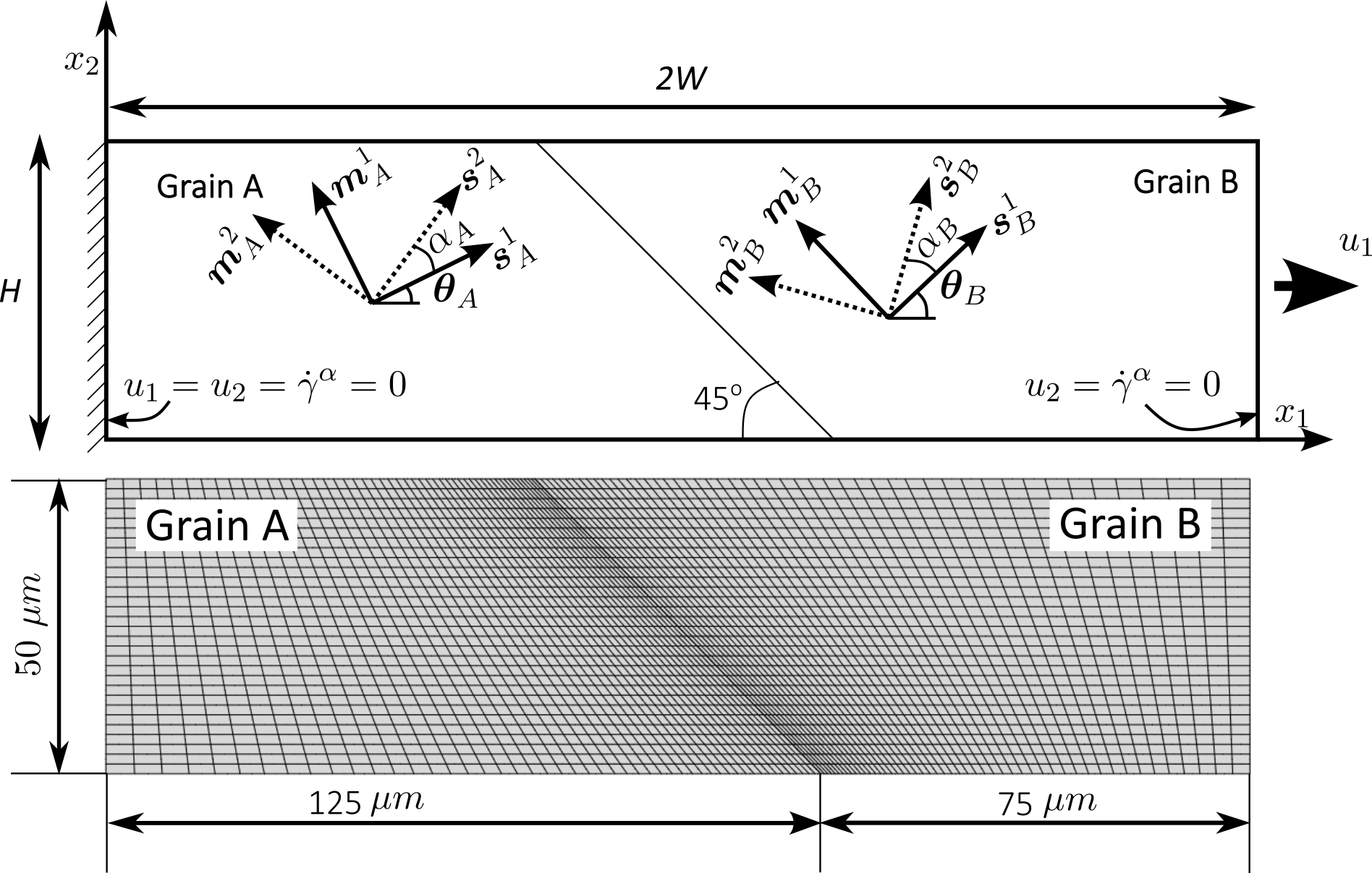}
		\caption{Schematic diagram of bicrystal tension with boundary conditions}
		\label{tension}
	\end{figure}
	To demonstrate the effectiveness and consistency of the proposed grain boundary formulation for a double slip system, an example of bicrystal tension \citep{ozdemir2014modeling} is presented. One bicrystal rectangular micro-sample is applied with tension as shown in figure \ref{tension}. The two grains are separated by an oblique grain boundary, which is $45^o$ aligned with a negative $x_1$ axis. The sample has slip angle $\theta_A$ and $\theta_B$ for the first slip plane, and the second plane is aligned with   $\alpha_A$ and $\alpha_B$ angle with the first slip plane. The rectangular sample is discretized with 3600 elements, as shown in the figure.
	
	The left side of the sample is micro-clamped. The right side is subjected to a maximum displacement of 5\%; however, the $x_2$ directional displacement and plastic slip are kept zero on the right side. The boundary condition to this problem can be written as 
	\begin{eqnarray}
		\nonumber
		{u_i}(0, x_2,t)=\gamma^\alpha(0,x_2,t)=0	\qquad i=1,2,~~\alpha=1,2, \qquad\forall~x_2,t \\
		u_1(2W,x_2,t)=U^*, u_2(2W,x_2,t)=\gamma^\alpha(2W,x_2,t)=0,	~~\alpha=1,2,\qquad \forall x_2,t
	\end{eqnarray} 
	where $U^*$ is the specified displacement. The slip angles are selected in such a way that the second slip angles of both grains are identical and aligned with the grain boundary. The objective is to ensure that this particular slip system does not interact with the grain boundary, and our aim is to examine if the suggested model can uphold this uniformity. The material properties remain unchanged from the previous instance of bicrystal shear. The considered slip angles are $\theta_A^1 = 30^o,~\theta_A^2 = -45^o , ~\theta_B^1=-30^o, ~\theta_B^2 = -45^o$. The bulk length scale is considered as $L^*/H=10$. 
	
	\subsubsection{Influence of $c_s$ on the tensile response}
	\begin{figure}[h!]
		\centering
		\begin{subfigure}{0.45\textwidth}
			\includegraphics[width=\textwidth]{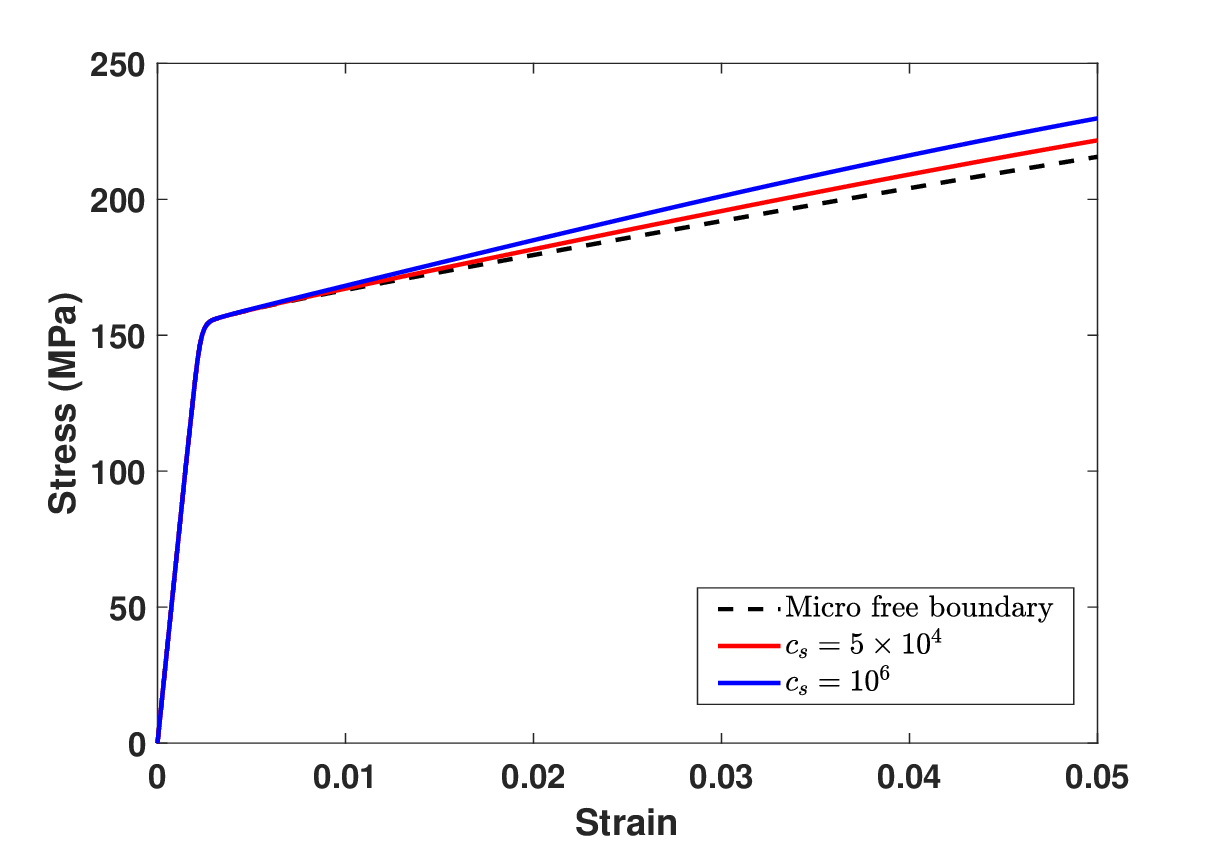}
			\caption{Tensile stress }
			\label{cs_effect_tension_fd}
		\end{subfigure} 
		\begin{subfigure}{0.45\textwidth}
			\includegraphics[width=\textwidth]{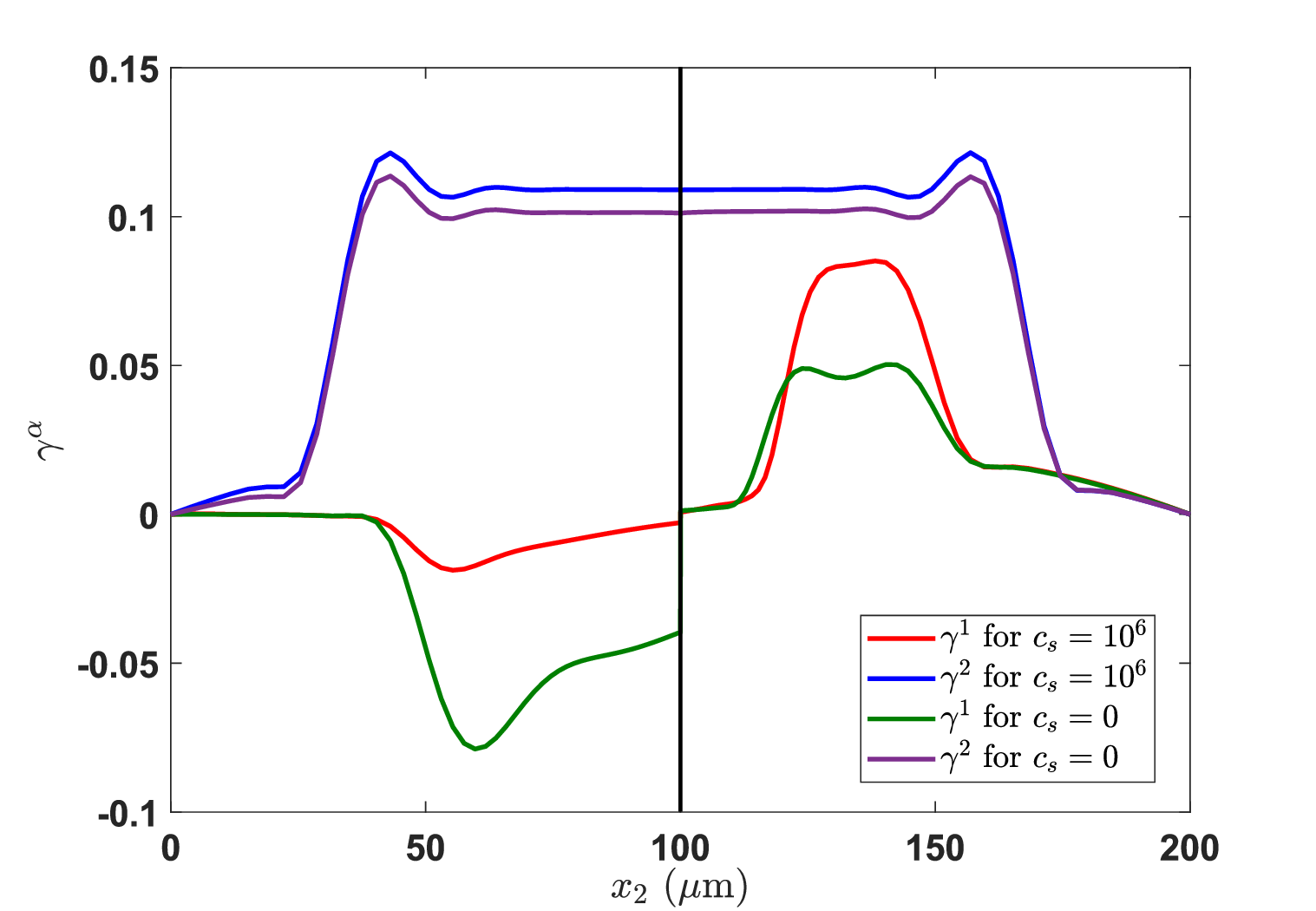}
			\caption{$\gamma^\alpha$ distribution along $x_2=25$ $\mu$m at $5\%$ strain}
			\label{cs_eff_slip_dist}
		\end{subfigure} 
		\caption{Effect of $c_s$ on the tensile stress and $\gamma^\alpha$ distribution along central line}
	\end{figure}
	\begin{figure}[h!]
		\centering
		\begin{subfigure}{0.75\textwidth}
			\includegraphics[width=\textwidth]{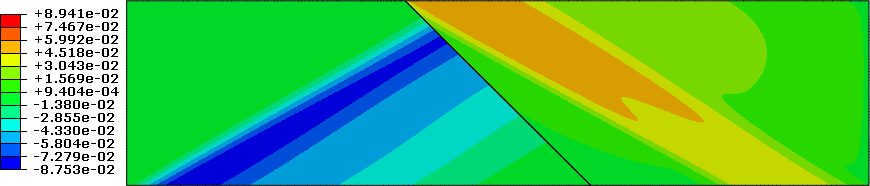}
			\caption{$\gamma^1$ distribution for micro free boundary condition }
			\label{cs_0_gamma1}
		\end{subfigure} 
		\begin{subfigure}{0.75\textwidth}
			\includegraphics[width=\textwidth]{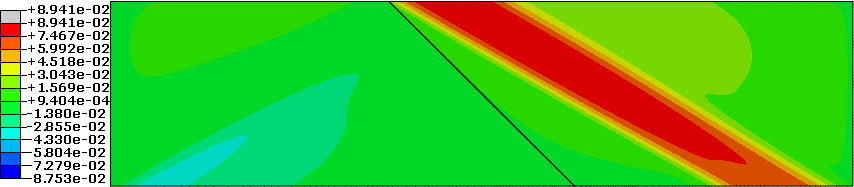}
			\caption{$\gamma^1$ distribution for $c_s = 10^6$}
			\label{cs_1_gamma1}
		\end{subfigure} 
		\begin{subfigure}{0.75\textwidth}
			\includegraphics[width=\textwidth]{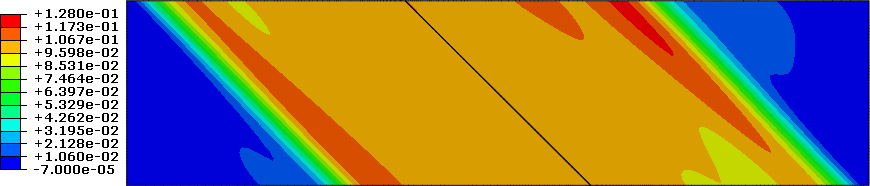}
			\caption{$\gamma^2$ distribution for micro free boundary condition}
			\label{cs_0_gamma2}
		\end{subfigure}
		\begin{subfigure}{0.75\textwidth}
			\includegraphics[width=\textwidth]{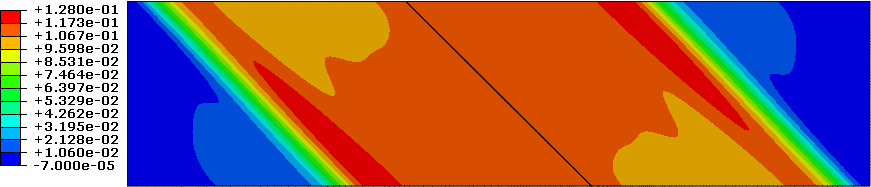}
			\caption{$\gamma^2$ distribution for $c_s=10^6$. }
			\label{cs_1_gamma2}
		\end{subfigure}
		\caption{Effect of $c_s$ on the plastic slip distribution at $5\%$ strain}
		\label{contour1}
	\end{figure}
	
	To evaluate the impact of the GB hardening coefficient $c_s$ on the tensile response, the $\zeta_s$ is maintained at zero while the bulk defect energy is assumed to be a quadratic functional of the total slip gradients. The stress-strain curve under tension is depicted in figure \ref{cs_effect_tension_fd}. As expected, the introduction of $c_s$ triggers grain boundary hardening, leading to a rise in hardening as $c_s$ increases in the stress-strain curve. The plastic slip contour at 5\% strain for micro free boundary condition and $c_s=10^6$ is presented in figure \ref{contour1}. Corresponding plastic slip distribution along a central line parallel to $x_1$ axis ($x_2=25~\mu$m) is plotted in figure \ref{cs_eff_slip_dist}. 
	
	Due to misorientation between two grains of the first slip system, a sharp jump in plastic slip is observed. With the increase in the $c_s$ value, the grain boundary energy brings down this gap, as observed in the figure \ref{cs_eff_slip_dist}. The second slip system is parallel to the grain boundary does not not interact with the grain boundary. The second plastic slip is continuous across the grain boundary. This observation is consistent with literature \citep{ozdemir2014modeling}.   
	\subsubsection{Influence of $\zeta_s$ on the tensile response}
	\begin{figure}[h!]
		\centering
		\begin{subfigure}{0.45\textwidth}
			\includegraphics[width=\textwidth]{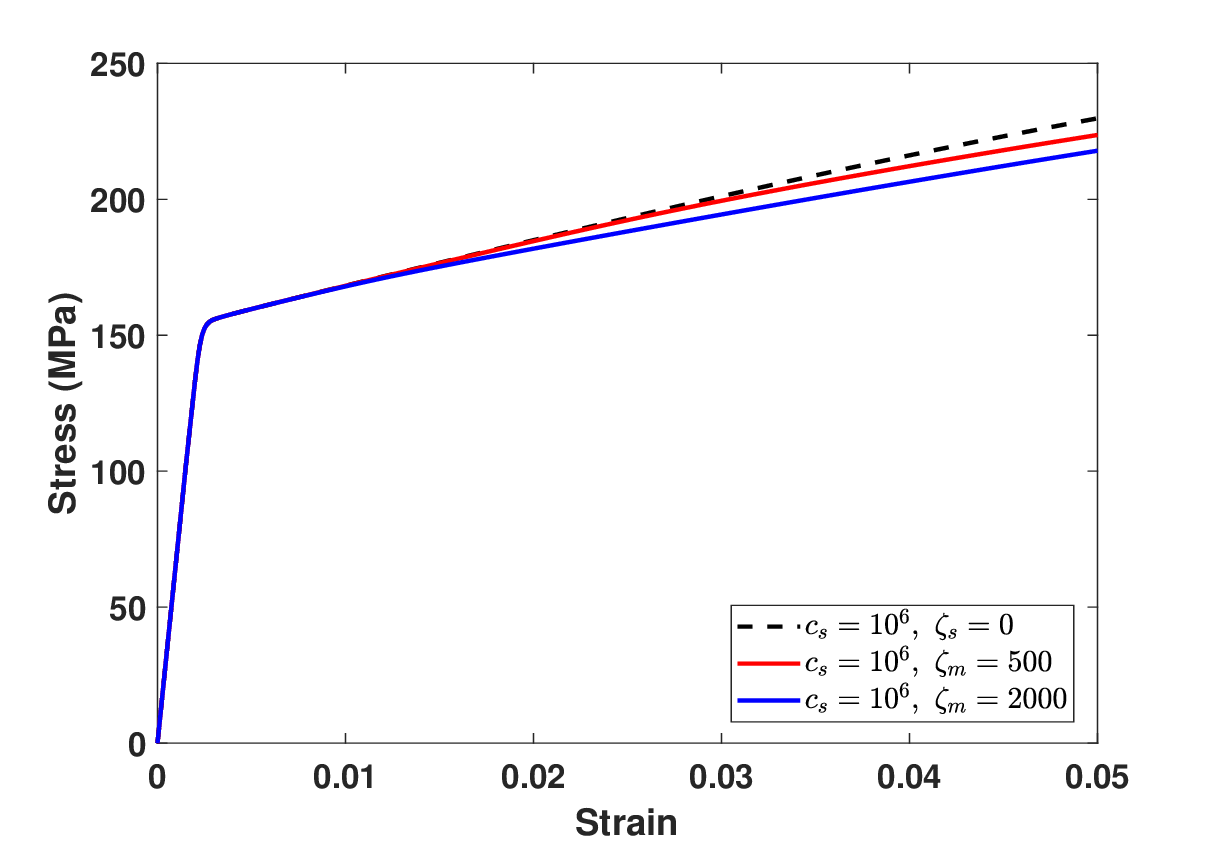}
			\caption{Tensile stress }
			\label{zeta_effect_tension_fd}
		\end{subfigure} 
		\begin{subfigure}{0.45\textwidth}
			\includegraphics[width=\textwidth]{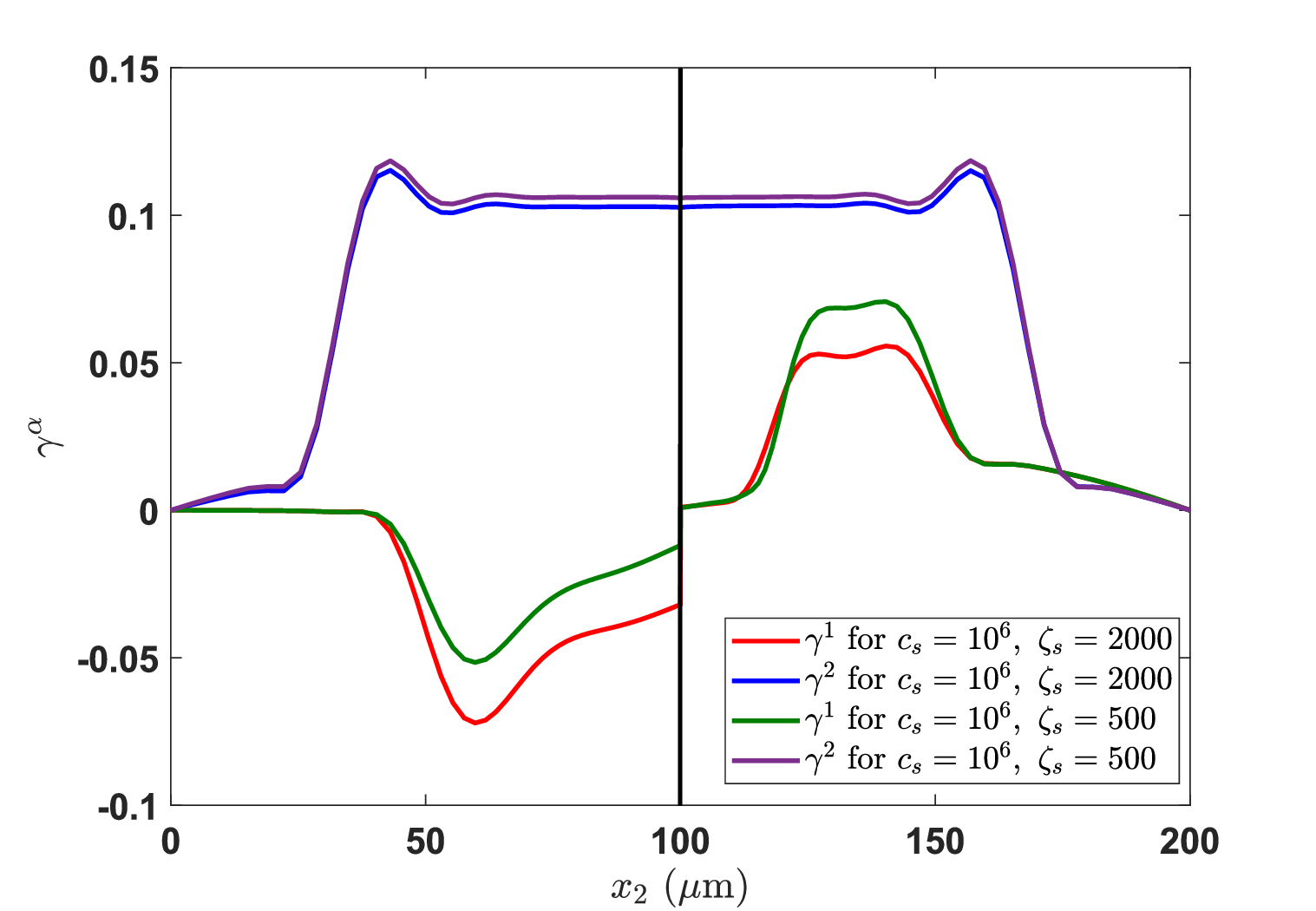}
			\caption{$\gamma^\alpha$ distribution along $x_2=25$ $\mu$m at $5\%$ strain}
			\label{zetas_eff_slip_dist}
		\end{subfigure} 
		\caption{Effect of $\zeta_s$ on the tensile stress and $\gamma^\alpha$ distribution along central line}
	\end{figure}
	
	\begin{figure}[h!]
		\centering
		\begin{subfigure}{0.75\textwidth}
			\includegraphics[width=\textwidth]{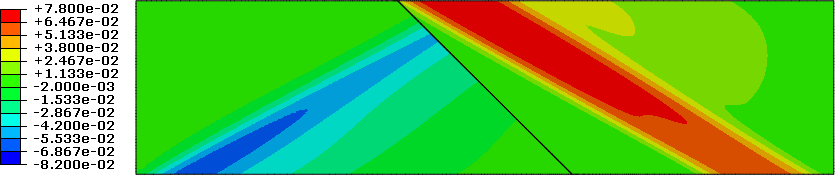}
			\caption{$\gamma^1$ distribution for $\zeta_s=500$ }
			\label{zs_500_g1}
		\end{subfigure} 
		\begin{subfigure}{0.75\textwidth}
			\includegraphics[width=\textwidth]{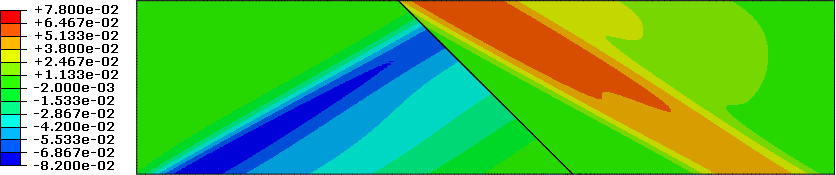}
			\caption{$\gamma^1$ distribution for $\zeta_s=2000$}
			\label{zs_2000_g1}
		\end{subfigure} 
		\begin{subfigure}{0.75\textwidth}
			\includegraphics[width=\textwidth]{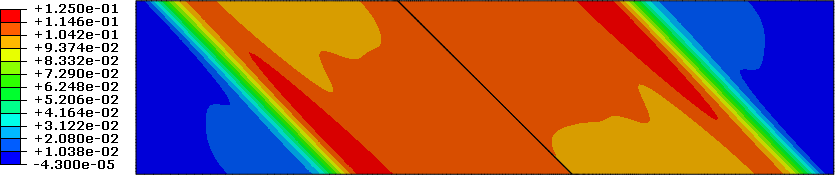}
			\caption{$\gamma^2$ distribution for $\zeta_s=500$}
			\label{zs_500_g2}
		\end{subfigure}
		\begin{subfigure}{0.75\textwidth}
			\includegraphics[width=\textwidth]{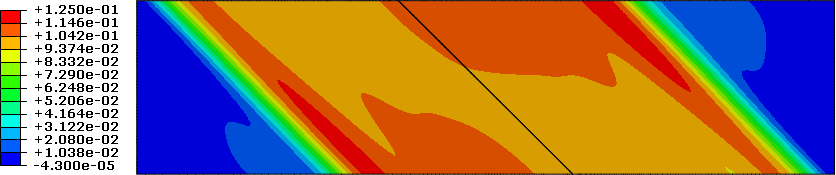}
			\caption{$\gamma^2$ distribution for $\zeta_s=2000$}
			\label{zs_2000_g2}
		\end{subfigure}
		\caption{Effect of $\zeta_s$ on the plastic slip distribution at $5\%$ strain ($c_s=10^6$)}
		\label{contour2}
	\end{figure}
	In order to assess the effect of $\zeta_s$ on the tensile response, we consider $c_s=10^6$ and $\zeta_s$ is varied from zero. The tensile stress-strain curve is plotted in figure \ref{zeta_effect_tension_fd}. In the absence of the $\zeta_s$, the post-yield hardening is linear; however, the $\zeta_s$ brings the nonlinearity to the GB hardening and causes dissipation. This dissipation relaxes some of the hardening to the post-yield portion.  
	
	The plastic slip contour for $\zeta_s=500$ and $\zeta_s=2000$ is presented in figure \ref{contour2}. Plastic slip variation along $x_2=25~\mu$m is plotted in figure \ref{zetas_eff_slip_dist}. The $\zeta_s$ relaxes some of the grain boundary energy; therefore, the plastic slip jump at the grain boundary increases with $\zeta_s$ due to the reduction in grain boundary energy. However, the plastic slip distribution pattern remains almost equal inside the grain. The second slip system, parallel to the grain boundary, does not interact with the boundary. This slip distribution is almost unaltered due to the introduction of $\zeta_s$.
	
	\section{Conclusion}
	\label{conculsion}
	The present study introduces a novel elastic gap-free formulation within the SGCP theory that effectively models dissipation associated to plastic slip gradient.  The theory proposes splitting the plastic slip gradient into two parts: the energetic part, which is associated with free energy, and the dissipative part, which contributes to higher-order dissipation. This decomposition also leads to a partition of GND densities. Higher-order stress quantities are obtained for each slip system by minimizing a dissipative potential employing necessary thermodynamic restrictions.  The resulting higher-order stress is nonlinear and evolves in a manner similar to the Armstrong-Fredrick type backstress equation. Unlike the conventional Gurtin-type model, the proposed model, as indicated by the microforce balance, does not exhibit dissipative hardening. However, it does demonstrate apparent strengthening, characterized by an apparent increase in the flow surface, due to the inclusion of nonlinear kinematic hardening. The model effectively captures plastic dissipation using a dissipative plastic slip gradient for both monotonic and cyclic loading scenarios. Under cyclic loading conditions, the stress-strain curve maintains its curvature during unloading.  In cases of non-proportional loading, no stress jump is observed, and the response asymptotically approaches that of the Gurtin-type dissipative model. Various combinations of bulk hardening and relaxation coefficients can be utilized to saturate the shear response at a specific stress value.
	
	In a manner analogous to the bulk constitutive proposal, the GB Burgers tensor is also separated into energetic and dissipative components. The GB energetic stress follows a nonlinear evolution equation similar to that of the bulk higher order stress. Numerical exploration with the bicrystal shear problem shows that the bulk relaxation coefficient drives the GND towards the boundary, leading to an increased GND pileup near the boundary. Conversely, the GB relaxation coefficient alleviates the GB energy, releasing some GND and thereby controlling the maximum values of the GND pile-up. The effectiveness of the proposed GB model is further demonstrated using a bicrystal tension problem, considering a double slip system with an oblique grain boundary. Finally, limited numerical exploration indicates the necessity of considering realistic microstructure within this framework and warrants additional investigation.
	
	\appendix
	\section{Two dimension Finite element implementation of the SGCP theory}
	\label{2dimplement}
	%
	\begin{figure}[h!]
		\centering
		\includegraphics[width=0.9\textwidth]{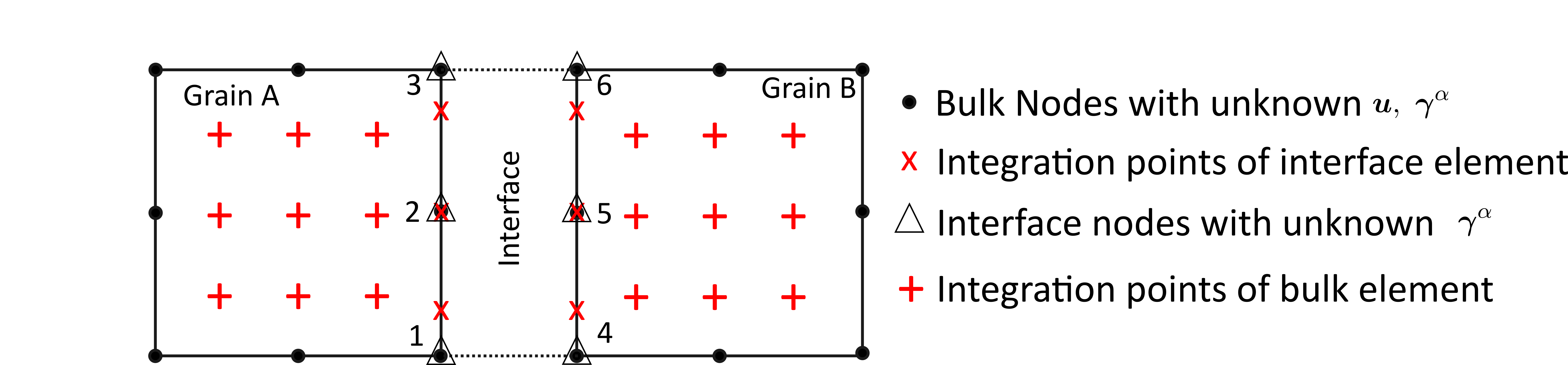}
		\caption{The interface element}
		\label{interface_lmn}
	\end{figure}
	In the finite element implementation, a 8-node serendipity element is used for the bulk part as shown in figure \ref{interface_lmn}. The element contains nodal displacement and plastic slip as primary variables. Suppose if we consider $k$ number of slip system then total unknown number per node is $2+k$ and total unknown per element is $8 \times(2+k)$. Full $3 \times 3$ integration is carried out during numerical integration.  Finite element formulation of the SGCP theory is presented here in matrix vector format. Here $\tilde{()}$ quantities are represented as the vector quantities evaluated at the integration point. $\hat{()}$ quantities are the vector containing information of nodal values, and $\tilde{\tilde{()}}$ quantities are signified for matrix, which is evaluated at the integration point.  
	
	Consider a particular element to form the stiffness matrix and residual vector. The nodal unknowns of displacement corresponding the $i^{th}$ node of the element are $ \hat{u}_1^{(i)},  \hat{u}_2^{(i)}$, and nodal unknowns corresponding to the plastic slip considering $k$ number of plastic slip are $ \gamma^{1,(i)},  \gamma ^{2,(i)}, ...  \gamma ^{k,(i)}$. Suppose $ \hat{\mathbf{u}}$ contain all the nodal unknown displacement of an element such that 
	\begin{equation}
		\hat{\mathbf{u}}= 
		\begin{bmatrix}
			\hat{u}_1^{(1)} &  \hat{u}_2^{(1)} & ... &  \hat{u}_2^{(8)} &  \hat{u}_2^{(8)}
		\end{bmatrix}^T
	\end{equation} 
	Vector $ \hat{\boldsymbol{\gamma}}$ contains all the nodal unknown slip of the element; that is, 
	\begin{equation}
		\hat{\boldsymbol{\gamma}}= \begin{bmatrix}
			\gamma^{1,(1)} &  \gamma^{2,(1)} & ... &  \gamma^{k,(1)} & ... &  \gamma^{1,(8)} &  \gamma^{2,(8)} & ... &  \gamma^{k,(8) }
		\end{bmatrix} ^T
	\end{equation}
	Displacement ($ \tilde{\mathbf{u}}$) and plastic slip ($ \tilde{\boldsymbol{\gamma}}$) of an integration point can be found from interpolation of the nodal variables. 
	Displacement and the plastic slip of the  Gauss point can be written as 
	\begin{equation}
		\tilde{\mathbf{u}} = \tilde{\tilde{\mathbf{N}}}_u  \hat{\mathbf{u}}, \qquad \text{and} \qquad
		\tilde{\boldsymbol{\gamma}} = \tilde{\tilde{\mathbf{N}}}_\gamma  \hat{\boldsymbol{\gamma}},
	\end{equation}
	where the matrix $\tilde{\tilde{\mathbf{N}}}_u$ and $\tilde{\tilde{\mathbf{N}}}_\gamma$ are the interpolation matrices for nodal displacement and plastic slip respectively. 
	The displacement gradient and plastic slip gradient can be derived from nodal displacement and plastic slip, with help of strain-displacement matrix $\tilde{\tilde{\mathbf{B}}}_u$ and slip-gradient-plastic-slip matrix  $\tilde{\tilde{\mathbf{B}}}_\gamma$ as follows 
	\begin{equation}
		\nabla \tilde{\mathbf{u}}= \tilde{\tilde{\mathbf{B}}}_u \hat{\mathbf{u}}, \qquad \text{and } \qquad \tilde{\boldsymbol{\kappa}} =  \tilde{\tilde{\mathbf{B}}}_\gamma  \hat{\boldsymbol{\gamma}},
	\end{equation}
	Now, the internal virtual power equation for bulk neglecting the GB for a particular element can be written using equation (\ref{w_int_gb}) as 
	\begin{equation}
		\label{virtual_work_mat}
		\delta \dot{W}^{int}_e = \int_{\Omega^e} \tilde{\boldsymbol{\sigma}}^T \tilde{\tilde{\mathbf{B}}}_u \delta \hat{\dot{\mathbf{u}}}~ d\Omega^e+ \int_{\Omega^e}\left[[\tilde{\boldsymbol{\pi}} -\tilde{\boldsymbol{\tau}}]^T \tilde{\tilde{\mathbf{N}}}_\gamma  + \tilde{\boldsymbol{\xi}}^T  \tilde{\tilde{\mathbf{B}}}_\gamma \right] \delta\hat{\dot{\boldsymbol{\gamma}}}~ d\Omega^e
	\end{equation}
	In the last equation $\tilde{\boldsymbol{\pi}}$, $\tilde{\boldsymbol{\tau}}$  contain scalar micro stress and resolved stress of all slip planes; that is 
	\begin{equation}
		\tilde{\boldsymbol{\pi}} = \begin{bmatrix}
			\pi^1 & \pi^2 & ... & \pi^k
		\end{bmatrix}^T, \qquad \text{and} \qquad  \tilde{\boldsymbol{\tau}} = \begin{bmatrix}
			\tau^1 & \tau^2 & ... & \tau^k
		\end{bmatrix}^T
	\end{equation}
	$	\tilde{\boldsymbol{\xi}} $ contains the vector micro stress of all slip planes; that is
	\begin{equation}
		\tilde{\boldsymbol{\xi}} = \begin{bmatrix}
			\boldsymbol{\xi}^1 & \boldsymbol{\xi}^2 & ... & \boldsymbol{\xi}^k
		\end{bmatrix}^T
	\end{equation}
	The internal force can be found by taking partial derivative of  the virtual work equation (\ref {virtual_work_mat}) with respect to the degrees of freedom as
	\begin{subequations}
		\begin{align}
			\tilde{\mathbf{f}}^{int}_u &= \frac{\partial (\delta \dot{W}^{int}_e)}{\partial (\delta \hat{\dot{\mathbf{u}}}) }= \int_{\Omega^e}  \tilde{\tilde{\mathbf{B}}}_u^T \tilde{\boldsymbol{\sigma}} ~d\Omega^e \\ 
			\tilde{\mathbf{f}}^{int}_\gamma &= \frac{\partial(\delta \dot{W}^{int}_e)}{\partial( \delta \hat{\dot{\boldsymbol{\gamma}}})} = \int_{\Omega^e} \left[\tilde{\tilde{\mathbf{N}}}_\gamma^T [\tilde{\boldsymbol{\pi}} - \tilde{\boldsymbol{\tau}}] + \tilde{\tilde{\mathbf{B}}}_\gamma^T\tilde{\boldsymbol{\xi}}\right] ~d\Omega^e \\ 
			\tilde{\mathbf{f}}^{int} &= \begin{bmatrix}
				\tilde{\mathbf{f}}_u^{int} &  \tilde{\mathbf{f}}_\gamma^{int} \label{f_bulk}
			\end{bmatrix}^T
		\end{align} 
	\end{subequations} 
	The nodal unknown degrees of freedom are evaluated by minimizing the residual between internal and external force vector with help of Newton-Raphson method. The residual vector can be written as 
	\begin{equation}
		\tilde{\mathbf{r}}(\tilde{\mathbf{d}}) = \tilde{\mathbf{f}}^{int}(\tilde{\mathbf{d}}) - \tilde{\mathbf{f}}^{ext}. 
	\end{equation} 
	In the last equation $\tilde{\mathbf{d}}$ contain the incremental generalized displacement quantity of the nodes; that is  $\tilde{\mathbf{d}}= \begin{bmatrix}
		\Delta \hat{\mathbf{u}}  & \Delta \hat{\boldsymbol{\gamma}}
	\end{bmatrix}^T$. The increment of $\tilde{\mathbf{d}}$ can be found by evaluating the consistent stiffness matrix $\tilde{\tilde{\mathbf{K}}}$ of the element as 
	\begin{equation}\label{k_bulk}
		\Delta \tilde{\mathbf{d}} =\tilde{\tilde{\mathbf{K}}}^{-1} \tilde{\mathbf{r}}(\tilde{\mathbf{d}}) \qquad \text{with} \qquad \tilde{\tilde{\mathbf{K}}}=\begin{bmatrix}
			\frac{\partial \tilde{\mathbf{f}}^{int}_u}{\partial \Delta \hat{\mathbf{u}}} & \frac{\partial \tilde{\mathbf{f}}^{int}_u}{\partial \Delta \hat{\boldsymbol{\gamma}}}\\
			\frac{\partial \tilde{\mathbf{f}}^{int}_\gamma}{\partial \Delta \hat{\mathbf{u}}} & \frac{\partial \tilde{\mathbf{f}}^{int}_\gamma}{\partial \Delta \hat{\boldsymbol{\gamma}}}
		\end{bmatrix} = \begin{bmatrix}
			\tilde{\tilde{\mathbf{K}}}_{ u u} & \tilde{\tilde{\mathbf{K}}}_{ u  \gamma} \\
			\tilde{\tilde{\mathbf{K}}}_{ \gamma  u} & \tilde{\tilde{\mathbf{K}}}_{ \gamma \gamma }
		\end{bmatrix}. 
	\end{equation}
	Now, the $\tilde{\mathbf{f}}_u$ and   $\tilde{\mathbf{f}}_\gamma$ can be arranged as the function of $\tilde{\mathbf{d}}$ as follows:
	\begin{subequations}
		\begin{align}
			\tilde{\mathbf{f}}_u^{int}&= \int_{\Omega^e}  \tilde{\tilde{\mathbf{B}}}_u^T \tilde{\boldsymbol{\sigma}} ~d\Omega^e = \int_{\Omega^e}\tilde{\tilde{\mathbf{B}}}_u^T \tilde{\tilde{\mathbf{C}}}[\nabla \tilde{\mathbf{u}} - \tilde{\tilde{\mathbf{T}}} \tilde{\boldsymbol{\gamma}}] ~ d\Omega^e =  \int_{\Omega^e}\tilde{\tilde{\mathbf{B}}}_u^T \tilde{\tilde{\mathbf{C}}}[\tilde{\tilde{\mathbf{B}}}_u \hat{\mathbf{u}}-\tilde{\tilde{\mathbf{T}}} \tilde{\tilde{\mathbf{N}}}_\gamma \hat{\boldsymbol{\gamma}}]~ d\Omega^e\\
			\nonumber
			\tilde{\mathbf{f}}_\gamma^{int}&=  \int_{\Omega^e} \left[\tilde{\tilde{\mathbf{N}}}_\gamma^T [\tilde{\boldsymbol{\pi}} - \tilde{\boldsymbol{\tau}}] + \tilde{\tilde{\mathbf{B}}}_\gamma^T\tilde{\boldsymbol{\xi}}\right] ~d\Omega^e = \int_{\Omega^e}\left[\tilde{\tilde{\mathbf{N}}}_\gamma^T \tilde{\boldsymbol{\pi}} + \tilde{\tilde{\mathbf{B}}}^T_\gamma \tilde{\boldsymbol{\xi}}\right] ~d\Omega^e -\int_{\Omega^e}\tilde{\tilde{\mathbf{N}}}_\gamma^T \tilde{\tilde{\mathbf{T}}}^T \tilde{\tilde{\mathbf{C}}}[\nabla \tilde{\mathbf{u}} - \tilde{\tilde{\mathbf{T}}} \tilde{\boldsymbol{\gamma}}]~d\Omega^e\\
			&= \int_{\Omega^e}\tilde{\tilde{\mathbf{N}}}_\gamma^T\tilde{\boldsymbol{\pi}} ~d\Omega^e - \int_{\Omega^e}\tilde{\tilde{\mathbf{N}}}_\gamma^T \tilde{\tilde{\mathbf{T}}}^T \tilde{\tilde{\mathbf{C}}}\tilde{\tilde{\mathbf{B}}}_u \hat{\mathbf{u}} ~d\Omega^e+ \int_{\Omega^e} \tilde{\tilde{\mathbf{N}}}_\gamma^T \tilde{\tilde{\mathbf{T}}}^T \tilde{\tilde{\mathbf{C}}} \tilde{\tilde{\mathbf{T}}} \tilde{\tilde{\mathbf{N}}}_\gamma \hat{\boldsymbol{\gamma}} ~d\Omega^e + \int_{\Omega^e}\tilde{\tilde{\mathbf{B}}}_\gamma^T \tilde{\boldsymbol{\xi}} ~d\Omega^e
		\end{align}
	\end{subequations}
	In the last equations $\tilde{\tilde{\mathbf{T}}}$ is the matrix containing the symmetrized Schmid tensor of all slip system.
	Now, the components of consistent stiffness matrix can be calculated as 
	\begin{subequations}
		\begin{align}
			\tilde{\tilde{\mathbf{K}}}_{uu} &= \int_{\Omega^e} \tilde{\tilde{\mathbf{B}}}^T_u \tilde{\tilde{\mathbf{C}}} \tilde{\tilde{\mathbf{B}}}_u ~d\Omega^e \qquad  
			\tilde{\tilde{\mathbf{K}}}_{u\gamma} = - \int_{\Omega^e} \tilde{\tilde{\mathbf{B}}}_u^T \tilde{\tilde{\mathbf{C}}} \tilde{\tilde{\mathbf{T}}} \tilde{\tilde{\mathbf{N}}}_\gamma ~d\Omega^e \\
			\tilde{\tilde{\mathbf{K}}}_{\gamma\gamma} &= \int_{\Omega^e}\left[\tilde{\tilde{\mathbf{N}}}_\gamma^T  \tilde{\tilde{\mathbf{T}}}^T \tilde{\tilde{\mathbf{C}}} \tilde{\tilde{\mathbf{T}}} \tilde{\tilde{\mathbf{N}}}_\gamma + \tilde{\tilde{\mathbf{N}}}_\gamma^T \frac{\partial \tilde{\boldsymbol{\pi}}}{\partial \Delta \hat{\boldsymbol{\gamma}}} + \tilde{\tilde{\mathbf{B}}}_\gamma^T \frac{\partial \tilde{\boldsymbol{\xi}}}{\partial \Delta \hat{\boldsymbol{\gamma}}}\right]~ d\Omega^e
		\end{align}
	\end{subequations} 
	\subsection{Calculation of matrix $\frac{\partial \tilde{\boldsymbol{\xi}}}{\partial \Delta  \hat{\boldsymbol{\gamma}}}$}
	Derivative of vector microscopic stress with respect to nodal incremental plastic slip can be calculated as follows:
	\begin{equation}
		\frac{\partial \tilde{\boldsymbol{\xi}}}{\partial \Delta \hat{\boldsymbol{\gamma}}}= \left[\frac{\partial \tilde{\boldsymbol{\xi}}}{\partial \Delta \tilde{\boldsymbol{\gamma}}}\right]\left[\frac{\partial \Delta \tilde{\boldsymbol{\gamma}}}{\partial \Delta \hat{\boldsymbol{\gamma}}}\right] +  \left[\frac{\partial \tilde{\boldsymbol{\xi}}}{\partial \Delta \tilde{\boldsymbol{\kappa}}}\right]\left[\frac{\partial\Delta \tilde{\boldsymbol{\kappa}}}{\partial \Delta \hat{\boldsymbol{\gamma}}}\right]= \left[\frac{\partial \tilde{\boldsymbol{\xi}}}{\partial \Delta \tilde{\boldsymbol{\gamma}}}\right] \tilde{\tilde{\mathbf{N}}}_\gamma +  \left[\frac{\partial \tilde{\boldsymbol{\xi}}}{\partial \Delta \tilde{\boldsymbol{\kappa}}}\right]  \tilde{\tilde{\mathbf{B}}}_\gamma
	\end{equation}
	\noindent
	Elements of $[{\partial \tilde{\boldsymbol{\xi}}}/{\partial \Delta \tilde{\boldsymbol{\gamma}}}]$ matrix can be derived  with the help of equation (\ref{dxi_dgamma}) as follows:
	\begin{eqnarray}
		\frac{\partial \boldsymbol{\xi}^\alpha_{n+1}}{\partial \Delta\gamma^\beta} = -\frac{\zeta}{1+ \zeta \Delta\bar{d}^\alpha} \frac{\partial \Delta \bar{d}^\alpha}{\partial \Delta\gamma^\beta} \boldsymbol{\xi}^\alpha_{n+1}=-\frac{\zeta}{1+ \zeta \Delta\bar{d}^\alpha} \frac{ \Delta {\gamma}^\alpha}{ |\Delta\gamma^\beta|} \boldsymbol{\xi}^\alpha_{n+1} \delta_{\alpha \beta},
	\end{eqnarray} 
	where $\delta_{\alpha\beta}$ is the Dirac-delta function. 
	Elements of  $[{\partial \tilde{\boldsymbol{\xi}}^\alpha}/{\partial \Delta  \tilde{\boldsymbol{\kappa}}^{t\beta}}]$ matrix can be derived with the help of equation (\ref{dxi_dkappa})
	\begin{equation}
		\frac{\partial \boldsymbol{\xi}^\alpha_{n+1}}{\partial \Delta \boldsymbol{\kappa}^\beta} = \frac{S_0L_*^2}{1+ \zeta \Delta \bar{d}^\alpha} \frac{\partial\Delta \boldsymbol{\kappa}^{t\alpha}}{\partial \Delta \boldsymbol{\kappa}^\beta} = \frac{S_0 L_*^2}{1+\zeta \Delta \bar{d}^\alpha} (\boldsymbol{s}^\alpha \otimes \boldsymbol{s}^\alpha) \delta_{\alpha\beta}
	\end{equation}   
	\subsection{calculation of ${\partial \tilde{\boldsymbol{\pi}}}/{\partial \Delta \hat{\boldsymbol{\gamma}}}$}
	Derivative of the scalar microscopic stress with respect to the incremental nodal plastic slip can be derived as follows:
	\begin{equation}
		\frac{\partial \tilde{\boldsymbol{\pi}}}{\partial \Delta \hat{\boldsymbol{\gamma}}} = \left[\frac{\partial \tilde{\boldsymbol{\pi}}}{\partial \Delta \tilde{\boldsymbol{\gamma}}}\right]\left[\frac{\partial \Delta \tilde{\boldsymbol{\gamma}}}{\partial \Delta \hat{\boldsymbol{\gamma}}}\right] = \left[\frac{\partial \tilde{\boldsymbol{\pi}}}{\partial \Delta \tilde{\boldsymbol{\gamma}}}\right] \tilde{\tilde{\mathbf{N}}}_\gamma 
	\end{equation}
	
	\noindent Calculation of $\left[\frac{\partial \tilde{\boldsymbol{\pi}}}{\partial \Delta \tilde{\boldsymbol{\gamma}}}\right]$: 
	%
	The microscopic  stress  stress quantity, equation (\ref{microstr}), can be written with help of incremental quantities as  follows
	\begin{equation}
		\pi^\alpha =  S^\alpha R(\dot{\bar{d}}^\alpha) \frac{\Delta \gamma^\alpha}{\Delta \bar{d} ^\alpha} = S^\alpha R(\dot{\bar{d}}^\alpha) \frac{\Delta\gamma^\alpha}{|\Delta \gamma^\alpha|}.
	\end{equation} 
	Partial derivative of the scalar microscopic stress with respected to plastic slip increment can be computed as 
	\begin{eqnarray}
		\label{dpdg}
		\frac{\partial \pi^\alpha}{\partial \Delta\gamma^\beta}
		=\frac{\Delta{\gamma}^\alpha}{|\Delta {\gamma}^\alpha|}\left[ R(\dot{\bar{d}}^\alpha)  \frac{\partial S^\alpha}{\partial \Delta \gamma^\beta} + S^\alpha  \frac{\partial R(\dot{\bar{d}}^\alpha)}{\partial \Delta \gamma^\beta}\right]
	\end{eqnarray}

	\subsubsection{Calculation of different derivatives used in equation (\ref{dpdg})}
%
		\noindent 
		\subsubsubsection{Derivatives of $R(\dot{\bar{d}}^\alpha)$}
		%
		Here we opt for the rate-sensitive function according to \cite{fuentes2020fracture} as follows: 
		\begin{equation}
			R(\dot{\bar{d}}^\alpha) = \begin{cases}
				\frac{\dot{\bar{d}}^\alpha}{\omega \dot{d}_0} \qquad \qquad ~~~\text{if} \qquad \dot{\bar{d}}^\alpha \leqslant \dot{d}^*\\
				\left(\frac{\dot{\bar{d}}^\alpha - \Theta}{\dot{d}_0}\right)^m \qquad \text{if} \qquad \dot{\bar{d}}^\alpha > \dot{d}^*
			\end{cases}
		\end{equation}
		The parameter $\omega$ is small positive constant ($\omega = 0.01$). Threshold value of the slip rate ($\dot{d}^*$) and $\Theta$ are obtained from the smooth transition of $R(\dot{\bar{d}}^\alpha)$ vs $\dot{\bar{d}}^\alpha$ curve at the critical $\dot{d}^*$. Following relation can be established by equating $R(\dot{\bar{d}}^\alpha)$ and its derivative at $\dot{d}^*$
		\begin{equation}
			\label{para}
			\dot{d}^* = \frac{\dot{d}_0}{m}\left(\frac{1}{\omega m}\right)^{\frac{1}{m-1}}  \qquad \text{and} \qquad
			\Theta = \dot{d}^*(1-m)
		\end{equation}
		Partial derivative of the rate-sensitive function with respect to  effective plastic slip rate can be written as
		\begin{equation}
			\frac{\partial R(\dot{\bar{d}}^\alpha)}{\partial \dot{\bar{d}}^\beta} = \begin{cases}
				\frac{1}{\omega \dot{d}_0} \delta_{\alpha\beta} \qquad \qquad  \qquad \qquad ~~\text{if} \qquad \Delta {\bar{d}}^\alpha \leqslant \dot{d}^* \Delta t,\\
				\frac{m}{\dot{d}_0}\left(\frac{\Delta{\bar{d}}^\alpha - \Theta \Delta t}{  \dot{d}_0 \Delta t}\right)^{m-1} \delta_{\alpha\beta} \qquad \text{if} \qquad \Delta{\bar{d}}^\alpha > \dot{d}^* \Delta t.
			\end{cases}
		\end{equation} 
		\noindent
		Partial derivative of rate-sensitive function $R(\dot{d}^\alpha)$ with respect to plastic slip increment can be written as 
			\begin{equation}
				\frac{\partial R(\dot{\bar{d}}^\alpha)}{\partial \Delta \gamma^\beta} = \frac{\partial R(\dot{\bar{d}}^\alpha)}{\partial \dot{\bar{d}}^\beta} \frac{\partial \dot{\bar{d}}^\beta}{\partial \Delta \gamma^\beta}= \frac{\partial R(\dot{\bar{d}}^\alpha)}{\partial \dot{\bar{d}}^\beta} \frac{\Delta \gamma^\beta}{|\Delta \gamma^\beta|} \frac{1}{\Delta t} 
			\end{equation}
		\subsubsubsection{Derivatives of $S^\alpha$}
		
		$\boldsymbol{S} = (S^1, S^2, ..., S^m)$ is the slip resistance, which is strictly positive stress-dimensioned quantity. Slip resistance of $\alpha$ slip plane  is governed by the following hardening equation 
		\begin{equation}
			\dot{S}^\alpha=\sum_\beta h^{\alpha \beta}(\boldsymbol{S}) (\dot{\bar{d}}^\beta), \qquad S^\alpha(0)=S_0>0
		\end{equation}
		The hardening moduli $h^{\alpha\beta}$ is presumed to have the following form according to \cite{gurtin2007gradient,ASARO1985923}
		\begin{equation}
			h^{\alpha\beta} (S^\beta)= \underbrace{s^{\alpha \beta} h(S^\beta)}_{\text{self hardening}} + \underbrace{(1-s^{\alpha\beta}) q h(S^\beta)} _{\text{latent hardening}},
		\end{equation}
		where $q > 0$ is the interaction constant, and   $s^{\alpha \beta}$ is defined by 
		\begin{equation}
			s^{\alpha\beta} = \begin{cases}
				1  \qquad\text{for coplaner slip system ($\boldsymbol{m}^\alpha \times \boldsymbol{m}^\beta = \boldsymbol{0}$)}
				\\0 \qquad \text{otherwise}
			\end{cases}
		\end{equation}
		Partial derivative of  slip resistance with respect to plastic slip and its gradient can be derived as
			\begin{equation}
				\frac{\partial S^\alpha}{\partial \Delta\gamma^\beta} = \frac{\partial}{\partial \Delta\gamma^\beta}(S^\alpha + \Delta S^\alpha) = \frac{\partial \Delta S^\alpha}{\partial \Delta\gamma^\beta}=\frac{\partial}{\partial \Delta\gamma^\beta} \left(\sum_\beta h^{\alpha \beta} \Delta \bar{d}^\beta\right)  = h^{\alpha \beta} \frac{\Delta \gamma^\beta}{|\Delta \gamma^\beta|}
			\end{equation}
		
		\subsection{Finite element implementation of the grain boundary model}
		\label{gb_model}
		In order to model the grain boundary within finite element framework, zero thickness element similar to \cite{ozdemir2014modeling} is used. The interface element is shown in figure  \ref{interface_lmn}. The interface element consists six number of nodes. The left and right sides nodes are actually overlapping; however, for clarity they are shown to be separate. In order to incorporate the zero slip condition, the left and right nodes are constrained such that they have equal amount of displacement  ($\boldsymbol{u}$). Plastic slip of the any sides can be interpolates from the interpolation of slip corresponding to that side.
		Full three-point quadrature is used during Gauss-integration.  
		
		Consider a particular interface element to form the element stiffness matrix and force vector. Suppose the vector $\hat{\boldsymbol{\gamma}}_s^\alpha$ contains all the nodal unknown plastic slips of $k$ number of slip plane as   
		\begin{equation}
			\hat{\boldsymbol{\gamma}}_s = \begin{bmatrix}
				\gamma^{1,(1)} &  \gamma^{2,(1)} & ... &  \gamma^{k,(1)} & ... &  \gamma^{1,(6)} &  \gamma^{2,(6)} & ... &  \gamma^{k,(6) }
			\end{bmatrix} ^T
		\end{equation}
		It to note that the plastic slip $\hat{\gamma}^{\alpha}_1,~\hat{\gamma}^{\alpha}_2, ~\hat{\gamma}^{\alpha}_3$ corresponds to grain A and $\hat{\gamma}^{\alpha}_4$, $\hat{\gamma}^{\alpha}_5$, $\hat{\gamma}^{\alpha}_6$ corresponds to grain B. Plastic slip at any point on interface toward grain A, and grain B side  can be calculated from the nodal interpolation as 
		\begin{equation}
			\tilde{\boldsymbol{\gamma}}_A  =\tilde{\tilde{\mathbf{N}}}_{A} \hat{\boldsymbol{\gamma}}_s, \qquad \text{and} \qquad \tilde{\boldsymbol{\gamma}}_B  =\tilde{\tilde{\mathbf{N}}}_{B} \hat{\boldsymbol{\gamma}}_s,
		\end{equation}
		where
		$\tilde{\tilde{\mathbf{N}}}_{A}, ~\tilde{\tilde{\mathbf{N}}}_{B}$ are the interpolation matrices.  
		Contribution of grain boundary to the internal power for a particular element can be written using equation as (\ref{internal_virt_gb})  
		\begin{equation}
			\delta\dot{W}_{s,e}^{int} = \int_{\mathcal{G}^e}( \tilde{\Pi}_B^T \tilde{\tilde{\mathbf{N}}}_{B}  - \tilde{\Pi}^T_A \tilde{\tilde{\mathbf{N}}}_{A} )\delta\hat{\boldsymbol{\gamma}}_s ~ d\mathcal{G}^e
		\end{equation} 
		The $\tilde{\Pi}_A$, $\tilde{\Pi}_B$ vectors are the energetic stress at the boundary for all $k$ number of slip system 
		\begin{equation}
			\tilde{\Pi}_A = [\Pi_A^1, ~\Pi_A^2, ~ ... , ~ \Pi_A^k]^T, \qquad \text{and} \qquad 	\tilde{\Pi}_B = [\Pi_B^1, ~\Pi_B^2, ~ ... , ~ \Pi_B^k]^T
		\end{equation}
		The internal force can be found form the internal virtual power by taking partial derivative about the nodal unknowns as 
		\begin{equation}\label{f_gb}
			\tilde{\mathbf{f}}^{int}_s = \frac{\partial (\delta  \dot{W}^{int}_{s,e})}{\partial (\delta\hat{\boldsymbol{\gamma}}_s)} =  \int_{\mathcal{G}^e}( \tilde{\tilde{\mathbf{N}}}_{B}^T \tilde{\Pi}_B   - \tilde{\tilde{\mathbf{N}}}_{A}^T \tilde{\Pi}_A  ) ~ d\mathcal{G}^e
		\end{equation}
		The consistent stiffness matrix can be found taking partial derivative of the internal force about nodal unknowns as 
		\begin{eqnarray}\label{k_gb}
			\nonumber
			\tilde{\tilde{\mathbf{K}}}_{s} = \frac{\partial \tilde{\mathbf{f}}^{int}_s}{\partial \Delta\hat{\boldsymbol{\gamma}}_s} = \int_{\mathcal{G}^e}\left( \tilde{\tilde{\mathbf{N}}}_{B}^T \frac{\partial \tilde{\Pi}_B}{\partial \Delta\hat{\boldsymbol{\gamma}}_s}   - \tilde{\tilde{\mathbf{N}}}_{A}^T \frac{\partial \tilde{\Pi}_A}{\partial \Delta\hat{\boldsymbol{\gamma}}_s}  \right) ~ d\mathcal{G}^e \\
			=\int_{\mathcal{G}^e} \left(\tilde{\tilde{\mathbf{N}}}^T_B \frac{\partial\tilde{\Pi}_B}{\partial \Delta\tilde{\boldsymbol{\gamma}}_A} \tilde{\tilde{\mathbf{N}}}_A + \tilde{\tilde{\mathbf{N}}}^T_B \frac{\partial \tilde{\Pi}_B}{\partial \Delta \tilde{\boldsymbol{\gamma}}_B} \tilde{\tilde{\mathbf{N}}}_B - \tilde{\tilde{\mathbf{N}}}^T_A \frac{\partial\tilde{\Pi}_A}{\partial \Delta\tilde{\boldsymbol{\gamma}}_A} \tilde{\tilde{\mathbf{N}}}_A - \tilde{\tilde{\mathbf{N}}}^T_A \frac{\partial\tilde{\Pi}_A}{\partial \Delta\tilde{\boldsymbol{\gamma}}_B} \tilde{\tilde{\mathbf{N}}}_B\right) ~d\mathcal{G}^e.
		\end{eqnarray}
		It is worth to note that the bulk stiffness matrix $\tilde{\tilde{\mathbf{K}}}$ (equation \ref{k_bulk}) and GB stiffness matrix $\tilde{\tilde{\mathbf{K}}}_s$ (equation \ref{k_gb}) should be added during the global stiffness matrix formation. Similarly, internal force of bulk $\tilde{\mathbf{f}}^{int}$ (equation \ref{f_bulk}) and GB internal force $\tilde{\mathbf{f}}^{int}_s$ (equation \ref{f_gb}) should be added to prepare the global internal force vector. The material tangent matrix $\partial \tilde{\Pi}_A/\partial\Delta\tilde{\boldsymbol{\gamma}}_A, ~ \partial \tilde{\Pi}_A/\partial\Delta\tilde{\boldsymbol{\gamma}}_B,~\partial \tilde{\Pi}_B/\partial\Delta\tilde{\boldsymbol{\gamma}}_A, ~ \partial \tilde{\Pi}_B/\partial\Delta\tilde{\boldsymbol{\gamma}}_B$ can be calculated as follows:
		\begin{eqnarray}
			\frac{\partial {\Pi}_I^\alpha}{\partial \Delta{\boldsymbol{\gamma}}_J^\beta} = \frac{\partial \boldsymbol{M}}{\partial \Delta{\boldsymbol{\gamma}}^\beta_J}: \boldsymbol{N}^\alpha_I, \qquad \forall ~ I=A,B ~\text{and} ~\text{J}=A,B.
		\end{eqnarray}
		Derivative of energetic stress with respect to plastic slip increment can be calculated using equation (\ref{moment_deriv}) as follows: 
		\begin{equation}
			\frac{\partial \boldsymbol{M}}{\partial \Delta {\gamma}^\beta_J} = \frac{1}{1+ \zeta_{s} |\Delta \boldsymbol{G}_s|}\left(c_s \underbrace{\frac{\partial\Delta\boldsymbol{G}_s}{\partial \Delta {\gamma}^\beta _J}}_{\mathbf{I}} - \zeta_{s} \boldsymbol{M} \underbrace{\frac{\partial |\Delta\boldsymbol{G}_s|}{\partial \Delta{\gamma}^\beta_J}}_{\mathbf{II}}\right)  \qquad \forall ~J = A,B
		\end{equation}
		derivative in $\mathbf{I}$ can be calculated as 
		\begin{equation}
			\frac{\partial\Delta\boldsymbol{G}_s}{\partial \Delta {\gamma}^\beta_A} = - \boldsymbol{N}^\beta_A, \qquad \text{and} \qquad \frac{\partial\Delta\boldsymbol{G}_s}{\partial \Delta {\gamma}^\beta_B} =  \boldsymbol{N}^\beta_B
		\end{equation}
		derivative in $\mathbf{II}$ can be calculated as follows:
		\begin{gather}
			\frac{\partial|\Delta \boldsymbol{G}_s|}{\partial \Delta\gamma^\beta_J} =  \frac{1}{|\Delta \boldsymbol{G}_s|}  \Delta \boldsymbol{G}_s : \frac{\partial \Delta\boldsymbol{G}_s}{\partial \Delta\gamma^\beta_J}  \qquad \forall ~J = A,B
		\end{gather} 

		\bibliographystyle{elsarticle-harv}
		\bibliography{mybib}
		
	\end{document}